\documentclass[twocolumn,secnumarabic,amssymb, nobibnotes, aps, prd]{revtex4-2}

\usepackage{subfig}
\usepackage{graphicx}
\usepackage{amsmath}
\usepackage{caption}
\usepackage{array} 
\newcolumntype{L}[1]{>{\raggedright\let\newline\\\arraybackslash\hspace{0pt}}m{#1}}
\newcolumntype{C}[1]{>{\centering\let\newline\\\arraybackslash\hspace{0pt}}m{#1}}
\newcolumntype{R}[1]{>{\raggedleft\let\newline\\\arraybackslash\hspace{0pt}}m{#1}}
\newcolumntype{N}{@{}m{0pt}@{}}
\usepackage{multirow}
\usepackage{color}
\begin{document}

\title{Modeling and Simulation of Transitional Rayleigh-Taylor Flow with Partially-Averaged Navier-Stokes Equations}%

\author{F.S. Pereira\textsuperscript{1}}\email[F.S. Pereira: ]{fspereira@lanl.gov}
\author{F.F. Grinstein\textsuperscript{2}}
\author{D.M. Israel\textsuperscript{2}}
\author{R. Rauenzahn\textsuperscript{2}}
\author{S.S. Girimaji\textsuperscript{3}}
\affiliation{\textsuperscript{1}Los Alamos National Laboratory, Theoretical Division, Los Alamos, New Mexico 87545, USA}
\affiliation{\textsuperscript{2}Los Alamos National Laboratory, X-Computational Physics Division, Los Alamos, New Mexico 87545, USA}
\affiliation{\textsuperscript{3}Texas A\&M University, Department of Ocean Engineering, College Station, Texas 77843, USA}


\begin{abstract}
The partially-averaged Navier-Stokes (PANS) equations are used to predict the variable-density Rayleigh-Taylor (RT) flow at Atwood number 0.5 and maximum Reynolds number $500$. This is a prototypical problem of material mixing featuring laminar, transitional, and turbulent flow, instabilities and coherent structures, density fluctuations, and production of turbulence kinetic energy by both shear and buoyancy mechanisms. These features pose numerous challenges to modeling and simulation, making the RT flow ideal to develop the validation space of the recently proposed PANS BHR-LEVM (Besnard-Harlow-Rauenzahn-linear eddy viscosity model) closure. The numerical simulations are conducted at different levels of physical resolution and test three approaches to set the parameters $f_\phi$ defining the range of physically resolved scales. The computations demonstrate the efficiency (accuracy vs. cost) of the PANS model predicting the spatio-temporal development of the RT flow. {\color{blue}Results comparable to large-eddy simulations and direct numerical simulations are obtained}, at significantly lower physical resolution without the limitations of the Reynolds-averaged Navier-Stokes equations in these transitional flows. The data also illustrate the importance of appropriate selection of the physical resolution and the resolved fraction of each dependent quantity $\phi$ of the turbulent closure, $f_\phi$. These two aspects determine the ability of the model to resolve the flow phenomena not amenable to modeling by the closure and, as such, the computations' fidelity.
\end{abstract}

\maketitle
\section{Introduction}
\label{sec:1}

The Rayleigh-Taylor (RT) \cite{RAYLEIGH_PLMS_1882,TAYLOR_PRSA_1950} flow and instability are of paramount importance to many engineering problems, e.g., astrophysics \cite{NITTMAN_MNRAS_1982,ISOBE_N_2005,CAPRONI_TAJ_2015}, inertial confinement fusion \cite{FREEMAN_NF_1977,BETTI_POP_1998,SRINIVASAN_PRL_2012}, geophysics \cite{SELIG_G_1966,DEBNATH_BOOK_1994,WUNSCH_ANFM_2004}, combustion \cite{VEYNANTE_JFM_1997,CHERTKOV_JFM_2009,ALMARCHA_PRL_2010}, and material mixing and transport \cite{WHITEHEAD_JGR_1975,LI_POFA_1991,CHOU_POF_2016}. This variable-density flow consists of two fluids of different densities initially separated by a perturbed interface. If the heavier fluid is above the light material, the configuration is unstable \cite{SHARP_PD_1984}.  The density difference is characterized by the Atwood number,
\begin{equation}
\label{eq:1_1}
\mathrm{At} \equiv \frac{\rho_h - \rho_l}{\rho_h + \rho_l} \;,
\end{equation}
where $\rho_h$ and $\rho_l$ are the densities of the heavy and light materials. Upon experiencing a body force (typically gravity), the materials immediately begin mixing, with the heavy material accelerating downward, and, the light material moving upward. The interface perturbations create a misalignment between the density gradient and the pressure fields that generates vorticity.  This is the mechanism for the RT instability, which leads to the formation of coherent structures with a characteristic mushroom-like shape. These vortical structures, named bubbles and spikes, exhibit Kelvin-Helmholtz instabilities along the shear layers on the sides of the structures. They are responsible for the onset and development of turbulence. Once in a turbulent state, the flow experiences larger mixing rates, and the mixing-layer grows rapidly. The development of the RT flow comprises of linear and nonlinear regimes. The first corresponds to early times, when the flow is laminar, and it can be described by linear stability theory \cite{RAYLEIGH_PLMS_1882,TAYLOR_PRSA_1950}, which predicts an exponential growth rate. In contrast, the nonlinear regime can comprise of laminar, transitional, and turbulent flow. For a comprehensive review of the RT instability and flow, the reader is referred to \citeauthor{SHARP_PD_1984} \cite{SHARP_PD_1984}, \citeauthor{KULL_PR_1991} \cite{KULL_PR_1991}, \citeauthor{BOFFETA_ARFM_2017} \cite{BOFFETA_ARFM_2017}, and \citeauthor{ZHOU_PR1_2017} \cite{ZHOU_PR1_2017,ZHOU_PR2_2017}.

The spatio-temporal development of the RT flow poses numerous challenges to modeling and simulation. This stems from its transient and transitional nature, multiple instabilities and coherent structures, correlations between distinct turbulence quantities, density fluctuations, and turbulence kinetic energy produced by shear and buoyancy mechanisms. Direct numerical simulation (DNS), large-eddy simulation (LES) \cite{SMAGORINSKY_MWR_1963}, and implicit LES (ILES) \cite{BORIS_FDR_1992,GRINSTEIN_BOOK_2010} have all been used to predict and investigate the physics of the RT flow. These scale-resolving simulation (SRS) \cite{PEREIRA_ACME_2021} models can accurately predict this class of flows, but incur high simulation costs as significant portion of the turbulence spectrum must be resolved. In contrast, the Reynolds-averaged Navier-Stokes (RANS) equations rely on a fully statistically approach in which all turbulence scales are represented through a constitutive relationship named the turbulence closure. Thus, RANS calculations are expected to be computationally less intensive than SRS models. RANS formulations can calculate mean-flow quantities accurately in fully-developed turbulent flow. Yet, existing RANS closures are inadequate for prediction of high-order statistics, transitional flow, and the complete spatio-temporal development of the RT flow. These are crucial for multiple applications, such as inertial confinement fusion, material mixing, and ocean circulation. For this reason, more efficient formulations are needed to predict the RT and similar types of variable-density flow. This has motivated the alpha-group initiative \cite{DIMONTE_POF_2004} and numerous numerical studies investigating the physics and efficient simulation of these flows {\color{blue}\cite{COOK_JFM_2001,DIMONTE_POF_2004,RISTORCELLI_JFM_2004,CABOT_N_2006,BANERJEE_IJHMT_2009,LIVESCU_JOT_2009,VLADIMIROVA_POF_2009,LIVESCU_PTRSA_2013,YOUNGS_PS_2017,KONNIKAKIS_PRE_2019,ZHOU_POF_2019,HILLIER_POF_2020,HAMZEHLOO_OIF_2021,SCHILLING_POF_2021,XIAO_POF_2021}}.

\begin{figure}[t!]
\centering
\includegraphics[scale=0.24,trim=0 0 0 0,clip]{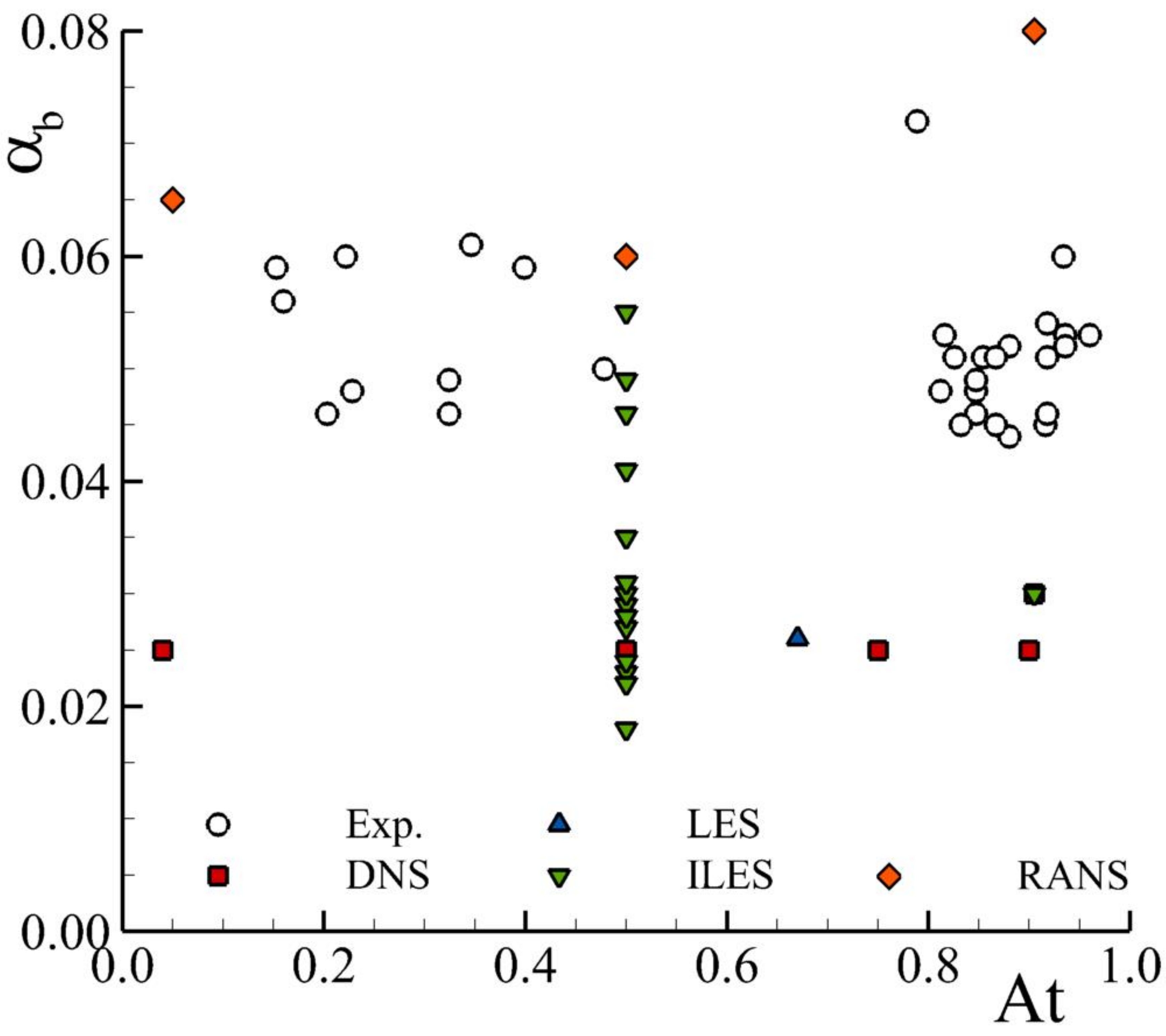}
\caption{Literature survey of experimental \cite{READ_PD_1984,YOUNGS_PD_1989,ANDREWS_POFA_1990,DIMONTE_POF_2000} and numerical \cite{DIMONTE_POF_2004,DIMONTE_POF_2006,BANERJEE_IJHMT_2009,LIVESCU_PS_2010,KOKKINAKIS_IJHFF_2015,YOUNGS_PS_2017,LIVESCU_PD_2021,SID_JFM_2021} results for the averaged growth rate of the mixing-layer half-width, $\alpha_b$, measured at different At. Numerical results presented as a function of the mathematical model.}
\label{fig:1_1}
\end{figure}

To illustrate the importance of these studies, figure \ref{fig:1_1} presents a literature survey of available experimental and numerical results for the growth rate of the lower interface, $\alpha_b$, of the RT mixing-layer. The data are depicted as a function of the At and turbulence model. As observed in previous studies \cite{DIMONTE_POF_2004},  figure \ref{fig:1_1} shows a large variability of experimental and numerical results. This behavior is caused by experimental, input (under-characterized initial and flow conditions), numerical, and modeling uncertainties \cite{ASME_BOOK_2009,PEREIRA_ACME_2021,PEREIRA_JCP_2021}. We emphasize that the RT is highly dependent on the initial flow conditions \cite{DIMONTE_PRE_2004,RAMAPRABHU_JFM_2005,ROZANOV_JRLR_2015}. Nonetheless, it is possible to infer that the experimental measurements lead to values of $\alpha_b$ significantly larger than those predicted by the simulations. Also, the DNS results show a relatively small variability, while ILES a broad range of $\alpha_b$. In addition to the reasons mentioned above, especially initial flow conditions {\color{blue}\cite{GRINSTEIN_CAF_2017}}, we attribute the variability of ILES to modeling assumptions such as inviscid fluid \cite{PEREIRA_PRE_2021} or 1D/2D flow. Overall, the present literature survey demonstrates the complexity involved in the simulation of the RT flow and the importance of the experimental and numerical setup.

The challenges to modeling and simulation of RT type of flows with traditional mathematical models have recently motivated the authors to extend the partially-averaged Navier-Stokes (PANS) equations \cite{GIRIMAJI_JAM_2005,GIRIMAJI_AIAA43_2005,BASARA_IJHFF_2018} to variable-density flow and propose the scale-aware BHR-LEVM (Besnard-Harlow-Rauenzahn-linear eddy viscosity model \cite{BESNARD_TREP_1992,ZARLING_TREP_2011,BANERJEE_PRE_2010,SCHWARZKOPF_JOT_2011}) PANS closure \cite{PEREIRA_PRF_2021,PEREIRA_PRF2_2021}. We emphasize that the PANS model aims to operate at any degree of physical resolution (range of resolved scales or filter cut-off), from RANS to DNS, and strives to only resolve the flow scales not amenable to modeling \cite{PEREIRA_JCP_2018}. The remaining scales can be represented through an appropriate closure. Such a modeling strategy allows for  accuracy-on-demand, and can lead to significant efficiency improvements (accuracy vs. cost) in SRS computations. This work evaluates the accuracy of the PANS BHR-LEVM simulating the spatio-temporal development of the RT flow at $\mathrm{At}=0.5$ and maximum Reynolds number $(\mathrm{Re})_{\max}\approx 500$. Toward this end, we conduct a series of simulations at {\color{blue}five} different physical resolutions using three approaches to set the parameters controlling the physical resolution of the model \cite{PEREIRA_PRF2_2021}. {\color{blue}The importance of these parameters is assessed, and a high-fidelity simulation resolving all flow scales is used as reference to prevent input uncertainties \cite{ASME_BOOK_2009} due to mismatches in initial conditions (see figure \ref{fig:1_1}) and enable the precise evaluation of the accuracy of the PANS BHR-LEVM}. The fidelity of the PANS model is ascertained through the analysis of the mean-flow, coherent, and turbulent flow fields. {\color{blue} In addition, the numerical accuracy of the simulations is monitored through grid refinement studies using four spatio-temporal grid resolutions. These results are summarized in Appendix \ref{sec:B}.} 

The paper is structured as follows. The details of the numerical simulations are described in Section \ref{sec:2}, and Section \ref{sec:3} presents the governing equations of PANS BHR-LEVM and the strategies used to set the physical resolution of the computations. The numerical results are discussed and interpreted in Section \ref{sec:4}, and Section \ref{sec:5} concludes the paper by summarizing the main findings.
%
%
%
\section{Simulations Details}
\label{sec:2}

\begin{figure}[t!]
\centering
\includegraphics[scale=0.13,trim=4 4 4 4,clip]{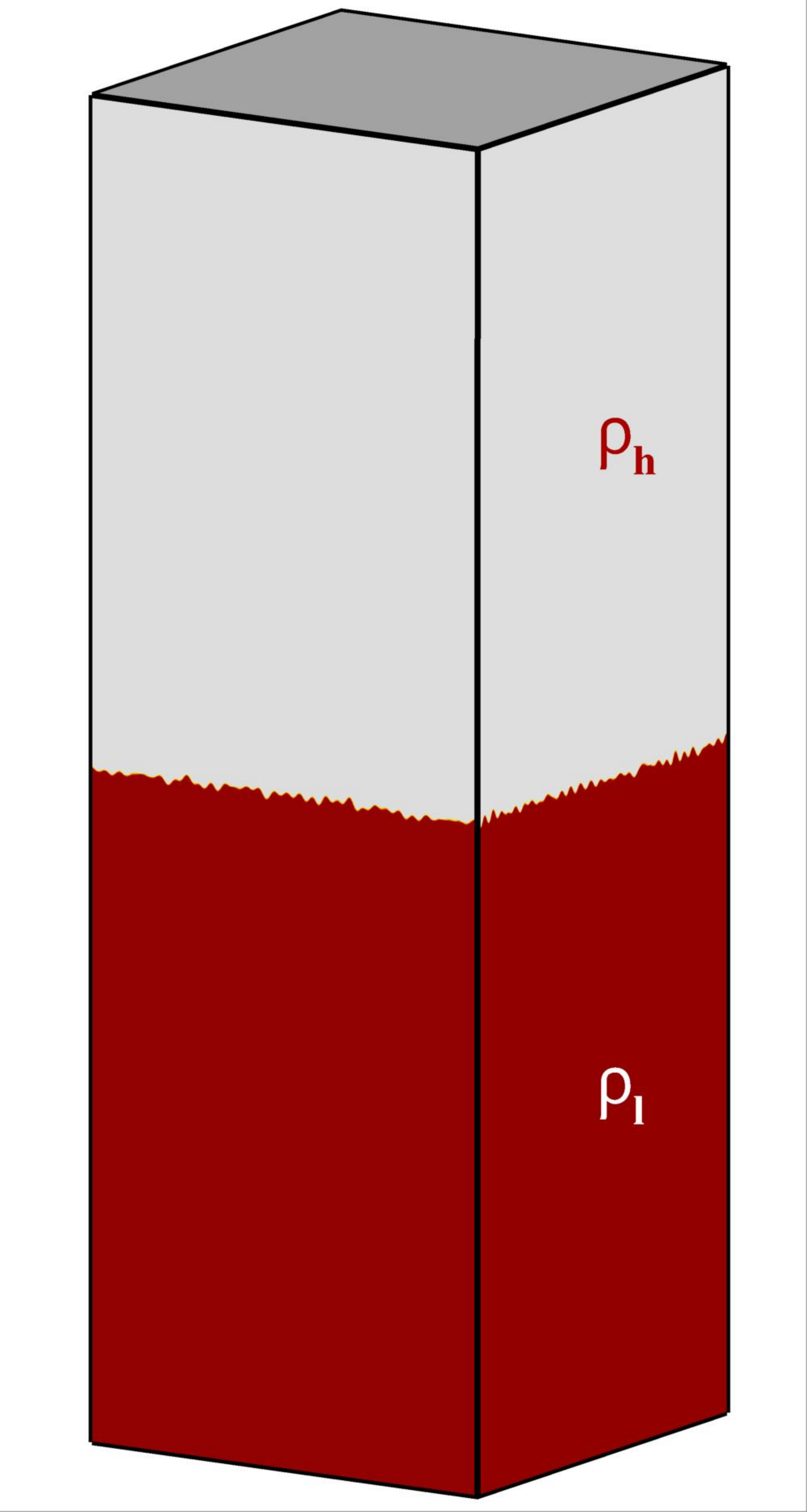}
\caption{Density field at $t=0$.}
\label{fig:2_1}
\end{figure}

The numerical simulations use the settings of \citeauthor{PEREIRA_PRF2_2021} \cite{PEREIRA_PRF2_2021}. They are performed in a prismatic domain defined in a Cartesian coordinate system $(x_1,x_2,x_3)$. This is illustrated in figure \ref{fig:2_1} and possesses a rectangular cross-section of width $L=2\pi$cm and a height of $3L$. The latter ensures a negligible effect of the top and bottom boundary conditions on the simulations during the simulated time $T=25$ time units. Periodic boundary conditions are prescribed on the lateral boundaries, whereas reflective conditions are applied on the top and bottom boundaries, $x_2=\pm 1.5L$. The $\mathrm{At}$ is set equal to $0.5$ and the $\mathrm{Re}$ can reach $500$ (equation \ref{eq:A_10} in Appendix \ref{sec:A}). {\color{blue}We simulate one single representative $\mathrm{At}$ and $\mathrm{Re}$ to enable the performance of  $f_k$ and grid refinement studies to assess and guarantee the accuracy of the simulations.}

\begin{figure}[t!]
\centering
\subfloat[Physical space.]{\label{fig:2_2a}
\includegraphics[scale=0.20,trim=4 4 4 4 ,clip]{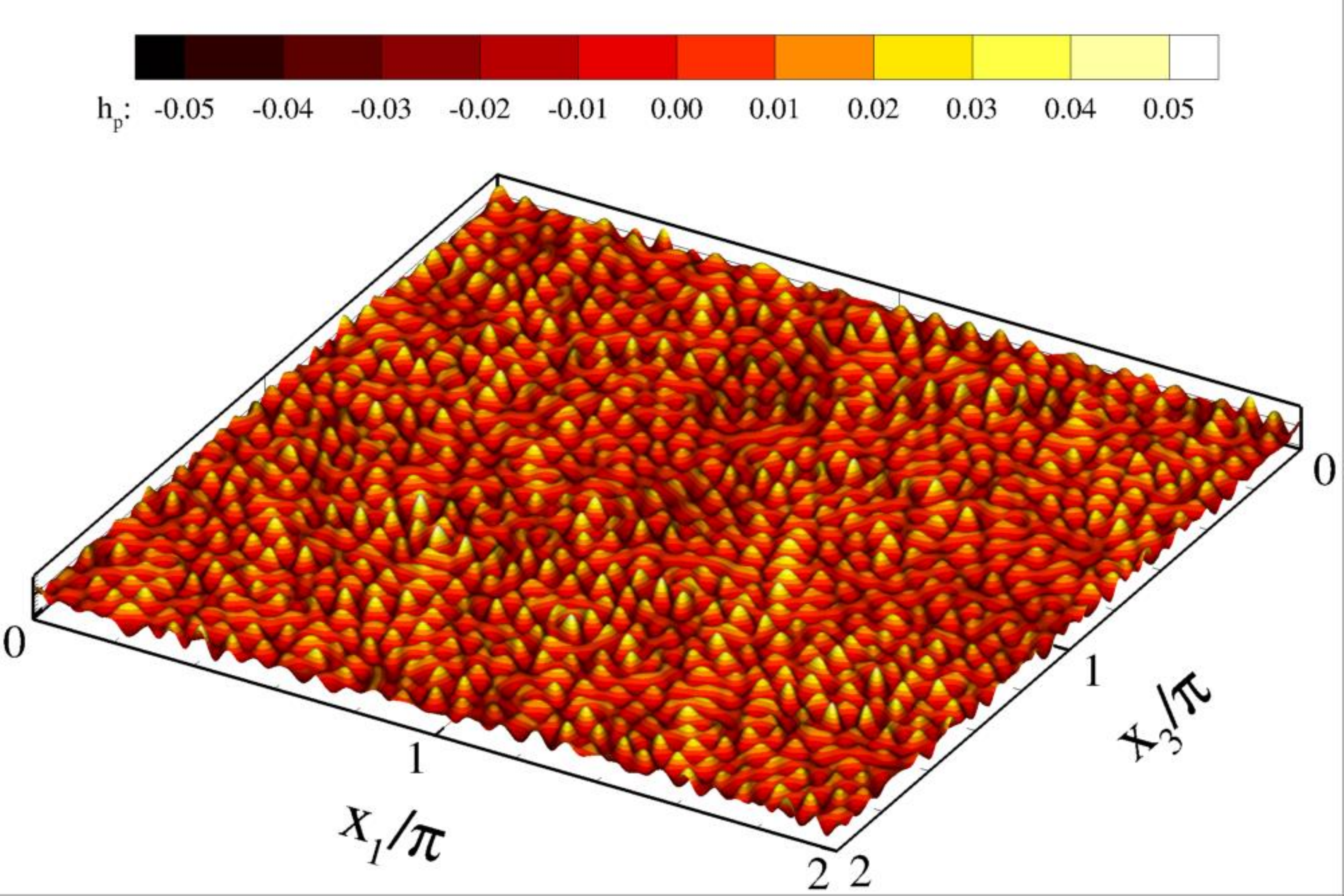}}
\\
\subfloat[Wavenumber space.]{\label{fig:2_2b}
\includegraphics[scale=0.19,trim=4 4 4 4,clip]{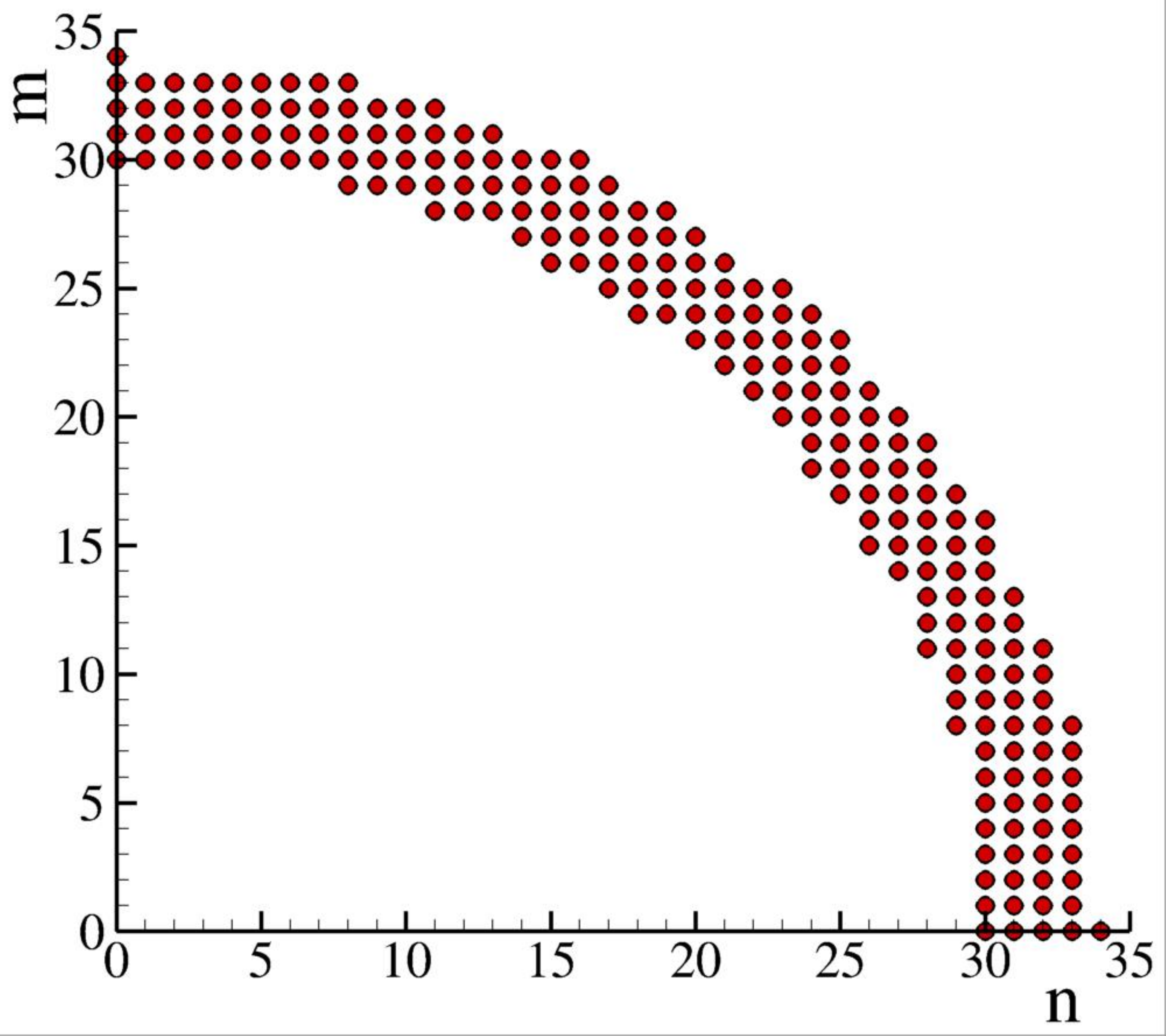}}
\caption{\color{blue}Initial perturbations at the interface ($x_2=0$) in physical and wavenumber (modes) space. Reproduced from F. S. Pereira, F. F. Grinstein, D. M. Israel, R. Rauenzahn, and S. S. Girimaji - Partially averaged Navier-Stokes closure modeling for variable-density turbulent flow. Phys. Rev. Fluids 6, 084602, with the permission of AIP Publishing \cite{PEREIRA_PRF2_2021}; (a) physical space, (b) wavenumber space.}
\label{fig:2_2}
\end{figure}

The interface separating the two fluids is initially perturbed with density fluctuations of height $h_p$,
\begin{equation}
\label{eq:2_1}
\begin{split}
h_p(x_1,x_3)= &\sum_{n,m} 	 \cos \left[ 2 \pi \left( n \frac{x_1}{L} + r_1 \right)\right]  \\
							&\times  \cos \left[ 2 \pi \left( m\frac{x_3}{L} + r_3 \right)\right]
\end{split}
\; .
\end{equation}
These fluctuations possess wavelengths ranging from modes $30$ to $34$ ($30\leq \sqrt{n^2+m^2}\leq 34$) and amplitudes featured by a maximum standard deviation equal to $0.04L$ \cite{BANERJEE_IJHMT_2009,GRINSTEIN_CMA_2019}. In equation \ref{eq:2_1}, the modes $m$ and $n$ are selected to include the most unstable mode of the linearized problem \cite{DUFF_POF_1962,LIVESCU_PD_2021}, and $r_1$ and $r_3$ are random numbers varying between $0$ and $1$. Figure \ref{fig:2_2} illustrates the perturbations in physical and wavenumber space. The simulations assume an ideal gas, and the initial temperature is set to maintain the flow Mach number $\mathrm{Ma}<0.10$. The values prescribed for the fluids' molecular viscosity, $\mu$, ratio between specific heats, $\gamma$, gravitational acceleration constant, $g$, initial specific turbulence kinetic energy and length scale, $k_o$ and $S_o$, Schmidt number, $\mathrm{Sc}$, and Prandtl number, $\mathrm{Pr}$, are summarized in table \ref{tab:2}.

The calculations are three-dimensional and conducted with the flow solver xRAGE \cite{GITTINGS_CSD_2008}. This code utilizes a finite volume approach to solve the compressible and multi-material conservation equations for mass, momentum, energy, and species concentration. The resulting system of governing equations is resolved through the Harten-Lax-van Leer-Contact (HLLC) \cite{TORO_SW_1994} Riemann solver using a directionally unsplit strategy, direct remap, parabolic reconstruction \cite{COLLELA_JCP_1987}, and the low Mach number correction proposed by \citeauthor{THORNBER_JCP_2008} \cite{THORNBER_JCP_2008}. The equations are discretized with second-order accurate methods: the spatial discretization is based on a Godunov scheme, whereas the temporal discretization relies on an explicit Runge-Kutta scheme known as Heun's method. The time-step, $\Delta t$, is set by prescribing the maximum instantaneous CFL (Courant–Friedrichs–Lewy) number,
\begin{equation}
\label{eq:2_2}
\Delta t = \frac{ \Delta x \times \text{CFL}}{3(|V| + c)} \; ,
\end{equation}
equal to $0.5$. In equation \ref{eq:2_2}, $c$ is the speed of sound, and $|V|$ is the velocity magnitude. The effect of the numerical scheme was evaluated in \cite{PEREIRA_CAF_2020,GRINSTEIN_POF_2021}. The code can utilize an adaptive mesh refinement (AMR) algorithm for following waves, especially shock-waves and contact discontinuities. This option is not used in the current work to prevent hanging-nodes \cite{PEREIRA_MASTER_2012} and, as such, the simulations use three-dimensional, orthogonal, and uniform hexahedral grids. We computed the RT with three resolutions, having $64^2\times 192$, $128^2\times 384$, and $256^2\times 768$ cells, respectively. {\color{blue} Grid convergence studies including an extra grid with $512^2\times 1536$ cells have shown that the spatio-temporal grid resolution with $256^2\times 768$ cells is adequate for computing the quantities of interest of this study with the most demanding model, i.e., the case where all flow scales are resolved ($f_k=0.0$). The results are summarized in Appendix \ref{sec:B}}.

The xRAGE solver models miscible material interfaces and strongly convection-driven flows with a van-Leer limiter \cite{LEER_JCP_1997}, without artificial viscosity, and no material interface treatments \cite{GRINSTEIN_PF_2011,HAINES_PRE_2014}. The code uses the assumption that cells containing more than one material are in pressure and temperature equilibrium as a mixed cell closure. The effective molecular kinematic viscosity in multi-material problems \cite{PEREIRA_PRE_2021} is defined as
\begin{equation}
\label{eq:2_3}
\nu=\sum_{n=1}^{n_t} \nu_n f_n \; ,
\end{equation}
where $n$ is the material number, $n_t$ is the number of materials, and $f_n$ is the volume fraction of material $n$. The diffusivity, ${\cal{D}}$, and thermal conductivity, $\kappa$, of the computations are defined by imposing the $\mathrm{Sc}$ and $\mathrm{Pr}$ equal to one. Turbulence is modeled through the variable-density PANS model recently proposed by the authors \cite{PEREIRA_PRF_2021,PEREIRA_PRF2_2021}. This SRS formulation uses a scale-aware BHR-LEVM closure \cite{PEREIRA_PRF2_2021} to model the unresolved scales, which allows the model to operate at any range of resolved scales. The main features of the model are summarized below.

\begin{table*}
\centering
\setlength\extrarowheight{3pt}
\caption{Problem thermodynamic and flow properties.}
\label{tab:2}   
\begin{tabular}{C{1.2cm}C{2.3cm}C{2.3cm}C{1.2cm}C{1.2cm}C{2.0cm}C{2.0cm}C{1.8cm}C{1.2cm}C{1.2cm}}
\hline
$\mathrm{Ma}$	&	$\mu_l$			&	$\mu_h$	 		&	$\gamma_l$	&	$\gamma_h$	&$g$	&$k_o$&$S_o$&$\mathrm{Sc}$&$\mathrm{Pr}$\\ [3pt] \hline
$<0.10$			&	0.002g/(cm.s)		&	0.006g/(cm.s)		&	$1.4$			&	$1.4$	  		&$-980$cm/s$^2$&$10^{-6} \mathrm{cm^2/s^2}$&$10^{-6} \mathrm{cm}$&$1.0$&$1.0$\\[3pt] \hline 
\end{tabular}
\end{table*}
%
%
%
\section{Turbulence Modeling \& Physical resolution}
\label{sec:3}
The PANS BHR-LEVM closure \cite{PEREIRA_PRF2_2021} represents the dynamics of the unresolved scales through the turbulent stress tensor,
\begin{equation}
\label{eq:3_1}
\tau^1(V_i,V_j) = 2 \nu_u \{S_{ij}\} - \frac{2}{3} k_u \delta_{ij}\; ,
\end{equation}
where $\nu_u$ is the unresolved turbulent viscosity,
\begin{equation}
\label{eq:3_2}
\nu_u=c_\mu  S_u  \sqrt{k_u}\; ,
\end{equation}
$\{S_{ij}\}$ is the resolved strain-rate tensor, $k_u$ is the unresolved specific turbulence kinetic energy, $\delta_{ij}$ is the Kronecker delta, and $S_u$ is the unresolved dissipation length-scale. Equations \ref{eq:3_1} and \ref{eq:3_2} include the variables $k_u$ and $S_u$, which need modeling to close the governing equations of PANS. In the scale-aware BHR-LEVM, this is accomplished through six evolution equations
\begin{equation}
\label{eq:3_3}
\frac{\partial{\langle \rho \rangle} k_u}{\partial t} +   \frac{\partial {\langle \rho \rangle} k_u\{V_j \}}{\partial x_j}  =  {\cal{P}}_{b_u} +  {\cal{P}}_{s_u} - {{\cal{E}}_u} +\frac{\partial}{\partial x_j}\left( \frac{\langle \rho \rangle \nu_u}{\sigma_k} \frac{f_\varepsilon}{f_k^2} \frac{\partial k_u}{\partial x_j}\right)   \; ,
\end{equation}
\begin{equation}
\label{eq:3_4}
\begin{split}
\frac{\partial {\langle \rho \rangle} S_u}{\partial t} + \frac{\partial {\langle \rho \rangle} S_u \{V_j\}}{\partial x_j} &= \frac{\partial}{\partial x_j}\left( \frac{\langle \rho \rangle \nu_u}{\sigma_S} \frac{f_\varepsilon}{f_k^2} \frac{\partial S_u}{\partial x_j} \right) \\
&  - c_2^*{\langle \rho \rangle}\sqrt{k_u} + \frac{S_u}{k_u} \left( c_4 {\cal{P}}_{b_u} + c_1 {\cal{P}}_{s_u} \right)     
\end{split}
\; ,
\end{equation}
\begin{equation}
\label{eq:3_5}
\begin{split}
\frac{\partial {\langle \rho \rangle}a_{i_u}}{\partial t} + \frac{\partial{\langle \rho \rangle} a_{i_u}\{V_j\}}{\partial x_j} &= f_{a_i} \frac{b_u}{f_b} \frac{\partial \langle P \rangle}{\partial x_i}  \\
& + \frac{ f_{a_i} }{f_k} \tau^1(V_i,V_j) \frac{\partial \langle \rho \rangle }{\partial x_j} \\ 
& - \frac{f_{a_i}}{f_{a_j}} {\langle \rho \rangle}a_{j_u} \frac{\partial \langle V_i \rangle }{\partial x_j} + \frac{{\langle \rho \rangle}}{f_{a_{j}}} \frac{\partial \left( a_{i_u}a_{j_u} \right)}{\partial x_j} \\
& - c_{a_1} {\langle \rho \rangle}a_{i_u} \frac{\sqrt{k_u}}{S_u}\frac{f_k}{f_\varepsilon} \\
&+ \frac{\partial}{\partial x_j}\left( \frac{{\langle \rho \rangle}\nu_u}{\sigma_a} \frac{f_\varepsilon}{f_k^2} \frac{\partial a_{i_u}}{\partial x_j} \right)
\end{split}
\; ,
\end{equation}
\begin{equation}
\label{eq:3_6}
\begin{split}
\frac{\partial{\langle \rho \rangle} b_u}{\partial t} + \frac{\partial {\langle \rho \rangle} b_u\{V_j\}}{\partial x_j}&= 2{\langle \rho \rangle}\frac{{a_{j_u}}}{f_{a_j}} \frac{\partial b_u}{\partial x_j} - 2\frac{a_{j_u}}{f_{a_j}} \left( b_u + f_b \right) \frac{\partial \langle \rho \rangle }{\partial x_j} \\
& - c_{b}{\langle \rho \rangle}b_u \frac{f_k}{f_\varepsilon} \frac{\sqrt{k_u}}{S_u}  \\
&+ \langle \rho \rangle{^2}\frac{\partial}{\partial x_j}\left( \frac{\nu_u}{ \langle \rho \rangle \sigma_b}\frac{f_\varepsilon}{f_k^2}  \frac{\partial b_u}{\partial x_j} \right)
\end{split}
\; ,
\end{equation}
that model $k_u$, $S_u$, the velocity mass flux vector,
\begin{equation}
\label{eq:3_7}
a_{i_u}=\frac{\langle \rho' v_i' \rangle}{\langle \rho \rangle} \; ,
\end{equation}
and the density-specific volume correlation,
\begin{equation}
\label{eq:3_8}
b_u=-\left\langle \rho'\left(1/\rho\right)'\right\rangle \; .
\end{equation}
The last two quantities are needed to model the production of unresolved turbulence kinetic energy by buoyancy mechanisms. Throughout this paper, the filtering operators $\langle \ \cdot \ \rangle$ and $\{ \ \cdot \ \}$ denote the resolved (or filtered) and density-weighted resolved components of a variable $\Phi$, 
\begin{equation}
\label{eq:3_9}
\Phi = \langle \Phi \rangle + \phi \; ,
\end{equation}
\begin{equation}
\label{eq:3_10}
\Phi = \frac{\langle \rho \Phi \rangle}{\langle \rho\rangle} +  \phi^* = \{ \Phi \} + \phi^* \; ,
\end{equation}
where $\phi$ and $\phi^*$ are the unresolved (or modeled) components of $\Phi$. Note that in the limit of a RANS simulation where all turbulence scales are modeled, decompositions \ref{eq:3_9} and \ref{eq:3_10} are equivalent to Reynolds- \cite{REYNOLDS_PTRSL_1985} and Favre-averaging \cite{FAVRE_CRAS_1958,FAVRE_CRASP_1971,FAVRE_JM_1965,FAVRE_JM2_1965},
\begin{equation}
\label{eq:3_11}
\Phi = \overline{\Phi} + \phi' \; ,
\end{equation}
\begin{equation}
\label{eq:3_12}
\Phi = \frac{\overline{ \rho \Phi }}{\overline{\rho}} +  \phi{''} = \tilde{\Phi} + \phi{''} \; .
\end{equation}
In equations \ref{eq:3_3} to \ref{eq:3_8}, $V_i$ is the velocity vector, ${\cal{P}}_{s_u}$ and ${\cal{P}}_{b_u}$ are the production of unresolved turbulence kinetic energy by shear and buoyancy mechanisms,
\begin{equation}
\label{eq:3_13}
{\cal{P}}_{s_u} = -\langle \rho \rangle \tau^1(V_i,V_j) \frac{\partial \{V_i\}}{\partial x_j} \; ,
\end{equation}
\begin{equation}
\label{eq:3_14}
{\cal{P}}_{b_u} = a_{j_u} \frac{\partial \langle P \rangle}{ \partial x_j} \; ,
\end{equation}
${\cal{E}}_u=\langle \rho \rangle \varepsilon_u$ is the unresolved turbulence kinetic dissipation, and $c_\mu$, $c_2^*$, $c_{a_1}$, $c_b$, $\sigma_k$, $\sigma_S$, $\sigma_a$, and $\sigma_b$ are coefficients. The derivation of the PANS BHR-LEVM closure is given in \citeauthor{PEREIRA_PRF2_2021} \cite{PEREIRA_PRF2_2021}.

The physical resolution of the PANS method is set through a group of parameters defining the unresolved-to-total ratio of the dependent quantities $\phi$,
\begin{equation}
\label{eq:3_15}
f_\phi = \frac{\phi_u}{\phi_t} \; ,
\end{equation}
of the turbulent closure. The subscripts $u$ and $t$ denote unresolved (PANS) and total (RANS) turbulent quantities. In the limit of $f_\phi=1.0$, $\phi_u$ and $\phi_t$ are equivalent because all turbulent scales are represented by the closure in a RANS simulation. Conversely, $f_\phi=0.0$ leads to a DNS, in which $\phi_u=0.0$ and all flow scales are resolved.

Therefore, the physical resolution of the PANS BHR-LEVM model is defined by the parameters,
\begin{equation}
\label{eq:3_16}
f_k=\frac{k_u}{k_t} \; , \hspace{5mm} f_\varepsilon=\frac{\varepsilon_u}{\varepsilon_t}  \; , \hspace{5mm} f_{a_i}=\frac{a_{i_u}}{a_{i_t}} \; , \hspace{5mm} f_b=\frac{b_u}{b_t} \; ,
\end{equation}
where $\varepsilon$ is the specific turbulence dissipation, and 
\begin{equation}
\label{eq:3_17}
f_S=\frac{S_u}{S_t} = \left(\frac{k_u^{3/2}}{\varepsilon_u}\right) \left(\frac{\varepsilon_t}{k_t^{3/2}}\right)=\frac{f_k^{3/2}}{f_\varepsilon}  \; .
\end{equation}
These parameters are included in the governing equations of the original BHR-LEVM closure \cite{BESNARD_TREP_1992,ZARLING_TREP_2011,BANERJEE_PRE_2010,SCHWARZKOPF_JOT_2011} using the framework proposed by \citeauthor{PEREIRA_PRF_2021} \cite{PEREIRA_PRF2_2021}, leading to equations \ref{eq:3_3} to \ref{eq:3_6}. It is important to emphasize that $k_u$ and $\varepsilon_u$ are turbulent quantities, but $a_{i_u}$ and $b_u$ can be composed of a coherent (periodic, non-turbulent \cite{PALKIN_FCT_2016}) and turbulent component \cite{HUSSAIN_JFM_1970,SCHIESTEL_PF_1987}. The importance of this aspect is addressed later.

We have recently performed \textit{a priori} exercises \cite{PEREIRA_PRF2_2021} to investigate the variation of $f_\varepsilon$, $f_{a_i}$, and $f_b$ with $f_k$ and propose guidelines toward the efficient selection of these parameters. Starting with two DNS databases, we used explicit filters to obtain turbulent fields with different levels of physical resolution, from DNS to RANS, and utilized these to calculate $f_k$, $f_\varepsilon$, $f_{a_i}$, and $f_b$ at multiple filter cut-offs. The selected canonical flows were the incompressible forced homogeneous isotropic turbulence (FHIT) of \citeauthor{SILVA_JFM_2018} \cite{SILVA_JFM_2018} and the homogeneous variable-density turbulence (HVDT) at $\mathrm{At}=0.75$ of \citeauthor{ASLANGIL_PD_2020} \cite{ASLANGIL_PD_2020,ASLANGIL_JFM_2020}. Considering the outcome of such exercises summarized in figure \ref{fig:3_1}, we suggested three possible strategies to prescribe $f_\phi$:
\begin{figure}[t!]
\centering
\subfloat[$f_\varepsilon(f_k)$ using FHIT database.]{\label{fig:3_1a}
\includegraphics[scale=0.21,trim=0 0 0 0,clip]{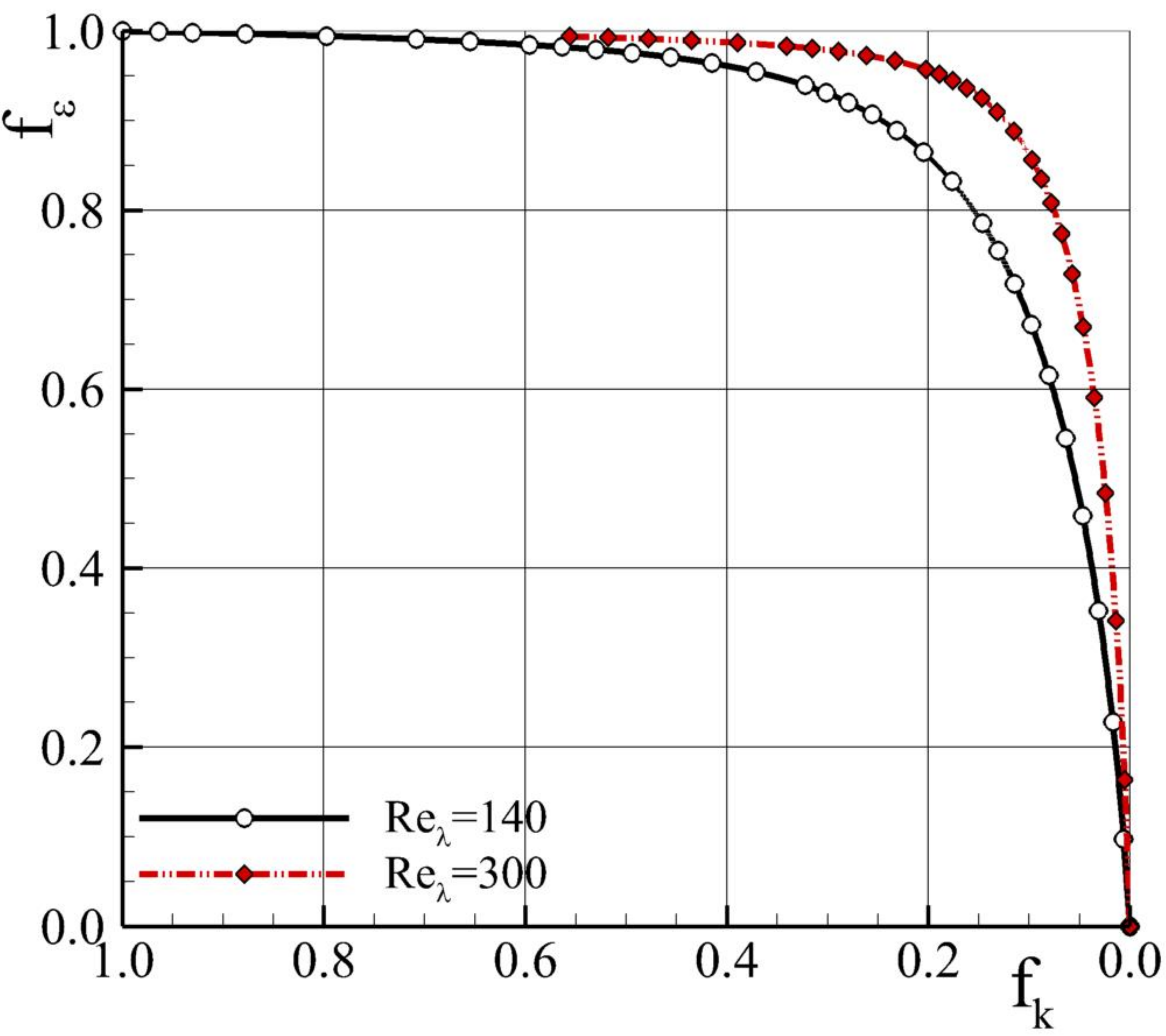}}
\\
\subfloat[$f_a(f_k)$ and $f_b(f_k)$ using HVDT database.]{\label{fig:3_1b}
\includegraphics[scale=0.21,trim=0 0 0 0,clip]{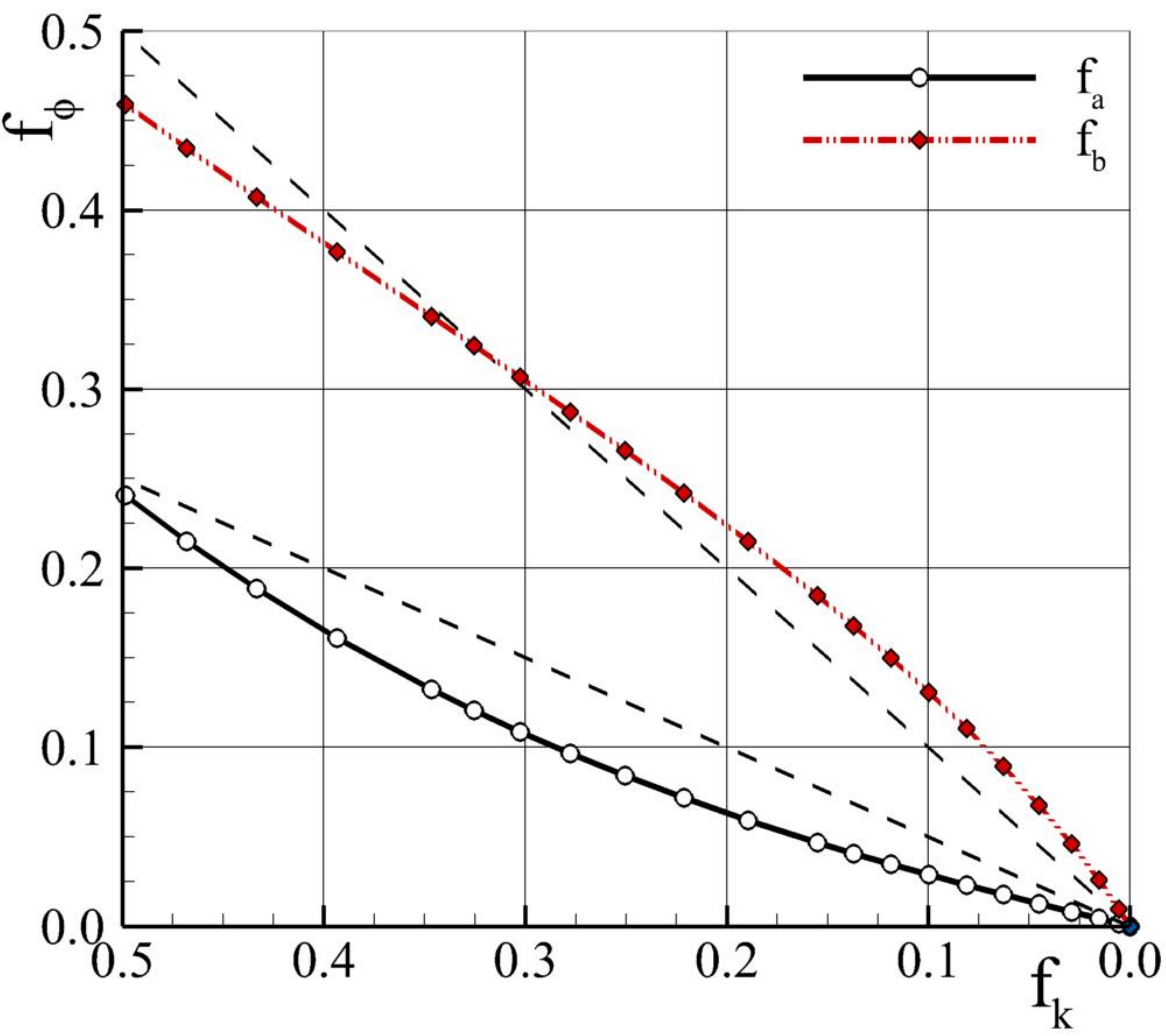}}
\caption{\color{blue}Variation of $f_\varepsilon$, $f_a$, and $f_b$ with $f_k$. Results obtained using FHIT and HVDT databases \cite{SILVA_JFM_2018,ASLANGIL_PD_2020,ASLANGIL_JFM_2020}; (a) $f_\varepsilon(f_k)$ using FHIT database, (b) $f_a(f_k)$ and $f_b(f_k)$ using HVDT database.}
\label{fig:3_1}
\end{figure}
\begin{itemize}
\item[--] $\mathbf{S_1}$: upon inspection of the governing equations of the PANS BHR-LEVM closure (equations \ref{eq:3_3} to \ref{eq:3_6}), we can infer that $f_k$ and $f_\varepsilon$ are main contributors to the magnitude of $k_u$ and, consequently, to $\tau^1(V_i,V_j)$. In addition, the results of figure \ref{fig:3_1a} indicate that assuming $f_\varepsilon=1.0$ is adequate for practical computations, i.e., $f_k\ge 0.20$. Hence, one can prescribe $f_k$ and set the remaining parameters equal to one - $f_\varepsilon=f_{a_i}=f_b=1.0$. This strategy does not require \textit{a priori} knowledge of the flow dynamics and has turned out to be numerically robust. However, it might require slightly smaller values of $f_k$ to compensate for possible calibration deficits.
\item[--] $\mathbf{S_2}$: assume that most flow density fluctuations are caused by the coherent field so that one can neglect the contribution of $\rho$ to $f_{a_i}$ and $f_b$. Similarly to $S_1$, define $f_\varepsilon=1.0$. Thus, prescribe $f_k$ and set $f_{a_i}=f_a=\sqrt{f_k}$ and $f_\varepsilon=f_b=1.0$. This approach is numerically robust and does not need \textit{a priori} knowledge of the flow dynamics.
\item[--] $\mathbf{S_3}$: based on the results of the \textit{a priori} exercises shown in figure \ref{fig:3_1} \cite{PEREIRA_PRF2_2021}, prescribe $f_k$ and set $f_\varepsilon=1.0$, $f_{a_i}=f_a\approx f_k/2$, and $f_b=f_k$. This strategy is best suited for HVDT class of flows and fully-developed turbulent flow.
\end{itemize}
The numerical simulations are conducted with these three approaches, using five values of $f_k$: $0.00$ (no closure), $0.25$, $0.35$, $0.50$, and $1.00$ {\color{blue}(equivalent to RANS)}. 
\begin{figure*}[t!]
\centering
\subfloat[$t=0.0$.]{\label{fig:4.1_1a}
\includegraphics[scale=0.08,trim=0 0 0 0,clip]{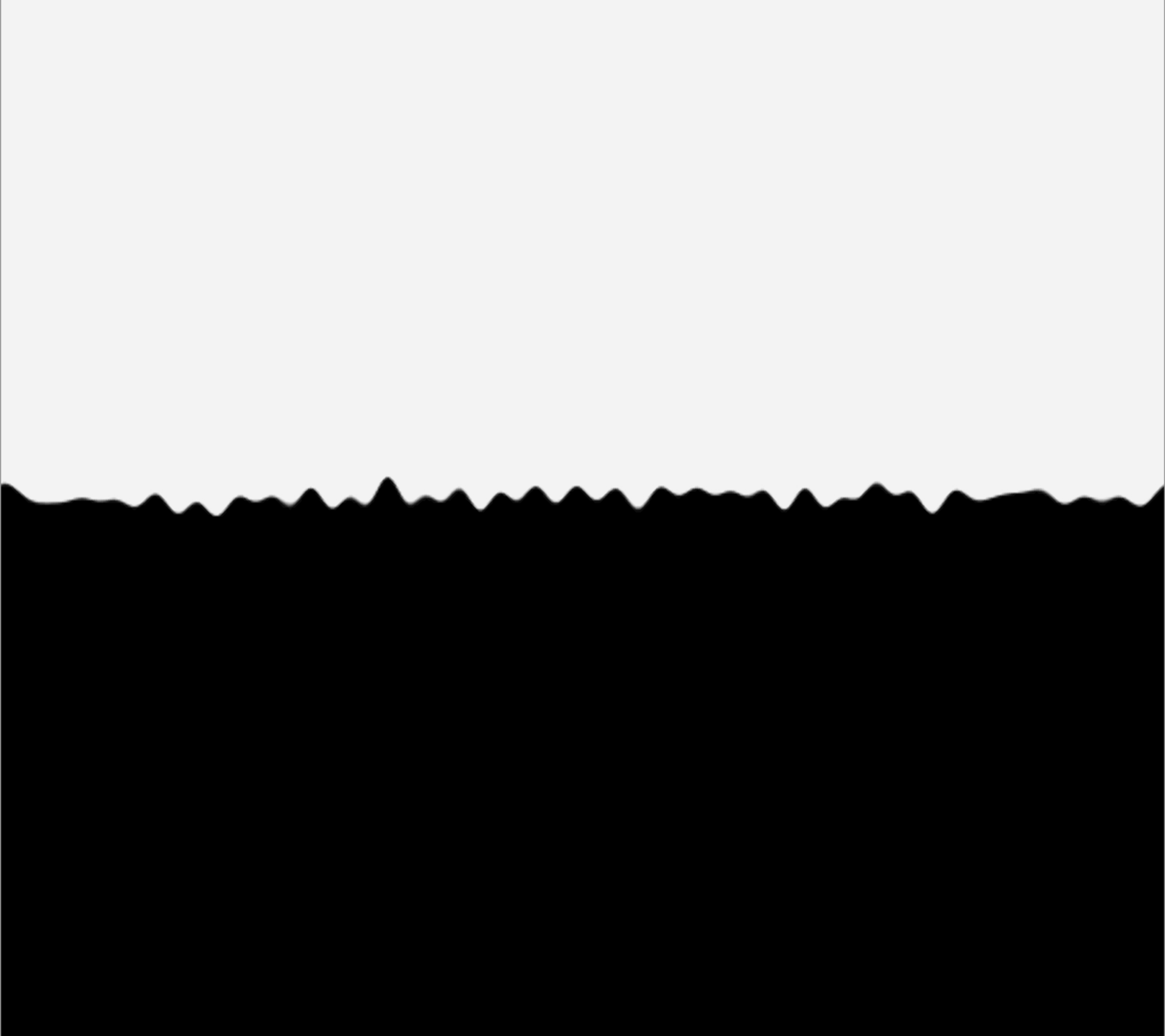}}
~
\subfloat[$t=1.0$.]{\label{fig:4.1_1b}
\includegraphics[scale=0.08,trim=0 0 0 0,clip]{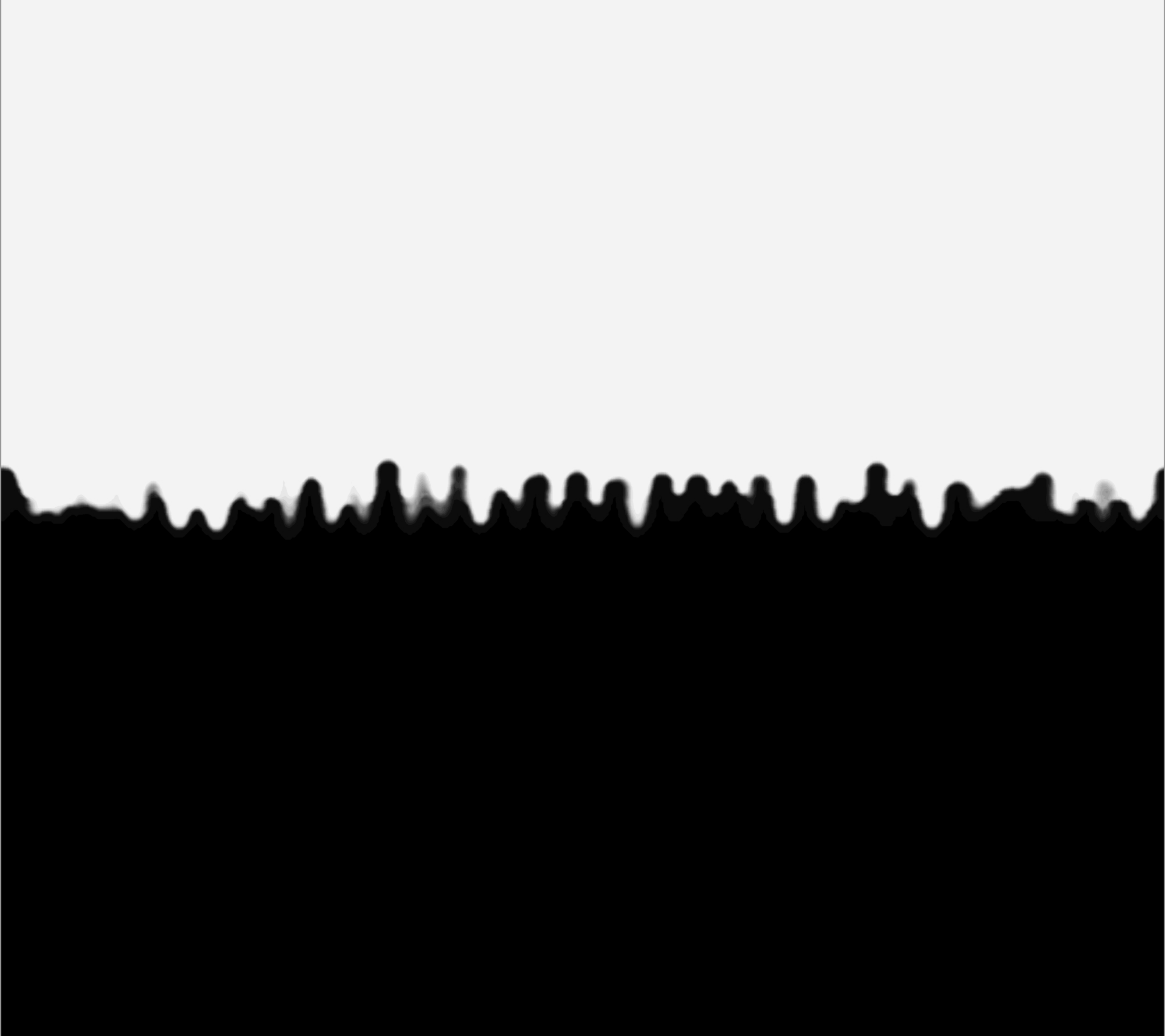}}
~
\subfloat[$t=1.5$.]{\label{fig:4.1_1c}
\includegraphics[scale=0.08,trim=0 0 0 0,clip]{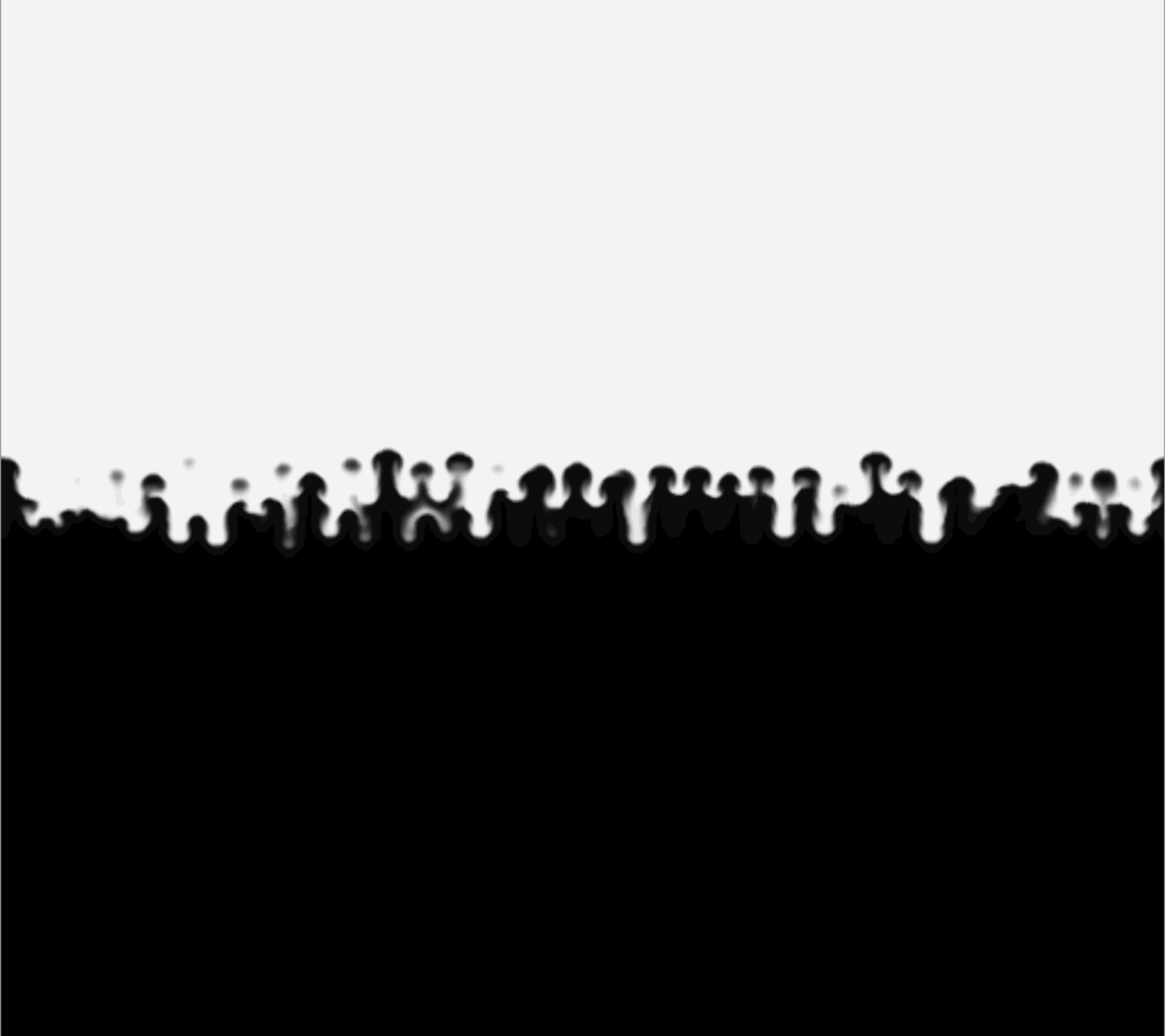}}
\\
\subfloat[$t=2.0$.]{\label{fig:4.1_1d}
\includegraphics[scale=0.08,trim=0 0 0 0,clip]{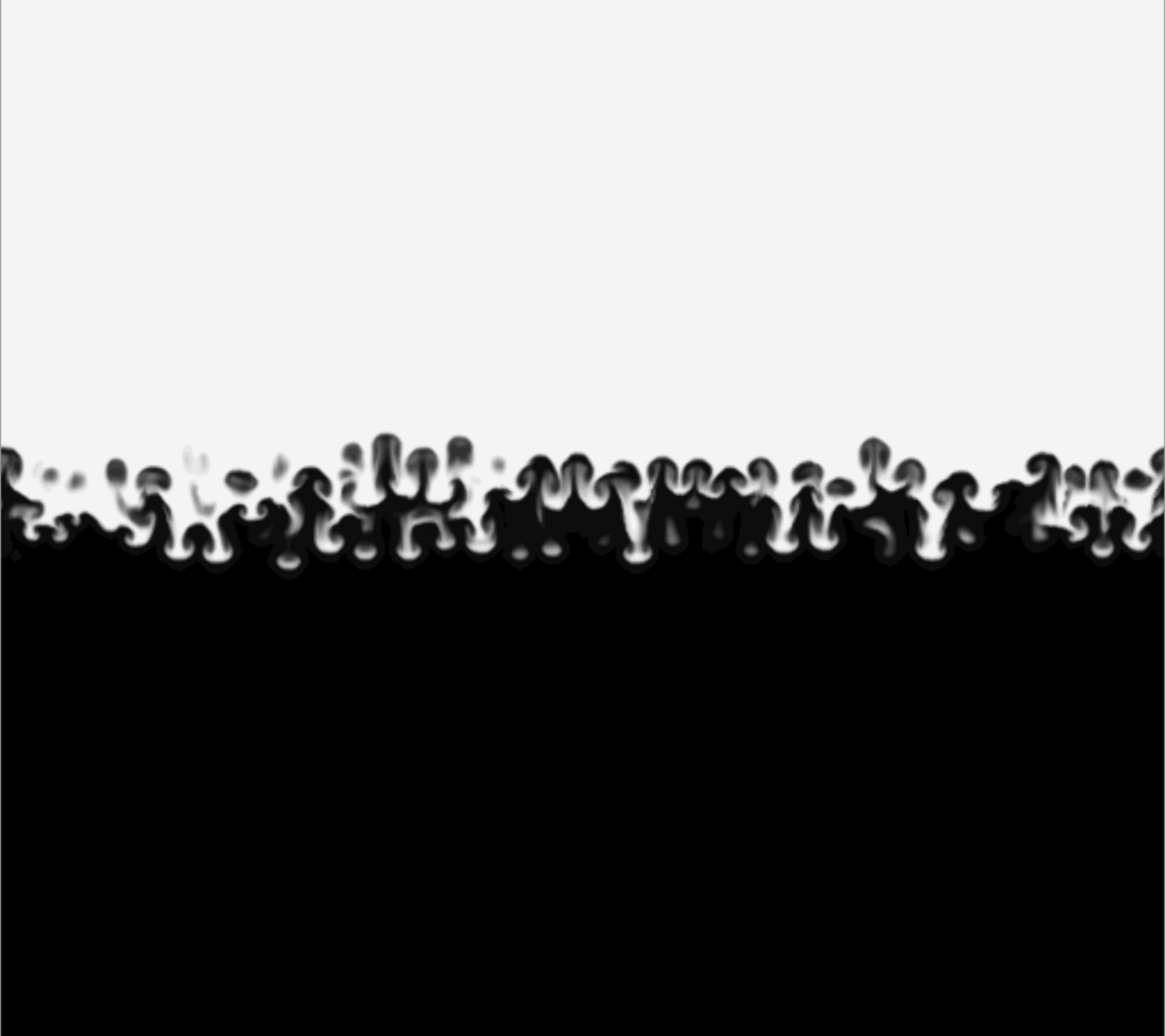}}
~
\subfloat[$t=2.5$.]{\label{fig:4.1_1e}
\includegraphics[scale=0.08,trim=0 0 0 0,clip]{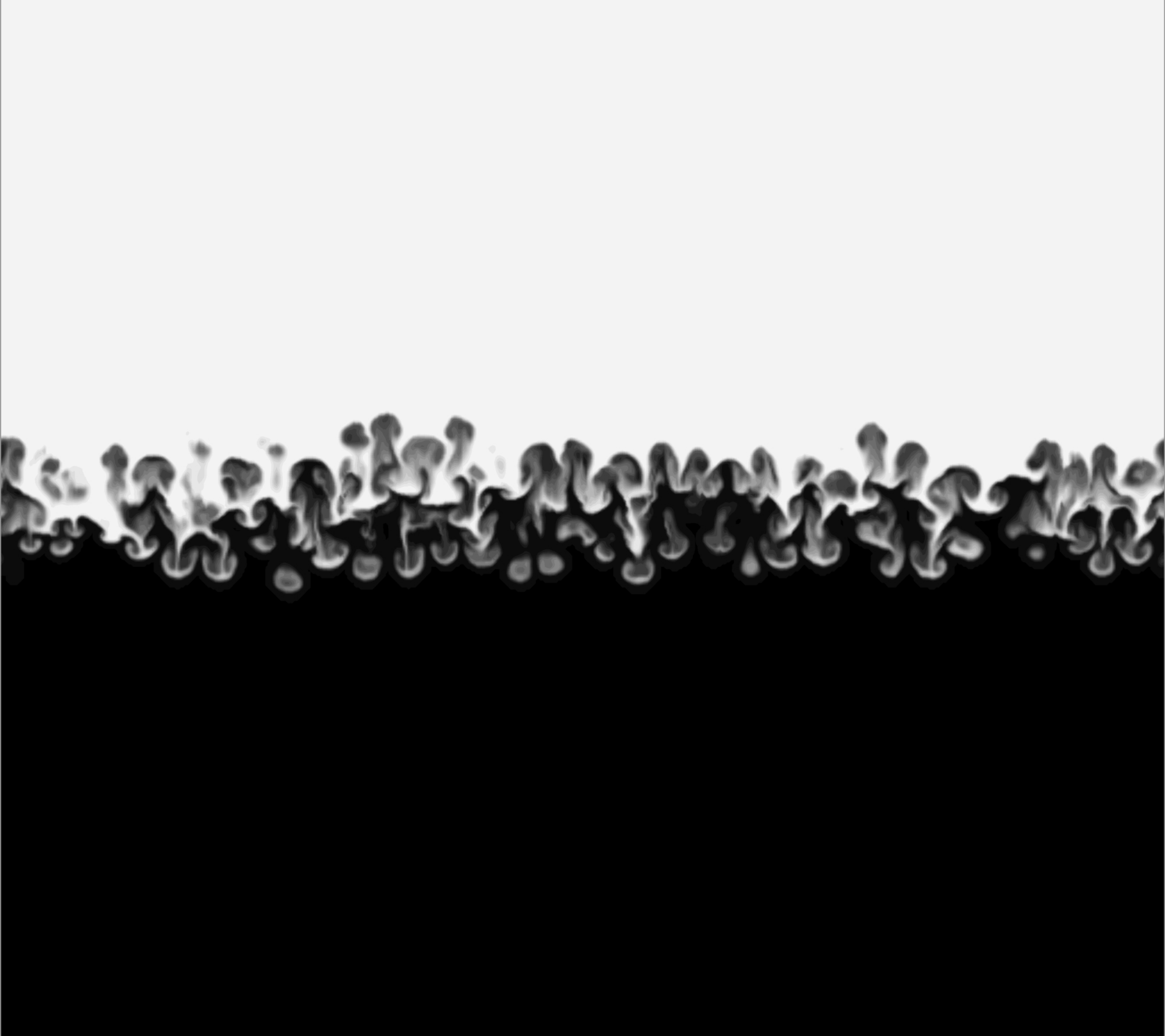}}
~
\subfloat[$t=3.0$.]{\label{fig:4.1_1f}
\includegraphics[scale=0.08,trim=0 0 0 0,clip]{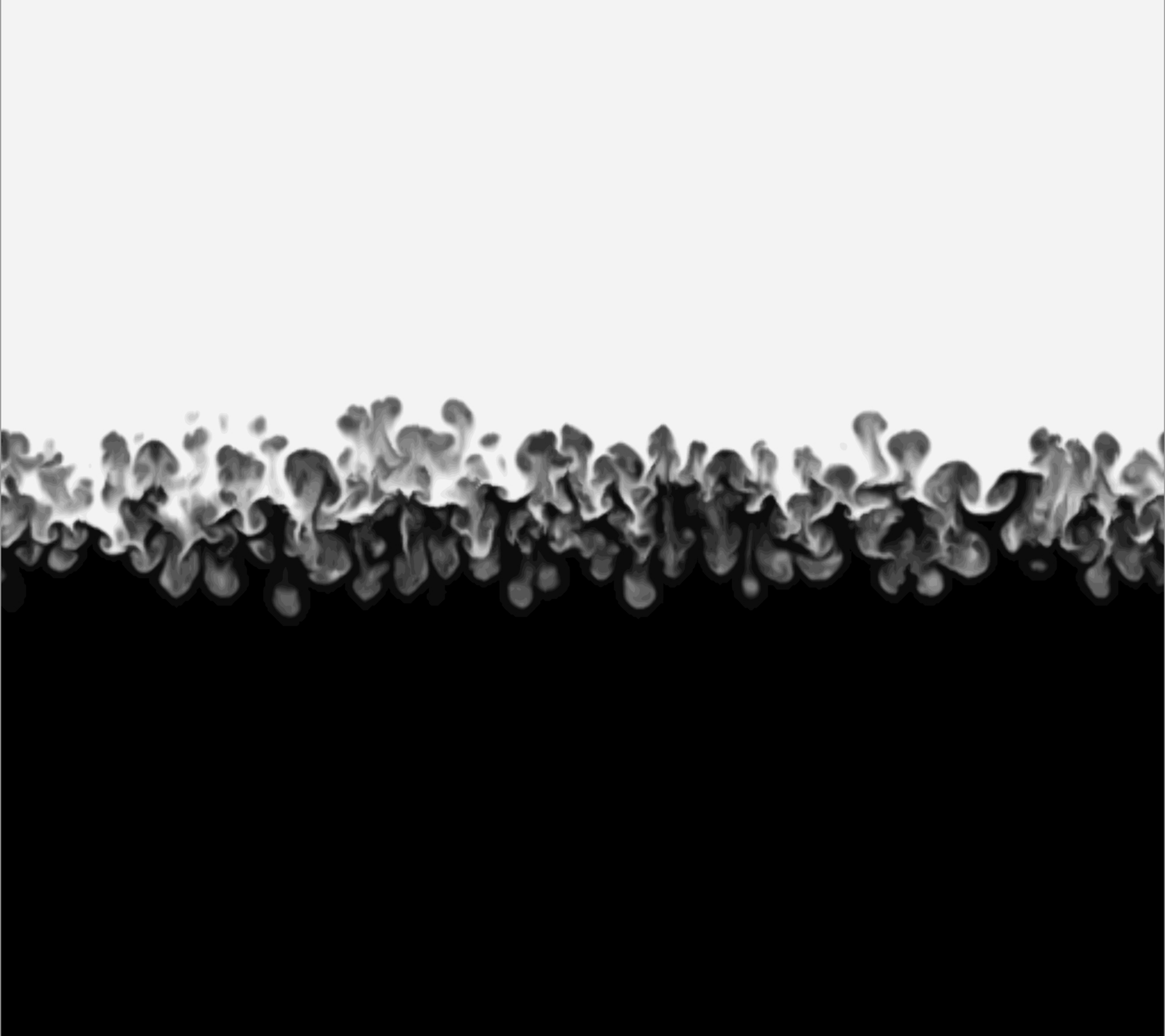}}
\\
\subfloat[$t=3.5$.]{\label{fig:4.1_1g}
\includegraphics[scale=0.08,trim=0 0 0 0,clip]{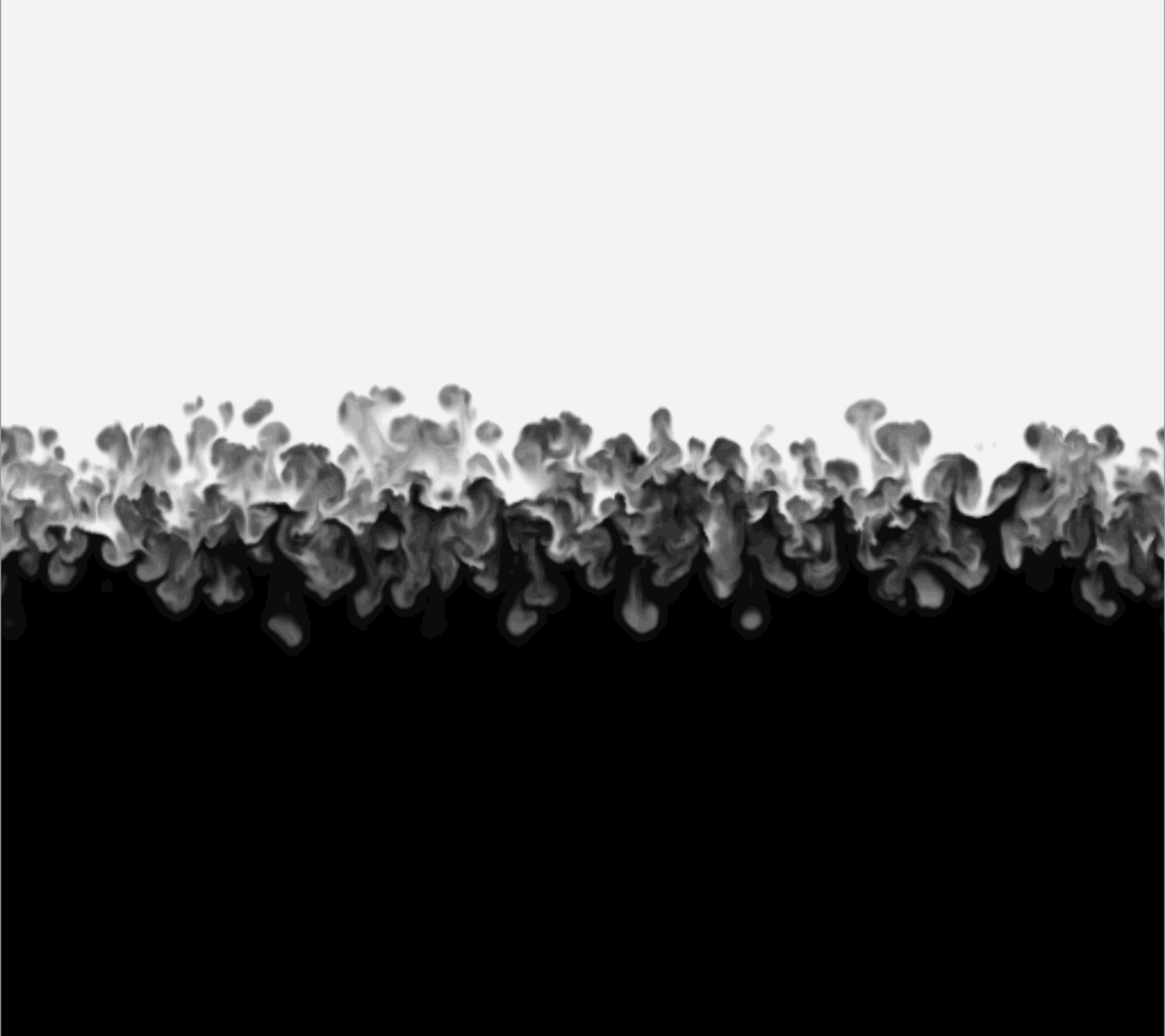}}
~
\subfloat[$t=4.0$.]{\label{fig:4.1_1h}
\includegraphics[scale=0.08,trim=0 0 0 0,clip]{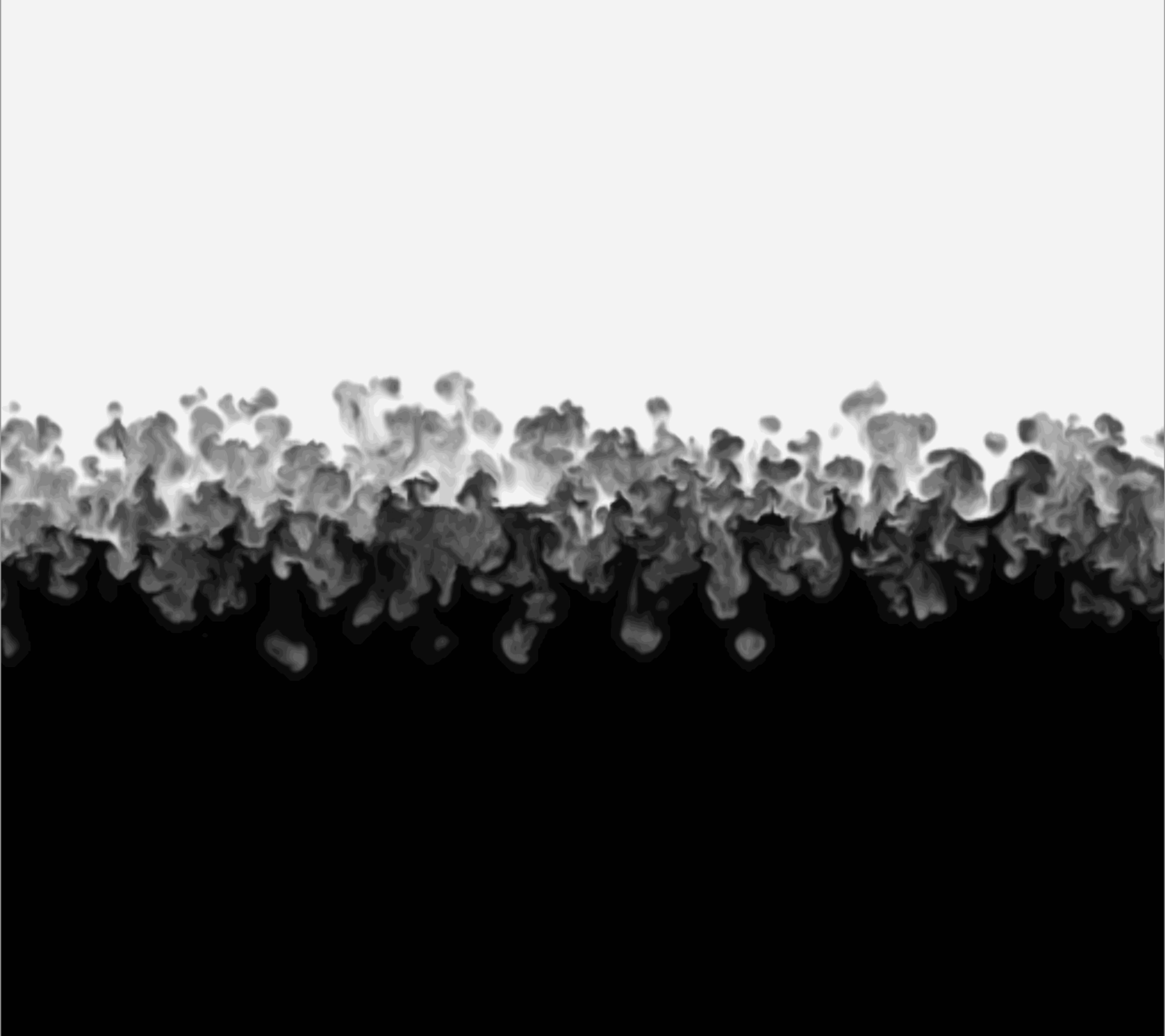}}
~
\subfloat[$t=5.0$.]{\label{fig:4.1_1i}
\includegraphics[scale=0.08,trim=0 0 0 0,clip]{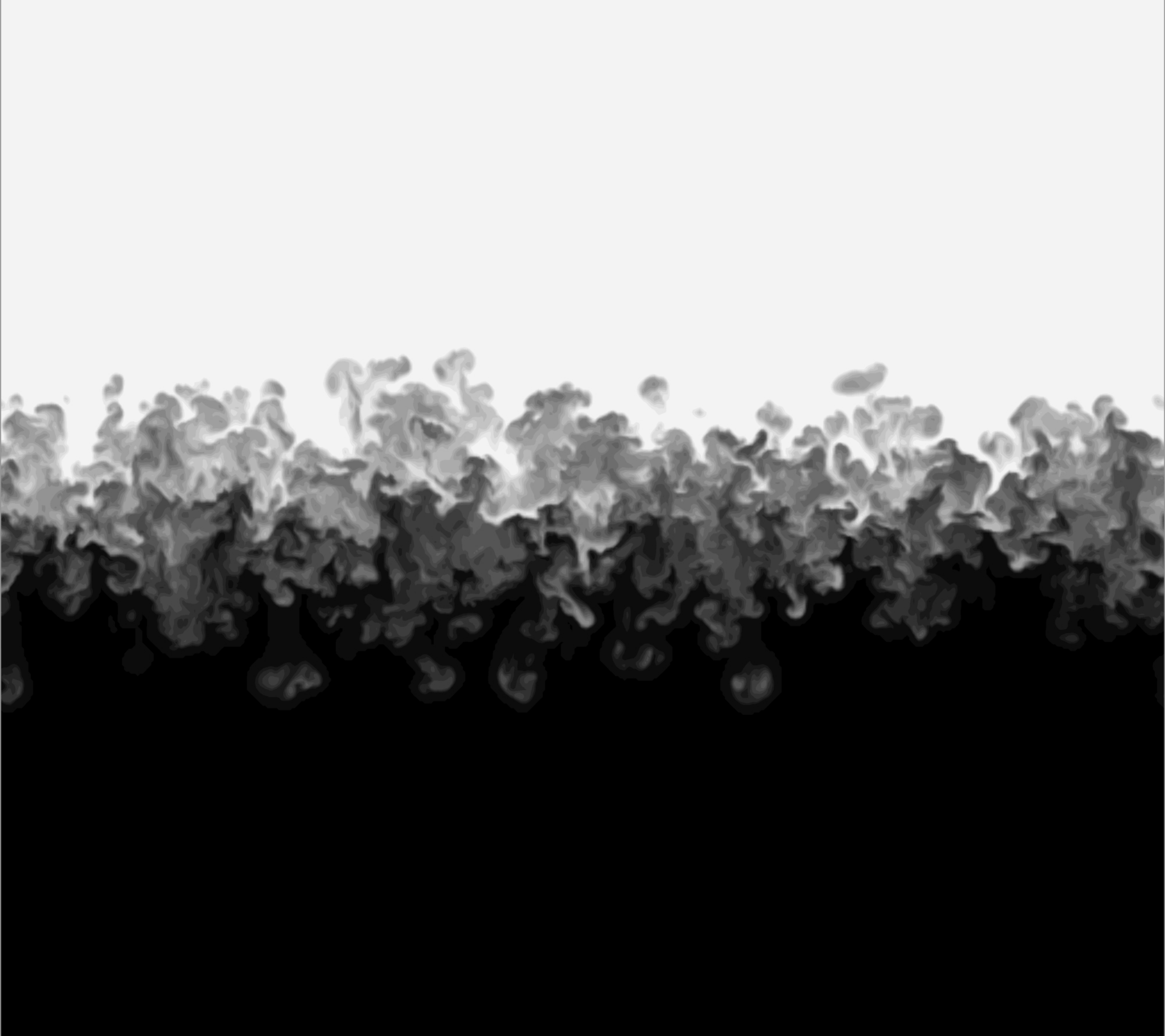}}
\caption{\color{blue}Temporal evolution of the density field, $\rho$, (2D slice) predicted at early flow times and $f_k=0.00$. Reproduced from F. S. Pereira, F. F. Grinstein, D. M. Israel, R. Rauenzahn, and S. S. Girimaji - Partially averaged Navier-Stokes closure modeling for variable-density turbulent flow. Phys. Rev. Fluids 6, 084602, with the permission of AIP Publishing  \cite{PEREIRA_PRF2_2021}. The animation is available online (Multimedia view); (a) $t=0.0$, (b) $t=1.0$, (c) $t=1.5$, (d) $t=2.0$, (e) $t=2.5$, (f) $t=3.0$, (g) $t=3.5$, (h) $t=4.0$, (i) $t=5.0$.}
\label{fig:4.1_1}
\end{figure*}
%
%
%
\section{Numerical Results}
\label{sec:4}
The discussion of the simulations starts with a succinct description of the temporal evolution of the flow in Section \ref{sec:4.1}. This step is crucial to understand the behavior of the PANS BHR-LEVM. Next, Section \ref{sec:4.2} evaluates the accuracy of the model and investigates the impact of the physical resolution and $f_\phi$ strategy on the computations. It includes comparisons against reference simulations obtained with $f_k=0.0$. {\color{blue}The use of the high-fidelity simulation resolving all flow scales as reference guarantees that all computations run on the same initial laminar flow conditions. This strategy prevents the solution mismatches observed in figure \ref{fig:1_1} caused by input uncertainties (see also \citeauthor{PEREIRA_JCP_2021} \cite{PEREIRA_JCP_2021}) and enables the robust evaluation of the modeling accuracy of the PANS BHR-LEVM.} Section \ref{sec:4.3} analyzes the physical rationale of the results. All flow quantities discussed in this paper are either defined in the text or in Appendix \ref{sec:A}. {\color{blue}They are computed on the grid resolution with $256^2\times 768$ cells}. The impact of the grid resolution on the computations is analyzed in Appendix \ref{sec:B}.
%
%
\subsection{Temporal flow evolution}
\label{sec:4.1}
%
\begin{figure*}[t!]
\centering
\subfloat[$t=2.5$.]{\label{fig:4.1_2a}
\includegraphics[scale=0.12,trim=5 7 5 5,clip]{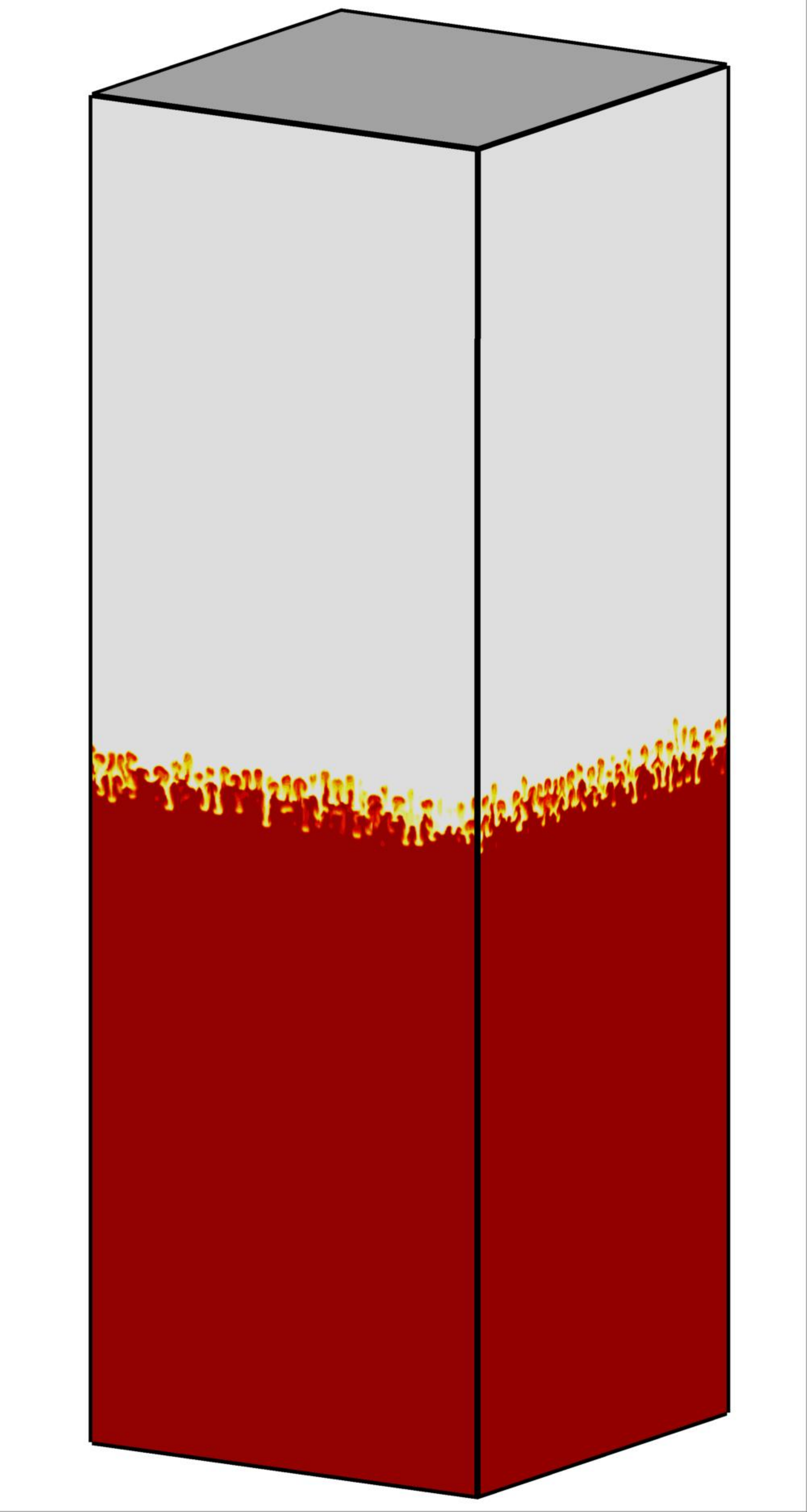}}
~
\subfloat[$t=5.0$.]{\label{fig:4.1_2b}
\includegraphics[scale=0.12,trim=5 7 5 5,clip]{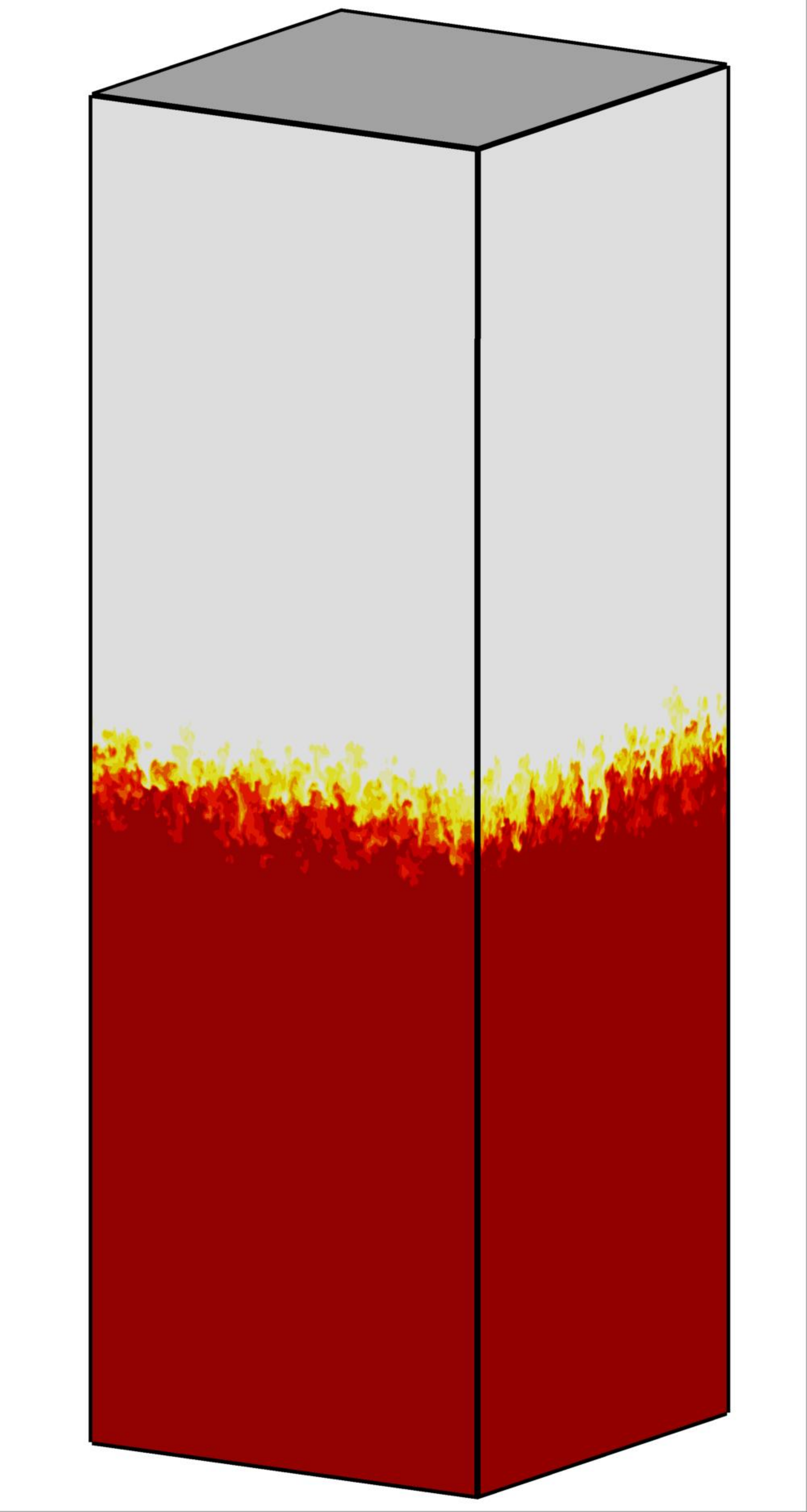}}
~
\subfloat[$t=10.0$.]{\label{fig:4.1_2c}
\includegraphics[scale=0.12,trim=5 7 5 5,clip]{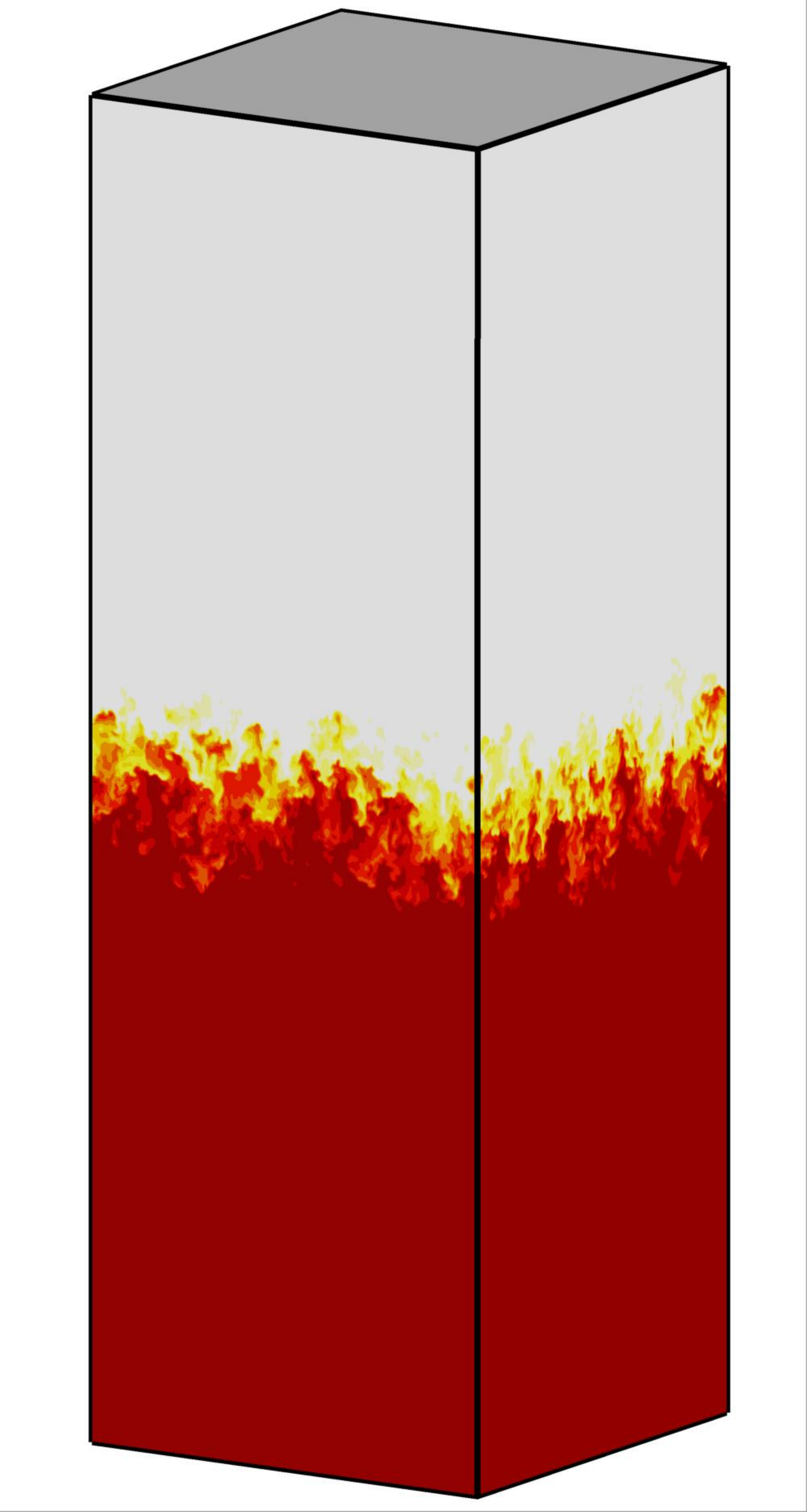}}
~
\subfloat[$t=20.0$.]{\label{fig:4.1_2d}
\includegraphics[scale=0.12,trim=5 7 5 5,clip]{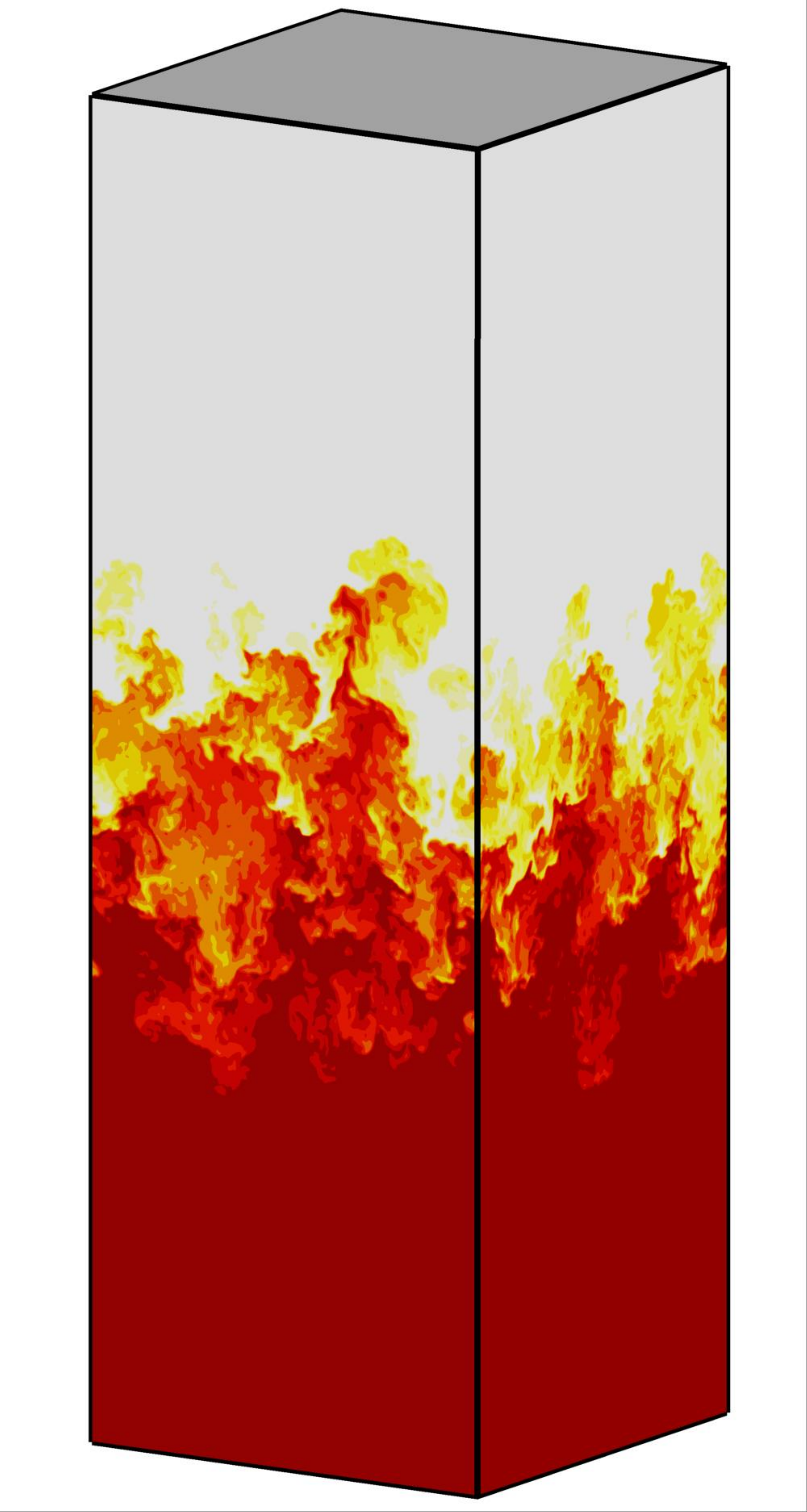}}
\caption{Temporal evolution of the density field, $\rho$, predicted with $f_k=0.00$; (a) $t=2.5$, (b) $t=5.0$, (c) $t=10.0$, (d) $t=20.0$.}
\label{fig:4.1_2}
\end{figure*}

\begin{figure}[t!]
\centering
\subfloat[$t=10.0$.]{\label{fig:4.1_3a}
\includegraphics[scale=0.12,trim=5 7 5 5,clip]{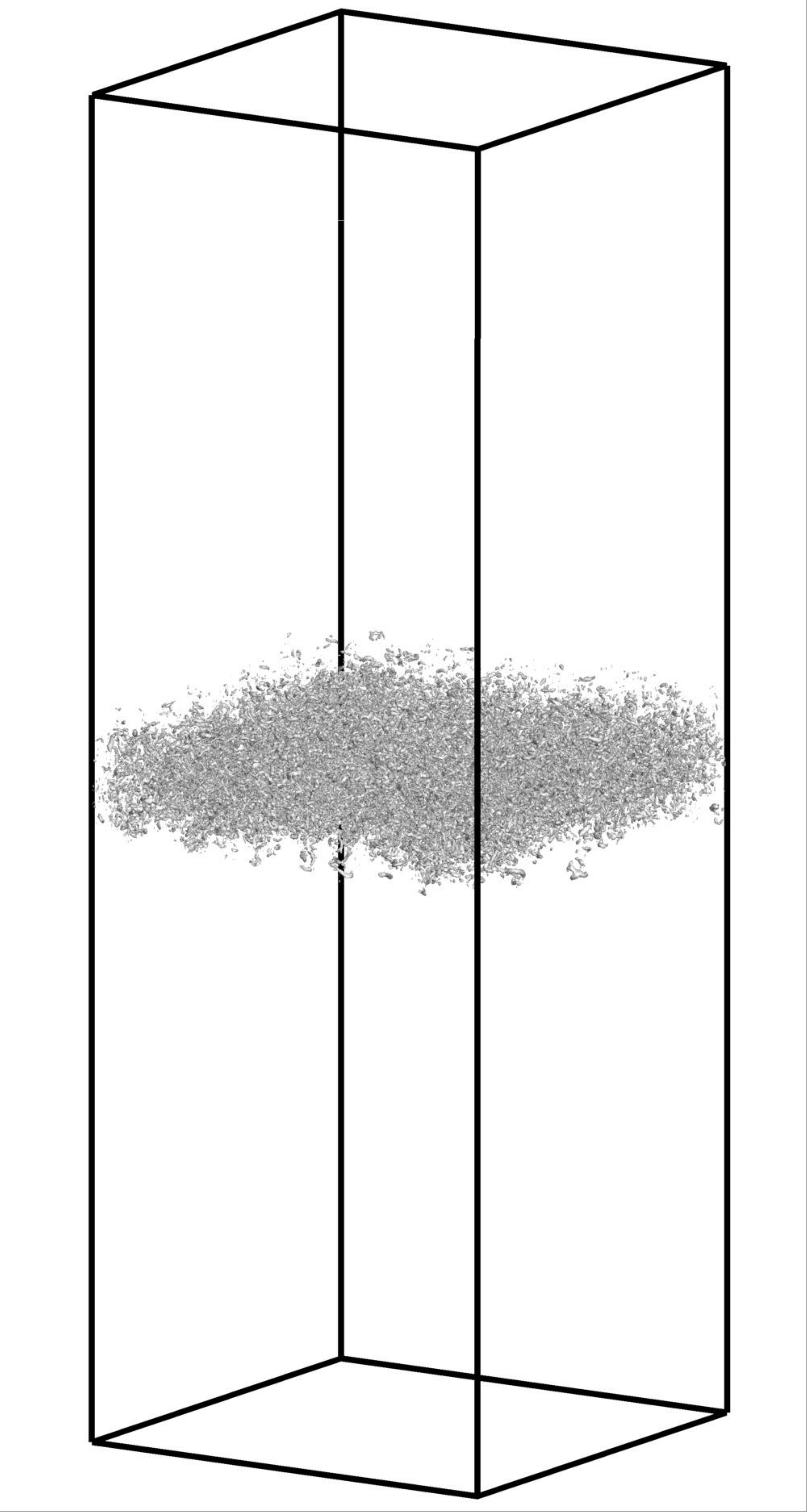}}
~
\subfloat[$t=20.0$.]{\label{fig:4.1_3b}
\includegraphics[scale=0.12,trim=5 7 5 5,clip]{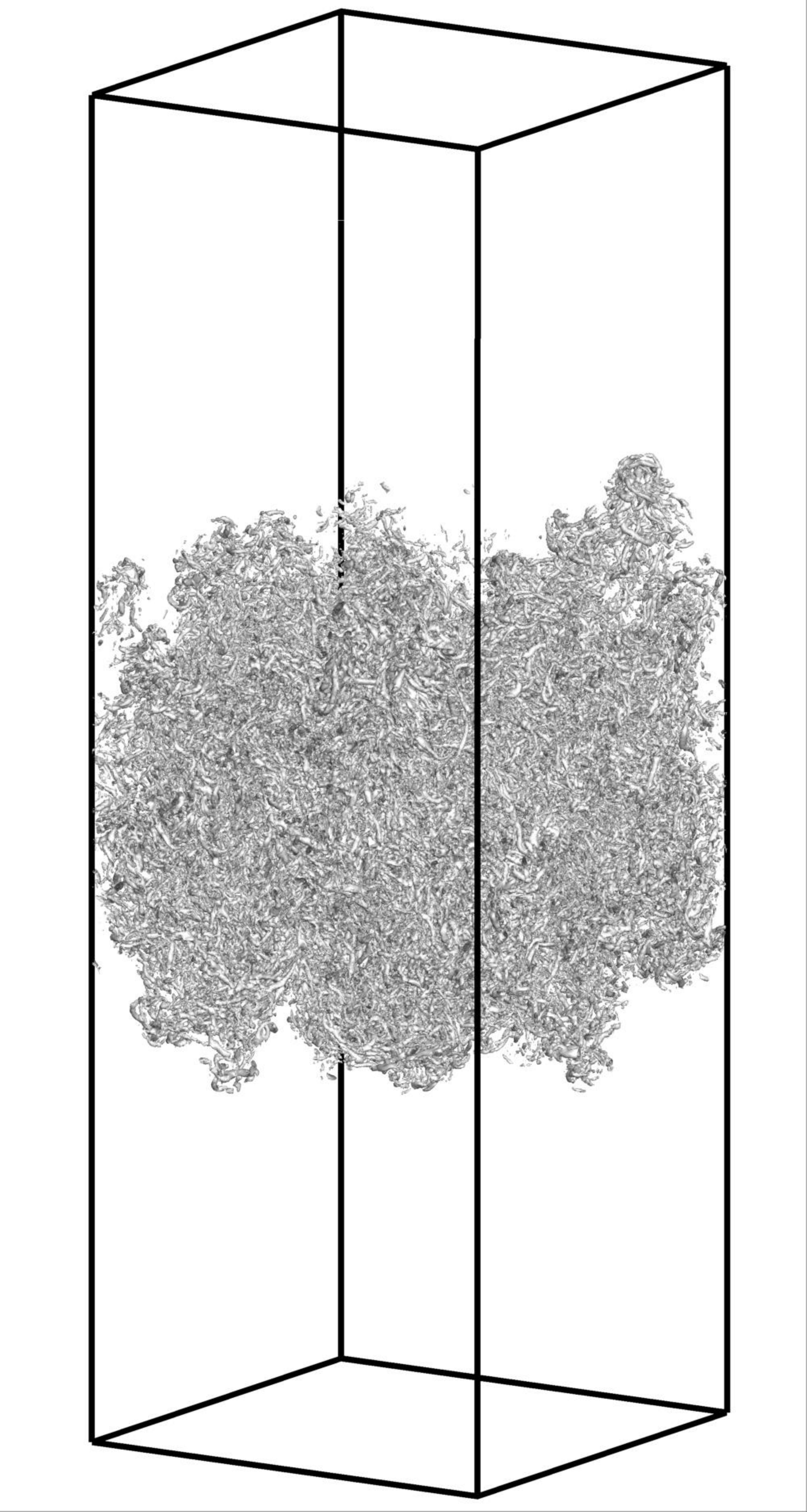}}
\caption{Temporal evolution of the flow coherent and turbulent structures predicted with $f_k=0.00$. Structures identified through the $\lambda_2$-criterion \cite{JEONG_JFM_1995} ($-\lambda_2=2.5$, $5.0$, and $10.0\times 10^4$); (a) $t=10.0$, (b) $t=20.0$.}
\label{fig:4.1_3}
\end{figure}

The RT flow is a benchmark variable-density problem where two materials of different densities mix. It features laminar flow, instabilities and coherent structures, density fluctuations, onset and development of turbulence, breakdown of coherent structures, high-intensity turbulence, and energy dissipation. To better understand the flow physics and the simulations, figures \ref{fig:4.1_1} (Multimedia view) to \ref{fig:4.1_3} present the temporal evolution of the coherent and turbulent structures of the flow. Figures \ref{fig:4.1_1} (Multimedia view) and \ref{fig:4.1_2} depict the density field, whereas figure \ref{fig:4.1_3} shows the $\lambda_2$-criterion proposed by \citeauthor{JEONG_JFM_1995} \cite{JEONG_JFM_1995}. The numerical results shown are obtained with $f_k=0.0$.

Figure \ref{fig:4.1_1a} shows that the two fluids are initially separated by the perturbed interface plotted in figure \ref{fig:2_2a}. In the present unstable flow configuration, the heavy fluid is located on top of the light material. Immediately after the initial time, the heavy material accelerates downward due to gravity, whereas the light material moves upward. Material mixing starts due to molecular diffusivity (viscosity and diffusivity). The interface perturbations create a misalignment between the density gradient and the pressure, leading to the production of vorticity by baroclinic processes, which amplify the perturbations. These perturbations grow into coherent structures.  The upward moving structures, which are the penetration of light fluid into heavy fluid, are called bubbles, and the downward moving structures of heavy fluid penetrating into light fluid are called spikes. Such laminar vortical structures have a characteristic three-dimensional mushroom-like form. Figures \ref{fig:4.1_1a}-\ref{fig:4.1_1e} and \ref{fig:4.1_2a} illustrate the onset and development of such coherent structures. 

After this period ($t>2.5$), the shearing motion of the edges of the coherent structures triggers the Kelvin-Helmholtz instability, responsible for the onset of turbulence and breakdown of the vortical structures. These processes increase the mixing rate of the two materials and the mixing-layer height. This development can be seen in figures \ref{fig:4.1_1f} to \ref{fig:4.1_1i} and figure \ref{fig:4.1_2}. Afterward, turbulence further develops, increasing the mixing-layer height and homogeneity. The growth of the mixing-layer and fine-scale turbulence is visible in figures \ref{fig:4.1_2c}-\ref{fig:4.1_2d} and \ref{fig:4.1_3}.

It is crucial to emphasize that the spatio-temporal development of the RT problem entails a linear and a nonlinear regime \cite{ZHOU_PR1_2017,ZHOU_PR2_2017}. The first consists of laminar flow, whereas the second includes laminar, transient and turbulent flow. From a modeling and simulation perspective, these regimes pose different challenges:
\begin{itemize}
\item[$i)$] Until $t \approx 2.5$, the flow evolves rapidly and undergoes multiple morphological changes. It features coherent structures with a mushroom-type shape and instabilities, which later lead to the onset and development of turbulence;
\item[$ii)$] After $t \approx 2.5$, the coherent structures break down, and turbulence develops. Fine-scale turbulence is visible at $t\ge 5$. This enhances the homogeneity of the mixture and increases the mixing-layer growth rate.
\end{itemize}
Comparing the two regimes, the first and the early stages of the second are expected to be difficult to fully model with one-point turbulence closures since these have not been designed to predict such flow phenomena.
%
%
%
\subsection{Effect of physical resolution}
\label{sec:4.2}
%
\begin{figure}[t!]
\centering
\subfloat[$S_1$.]{\label{fig:4.2_1a}
\includegraphics[scale=0.208,trim=0 0 0 0,clip]{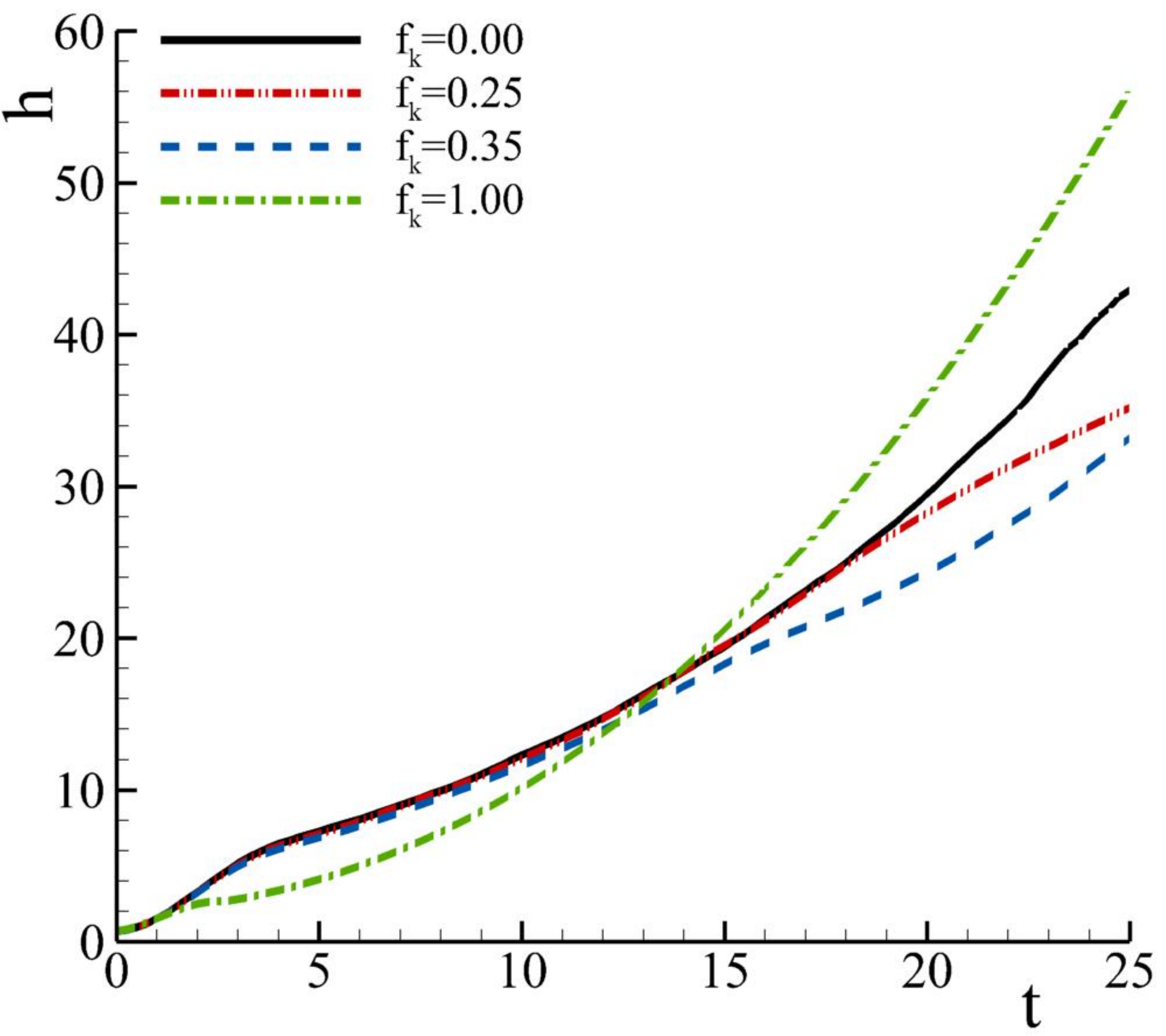}}
\\
\subfloat[$S_2$.]{\label{fig:4.2_1b}
\includegraphics[scale=0.208,trim=0 0 0 0,clip]{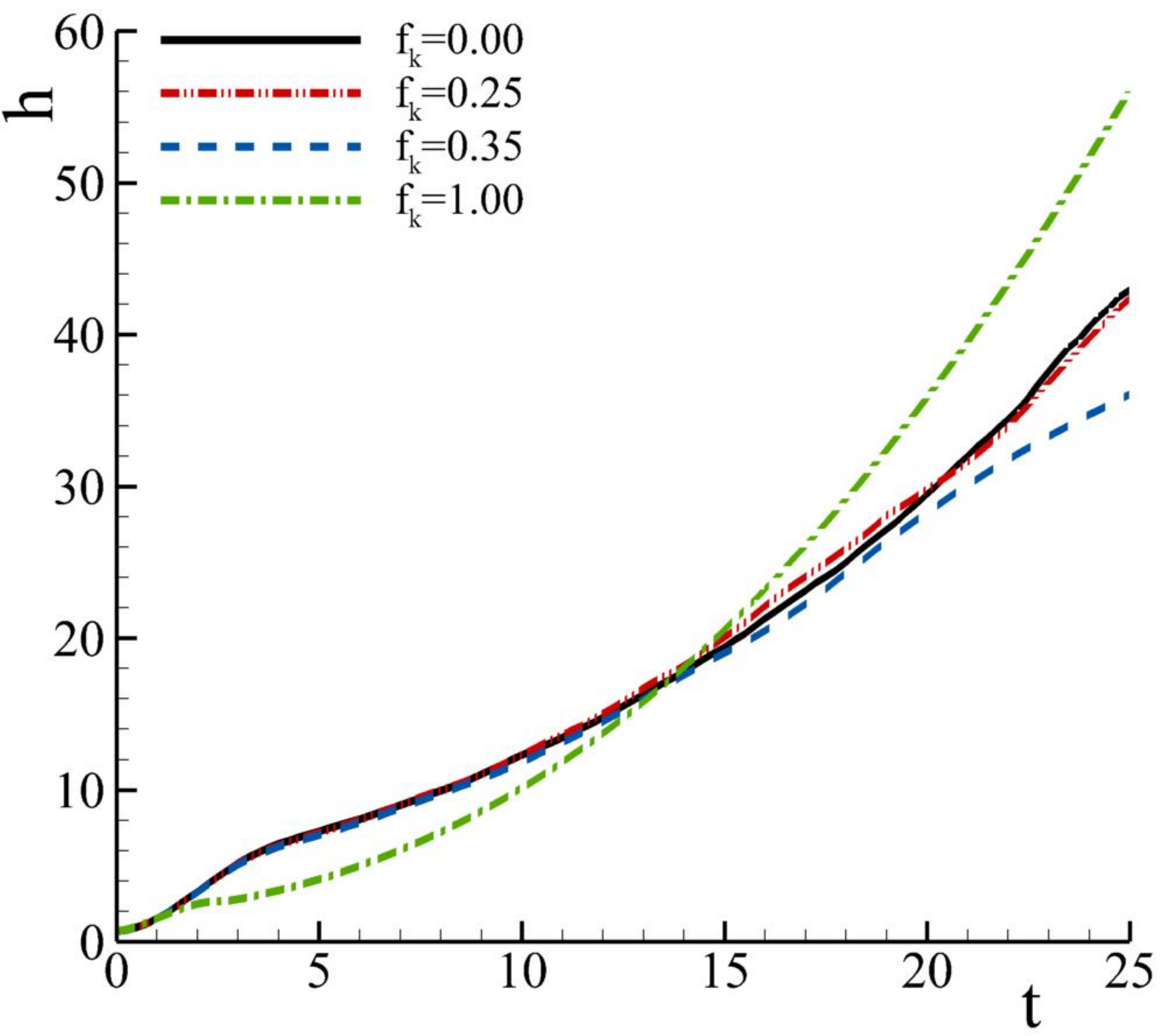}}
\\
\subfloat[$S_3$.]{\label{fig:4.2_1c}
\includegraphics[scale=0.208,trim=0 0 0 0,clip]{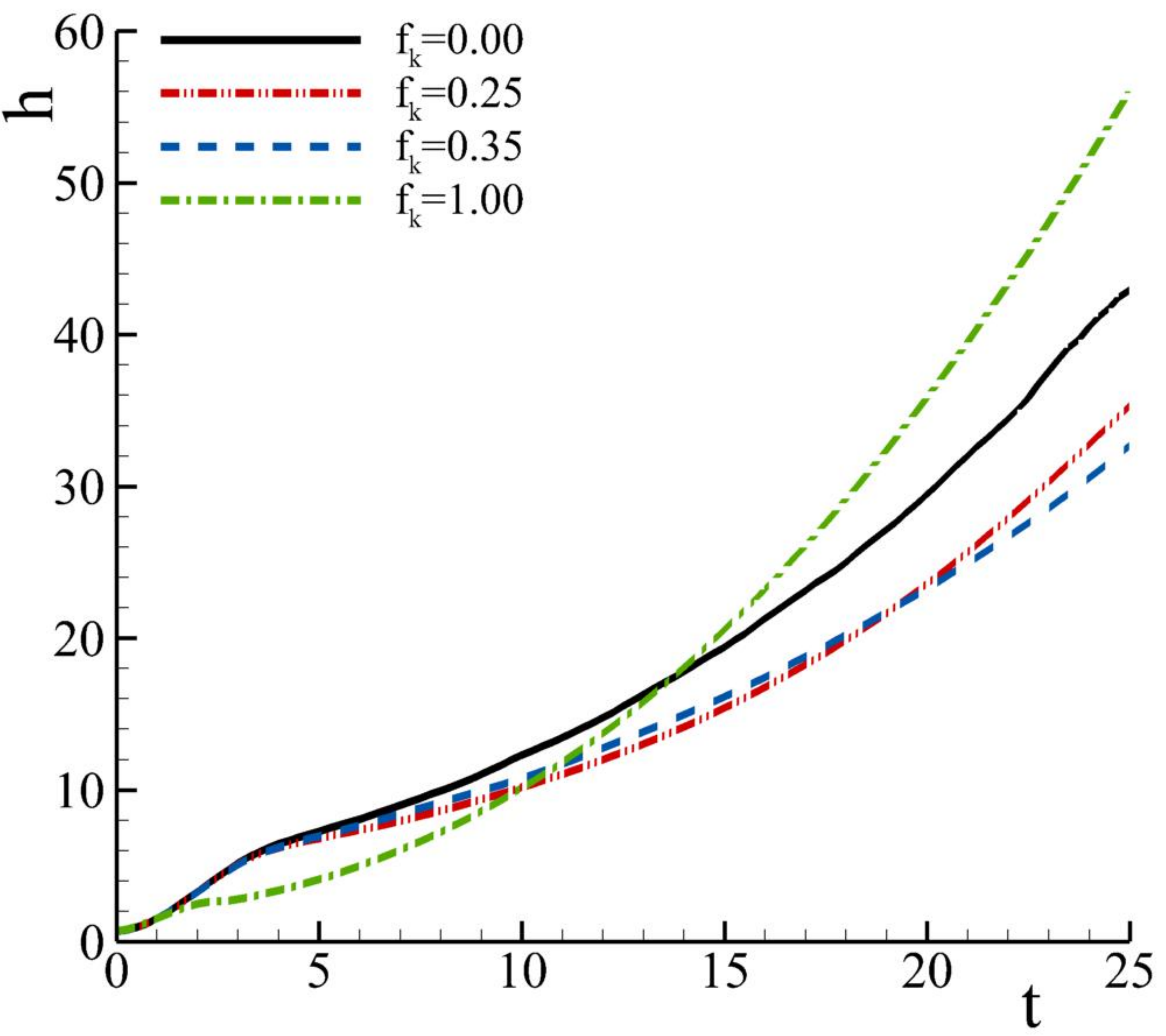}}
\caption{\color{blue}Temporal evolution of the mixing-layer height, $h$, predicted with different $f_k$ and $S_i$; (a) $S_1$, (b) $S_2$, (c) $S_3$.}
\label{fig:4.2_1}
\end{figure}

The modeling accuracy of SRS models is dictated by the physical resolution, its impact on the unresolved-to-total fraction of each dependent quantity of the closure, and the fidelity of the closure. This section evaluates the accuracy of the PANS BHR-LEVM computations, focusing on its dependence on the first two aspects.

Figure \ref{fig:4.2_1} presents the temporal evolution of the mixing-layer height, $h$, obtained with different $f_k$ and $S_i$. For the sake of clarity, simulations with $f_k=0.50$ are not shown because their accuracy lies between that of the computations with $f_k=0.35$ and $1.00$. The results of simulations using $f_k=0.00$ (reference solution) indicate the existence of two distinct growth rates of the mixing-layer height: $i)$ until $t \approx 2.5$; and $ii)$ after $t \approx 2.5$. The first period corresponds to the linear regime, in which the flow is laminar, it can be studied through linear stability theory, and the mixing-layer height grows exponentially. The second interval belongs to the nonlinear regime and, as such, features laminar, transitional, and turbulent flow. The mixing-layer height grows nonlinearly, and more rapidly at late times when the flow features high-intensity turbulence (with $\sim t^2$). This can be measured through the averaged growth rate of the mixing-layer half-width, $\alpha_b$, presented in table \ref{tab:4.2_1}. It shows that this quantity is approximately $0.025$, matching the value obtained in the DNS study of \citeauthor{LIVESCU_PD_2021} \cite{LIVESCU_PD_2021}.

The comparison of the simulation using $f_k=1.00$ against the reference solution indicates that the BHR-LEVM closure does not accurately represent the spatio-temporal development of all flow scales since $h$ is poorly predicted after $t\approx 2$. Despite being initially thinner, $h$ is overpredicted in the late stages of the nonlinear regime. As a result, $\alpha_b$ grows from $0.025$ ($f_k=0.00$) to $0.043$ ($f_k=1.00$). The results also point out that the turbulent closure has a meaningful impact on the flow dynamics in the linear regime due to the differences between the solutions obtained with $f_k=0.00$ and $1.00$ at $t\approx 2$. In this regime, the flow is laminar and, as such, the total turbulent stresses should be negligible ($\tau^1(V_i,V_j)\approx 0$) and $h(f_k=0.0)\approx h(f_k=1.0)$. It is important to recall that most RANS closures have not been designed to predict transient and transitional flows.

As expected, refining the physical resolution of the computations ($f_k \rightarrow 0$) enhances the accuracy of the results by reducing the comparison error, i.e., the difference between a given and reference solution \cite{ASME_BOOK_2009}. The data also show that the computations converge toward the reference solution upon $f_k$ refinement and lead to minor comparison errors at sufficiently fine values of $f_k$. The exceptions are the calculations using $S_3$.

\begin{table}
\centering
\setlength\extrarowheight{3pt}
\caption{Averaged growth rate of the mixing-layer half-width, $\alpha_b$, predicted with different $f_k$ and $S_i$.}
\label{tab:4.2_1}   
\begin{tabular}{C{1.2cm}C{1.2cm}C{1.2cm}C{1.2cm}C{1.2cm}C{1.2cm}}
\hline
$f_k$			& $0.00$	&	$0.25$	 	&	$0.35$		&	$0.50$		&$1.00$	\\ [3pt] \hline
$S_1$			&				&	0.016		&	0.012		&	0.027	  	&				\\[3pt] \cline{1-1}\cline{3-5}
$S_2$			& 0.025		&	0.024		&	0.015		&	0.016  		&0.043		\\[3pt] \cline{1-1}\cline{3-5}
$S_3$			&				&	0.018		&	0.012		&	0.013	  	&				\\[3pt] \hline 
\end{tabular}
\end{table}

Turning our attention to the selection of $f_\phi$ and $S_i$, figure \ref{fig:4.2_1} and table \ref{tab:4.2_1} confirm the importance of considering the unresolved-to-total ratio of each dependent quantity of the closure to obtain efficient high-fidelity computations. Regarding strategy $S_1$, the results indicate that the accuracy of the simulations improves monotonically with $f_k$, and the computation using $f_k=0.25$ is in good agreement with the reference solution. Nonetheless, it is still possible to observe some discrepancies at $t>20$. These are likely caused by the assumption $f_a= 1.0$ because they are not observed in computations using $S_2$ and $f_k=0.25$. Besides improving the simulations upon $f_k$ refinement, strategy $S_2$ leads to the smallest comparison errors. Figure \ref{fig:4.2_1b} and table \ref{tab:4.2_1} illustrate that the simulation with $S_2$ and $f_k=0.25$ is in excellent agreement with the reference data. For example, $\alpha_b$ predicted at $f_k=0.00$ and $0.25$ are approximately $0.025$ and $0.024$. The results also show that even $f_k=0.35$ improves the computations considerably and leads to smaller comparison errors than $f_k=0.25$ and $S_1$. These results confirm the importance of considering $f_a$ at late times and the impact of $f_\phi$ and $S_i$ on the simulations.

On the other hand, computations relying on strategy $S_3$ exhibit significant comparison errors and do not converge toward the reference solution upon $f_k$ refinement. This behavior begins in the early stages of the nonlinear growth regime. We emphasize that computations using this approach to set $f_\phi$ and $f_k<0.50$ experienced robustness issues, requiring a CFL number not exceeding $0.5$. These numerical issues grew upon grid refinement, being most pronounced for the finest mesh, which possesses sufficient resolution to resolve the most unstable mode of the initial perturbations. We attribute these problems to the fully-developed turbulence assumption embedded in strategy $S_3$, which is not observed in early flow times. This aspect is analyzed in Section \ref{sec:4.3}.

\begin{figure}[t!]
\centering
\subfloat[$f_k=0.00$.]{\label{fig:4.2_2a}
\includegraphics[scale=0.208,trim=0 0 0 0,clip]{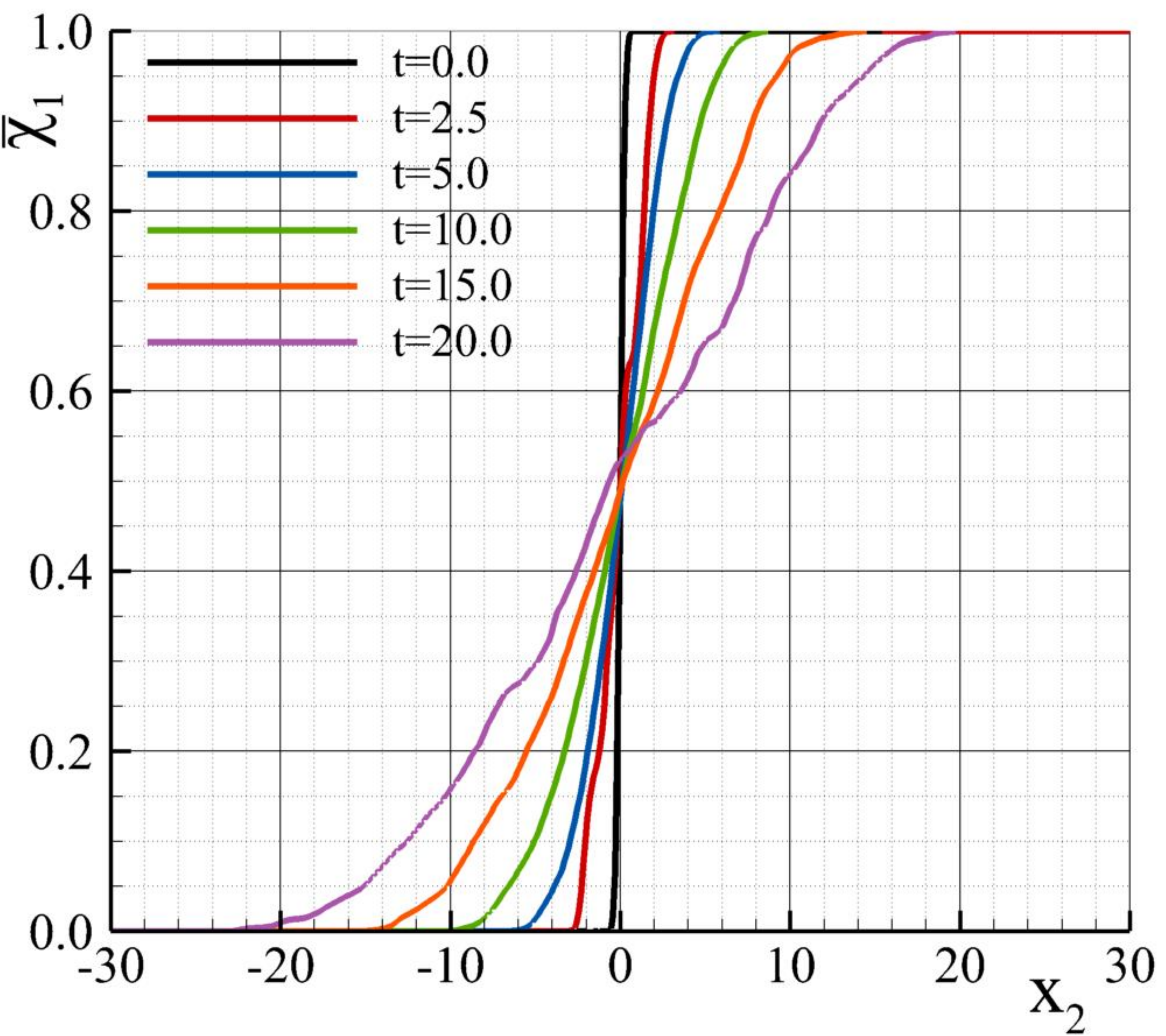}}
\\
\subfloat[$f_k=0.25$ and $S_2$.]{\label{fig:4.2_2b}
\includegraphics[scale=0.208,trim=0 0 0 0,clip]{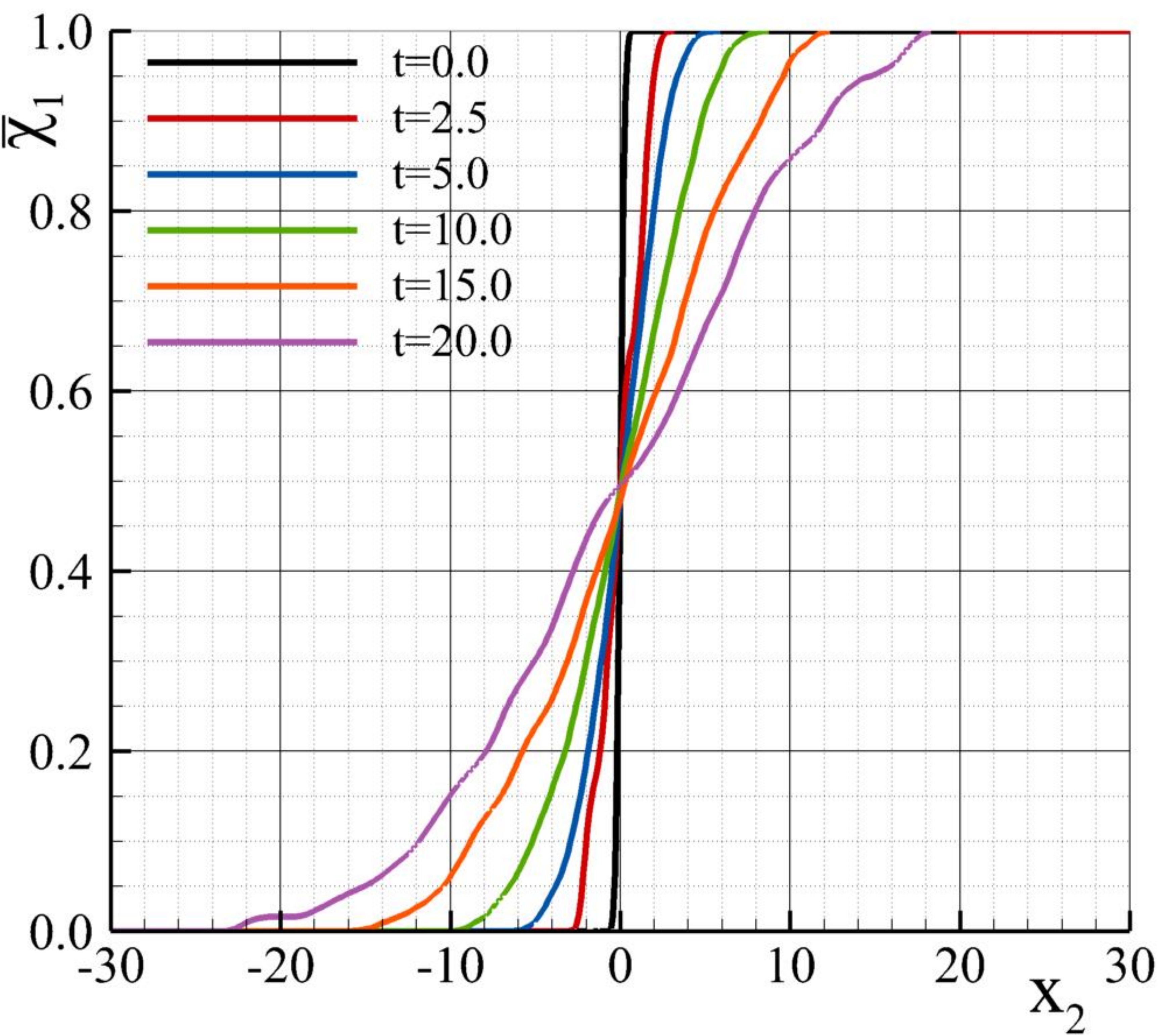}}
\\
\subfloat[$f_k=1.00$.]{\label{fig:4.2_2c}
\includegraphics[scale=0.208,trim=0 0 0 0,clip]{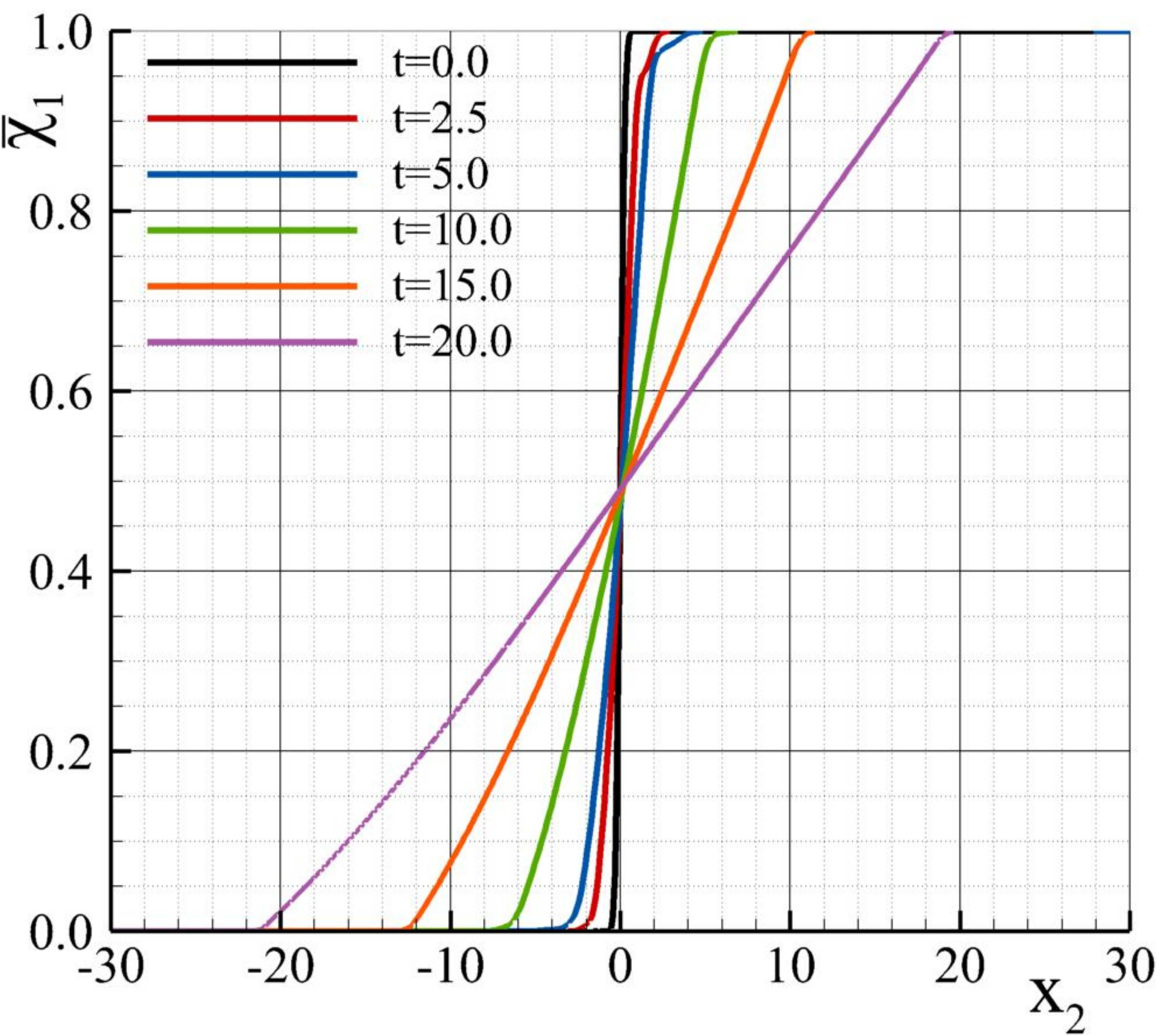}}
\caption{\color{blue}Temporal evolution of the mixture fraction, $\overline{\chi}_1$, predicted with different $f_k$; (a) $f_k=0.00$, (b) $f_k=0.25$ and $S_2$, (c) $f_k=1.00$.}
\label{fig:4.2_2}
\end{figure}

Next, figure \ref{fig:4.2_2} depicts the temporal evolution of the mixture fraction, $\overline{\chi}_1$ (equation \ref{eq:A_4}). Since the behavior of the results is similar to those of figure \ref{fig:4.2_1}, we only present the cases using $f_k=0.00$, $f_k=0.25$ and $S_2$, and $f_k=1.00$. The plots show that the calculations using $f_k=0.00$ and $0.25$ are in close agreement and lead to nonlinear profiles of $\overline{\chi}_1$. Their shape indicates that the mixing rate varies from the center to the boundaries of the mixing-layer. However, this is not observed in the simulation with $f_k=1.00$ (RANS).  The $\overline{\chi}_1$ profiles are linear, indicating a higher global mixing rate. This further suggests that $\nu_u$ and $\tau^1(V_i,V_j)$ are overpredicted with $f_k=1.00$. In addition, figure \ref{fig:4.2_2} shows that the $\overline{\chi}_1$ profiles are more irregular for finer $f_k$. This stems from the increased resolution of finer scale features in the resolved fields. A larger number of sampling points would be necessary to improve the planar averaging at late times and smooth out the profiles.

\begin{figure}[t!]
\centering
\subfloat[$S_1$.]{\label{fig:4.2_3a}
\includegraphics[scale=0.208,trim=0 0 0 0,clip]{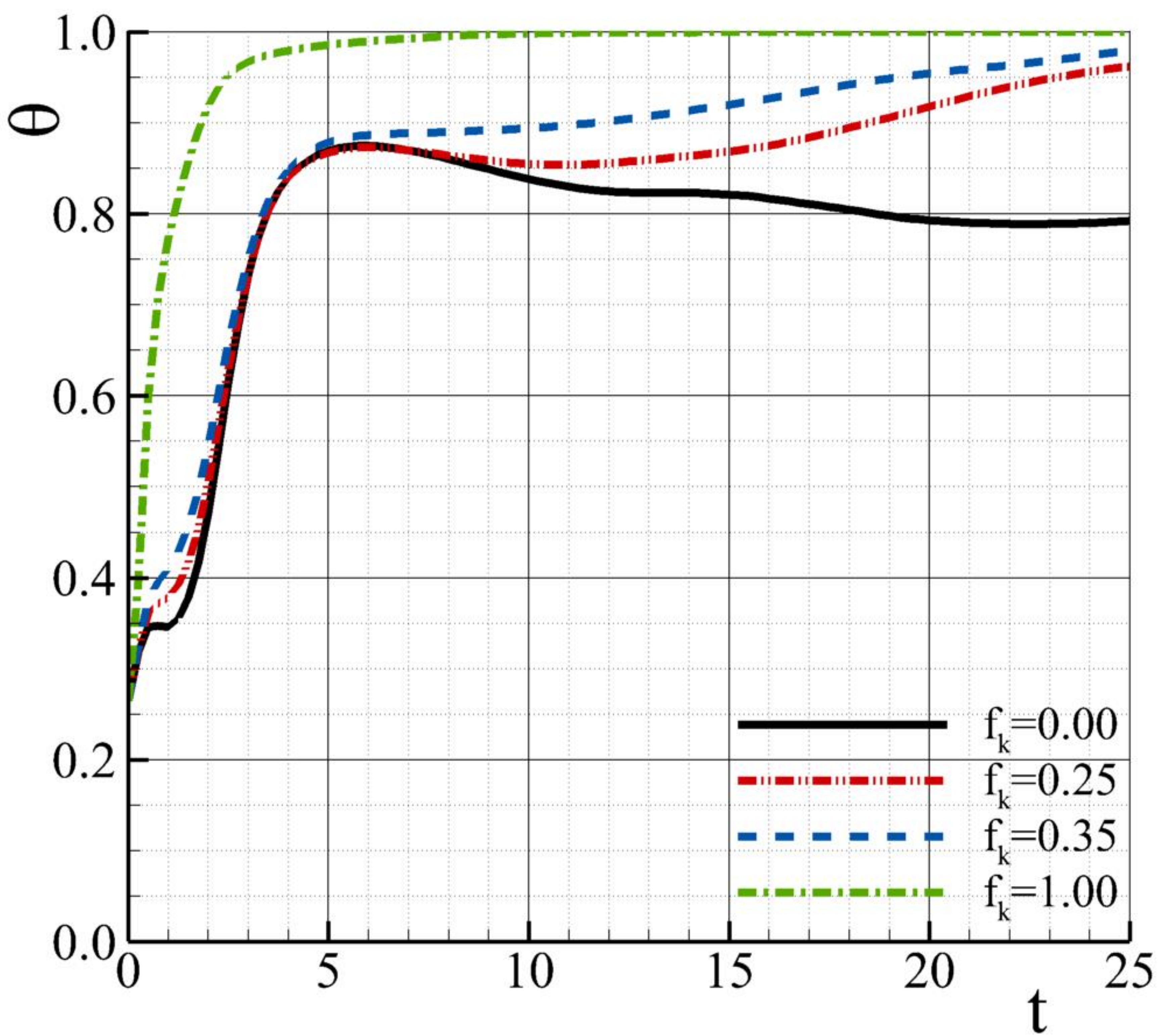}}
\\
\subfloat[$S_2$.]{\label{fig:4.2_3b}
\includegraphics[scale=0.208,trim=0 0 0 0,clip]{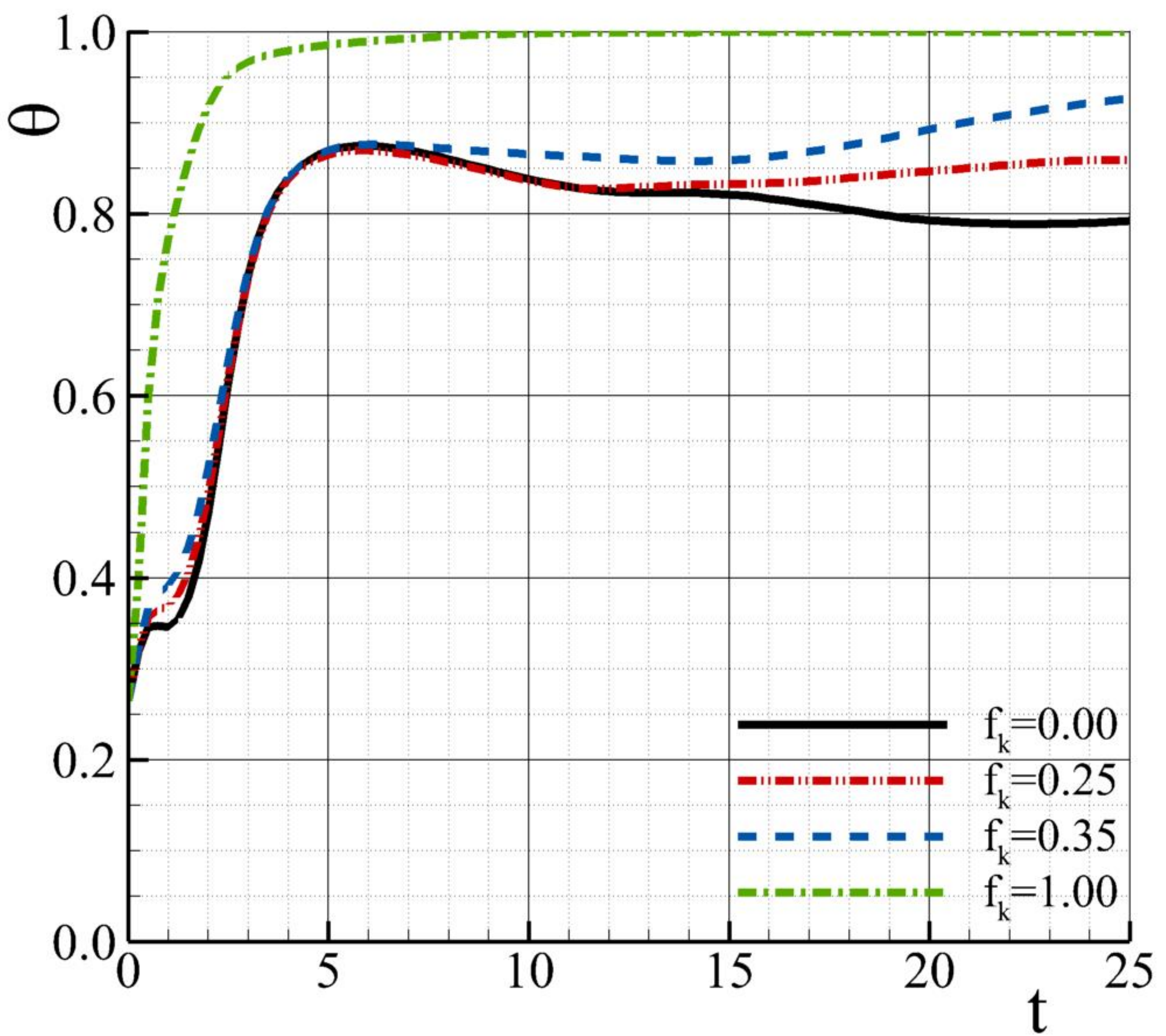}}
\\
\subfloat[$S_3$.]{\label{fig:4.2_3c}
\includegraphics[scale=0.208,trim=0 0 0 0,clip]{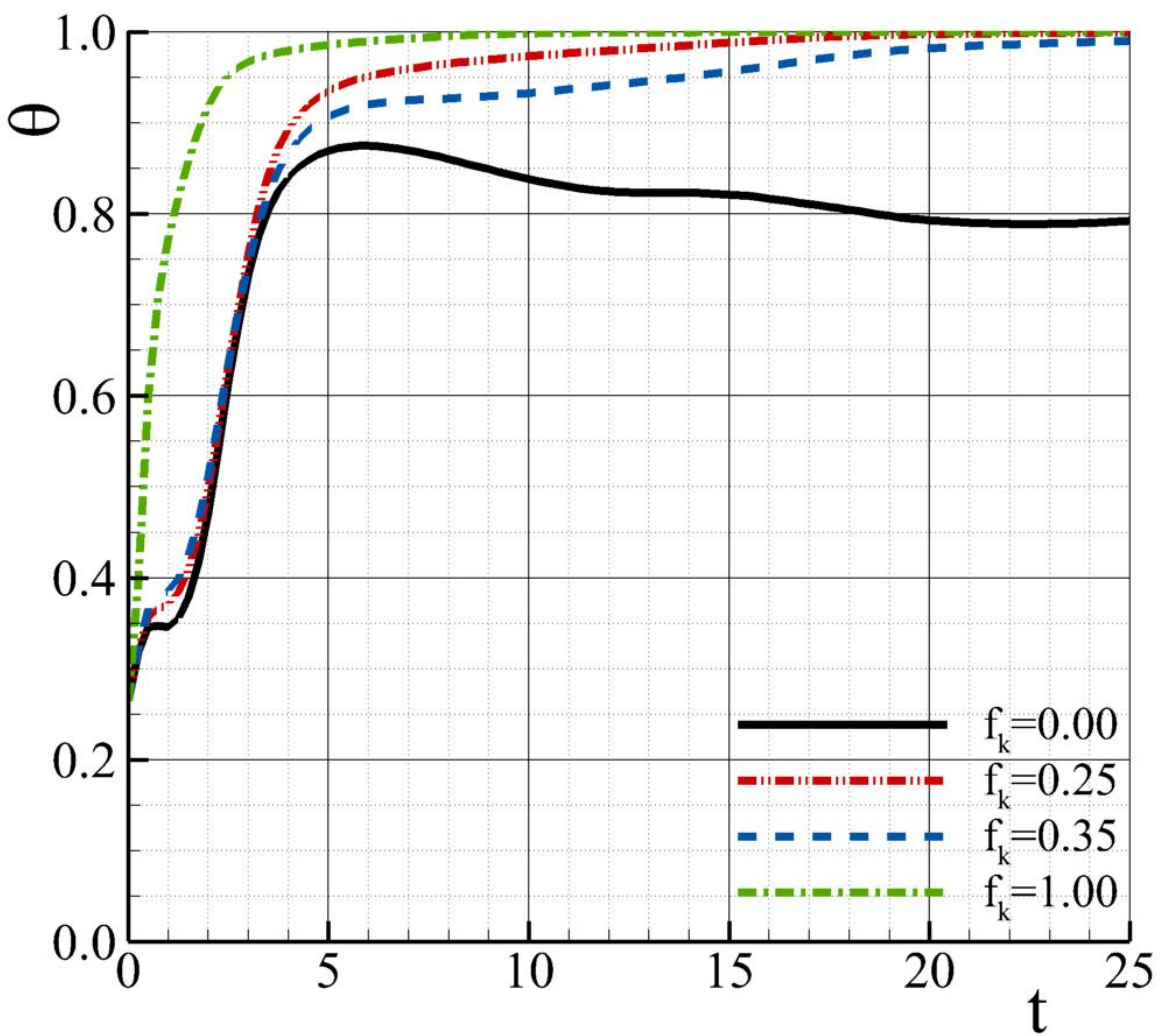}}
\caption{\color{blue}Temporal evolution of the molecular mixing parameter, $\theta$, predicted with different $f_k$ and $S_i$; (a) $S_1$, (b) $S_2$, (c) $S_3$.}
\label{fig:4.2_3}
\end{figure}

Figure \ref{fig:4.2_3} demonstrates that the former results are also noticed in the molecular mixing properties of the mixture. It depicts the temporal evolution of the molecular mixing parameter, $\theta$ (equation \ref{eq:A_8}), predicted with different $f_k$ and $S_i$. This parameter quantifies the homogeneity of the mixing-layer, with $\theta=0$ and $\theta=1$ referring to a heterogeneous and homogeneous mixture, respectively. We stress that the homogeneity degree of the mixture is crucial to numerous variable-density problems, such as combustion, dispersion of pollutants, and material mixing. 

The reference simulation shows $\theta \approx 0.26$ at $t=0$. Immediately after this time, the mixing parameter experiences a pronounced increase during the linear and early nonlinear regime ($t \leq 5$). After this period, $\theta$ converges toward a value of $0.8$ ($t \approx 25$). Note that this parameter should ideally be equal to zero at $t=0$, this meaning that the initial perturbations were fully resolved. However, this is computationally very intensive, and our grid refinement exercises have shown that the present mesh resolution is adequate for this work - see Appendix \ref{sec:B}. Initial values of $\theta$ this high or larger are not uncommon even in published DNS studies. Regarding the RANS ($f_k=1.00$) calculation, the results show that $\theta$ is initially equal to $0.26$ but reaches $0.95$ before $t=5$, converging to one during the simulated time. This points out that $\nu_u$ and $\tau^1(V_i,V_j)$ are overpredicted and responsible for the enhanced mixing degree of the mixture. On the other hand, the successive refinement of $f_k$ improves the simulations significantly, leading to a good agreement between simulations using $f_k=0.00$ and $f_k=0.25$ (in particular, with $S_2$). Once again, the exceptions are the cases using $S_3$. {\color{blue}It is important to emphasize that the calculation of $\theta$ only considers the resolved density field and, as such, the magnitude of this quantity should slightly increase with $f_k\rightarrow 0$. Yet, the results exhibit the opposite trend, further illustrating the limitations of the closure representing most flow scales.}

Overall, the results of figures \ref{fig:4.2_1} to \ref{fig:4.2_3} and table \ref{tab:4.2_1} demonstrate that the PANS BHR-LEVM can accurately predict the spatio-temporal development of the RT flow if $f_k$ and $S_i$ are adequate. The outcome of the PANS computations is now physically interpreted.
%
%
%
%
\subsection{Physical interpretation of the results}
\label{sec:4.3}
%
\subsubsection{Effective Reynolds number and viscosity}
\label{sec:4.3.1}
\begin{figure}[th!]
\centering
\subfloat[$S_1$.]{\label{fig:4.3.1_1a}
\includegraphics[scale=0.202,trim=0 0 0 0,clip]{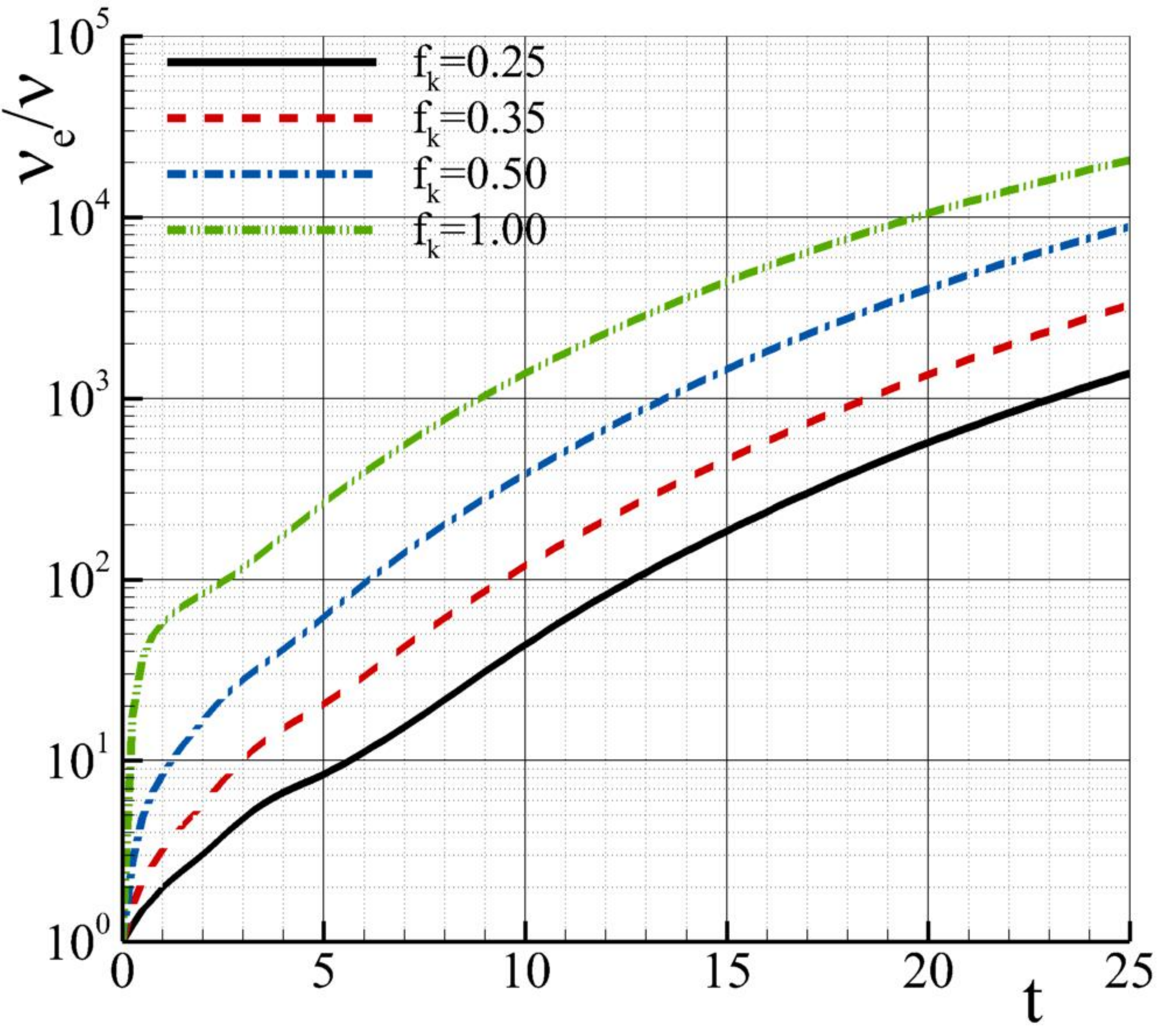}}
\\
\subfloat[$S_2$.]{\label{fig:4.3.1_1b}
\includegraphics[scale=0.202,trim=0 0 0 0,clip]{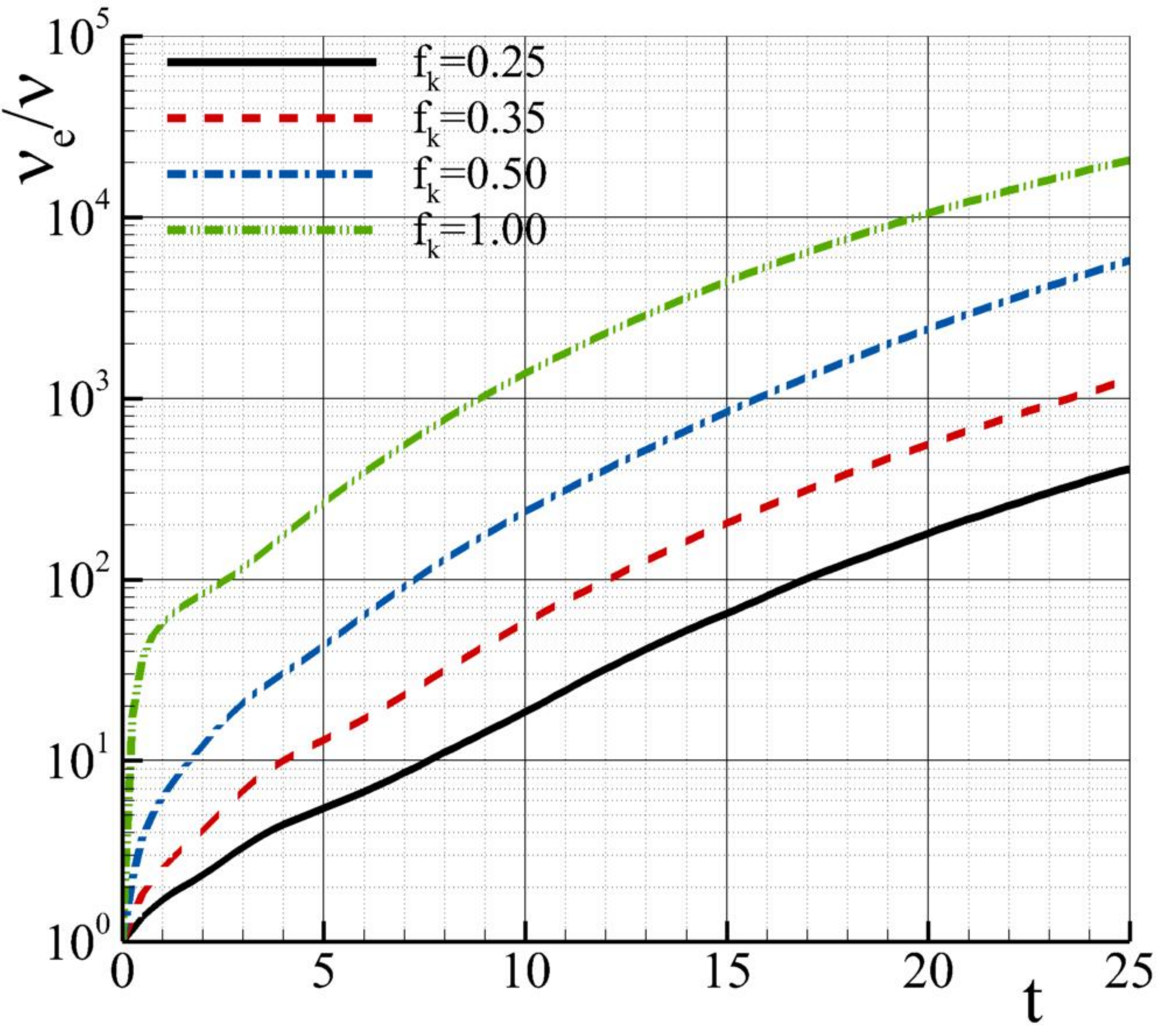}}
\\
\subfloat[$S_3$.]{\label{fig:4.3.1_1c}
\includegraphics[scale=0.202,trim=0 0 0 0,clip]{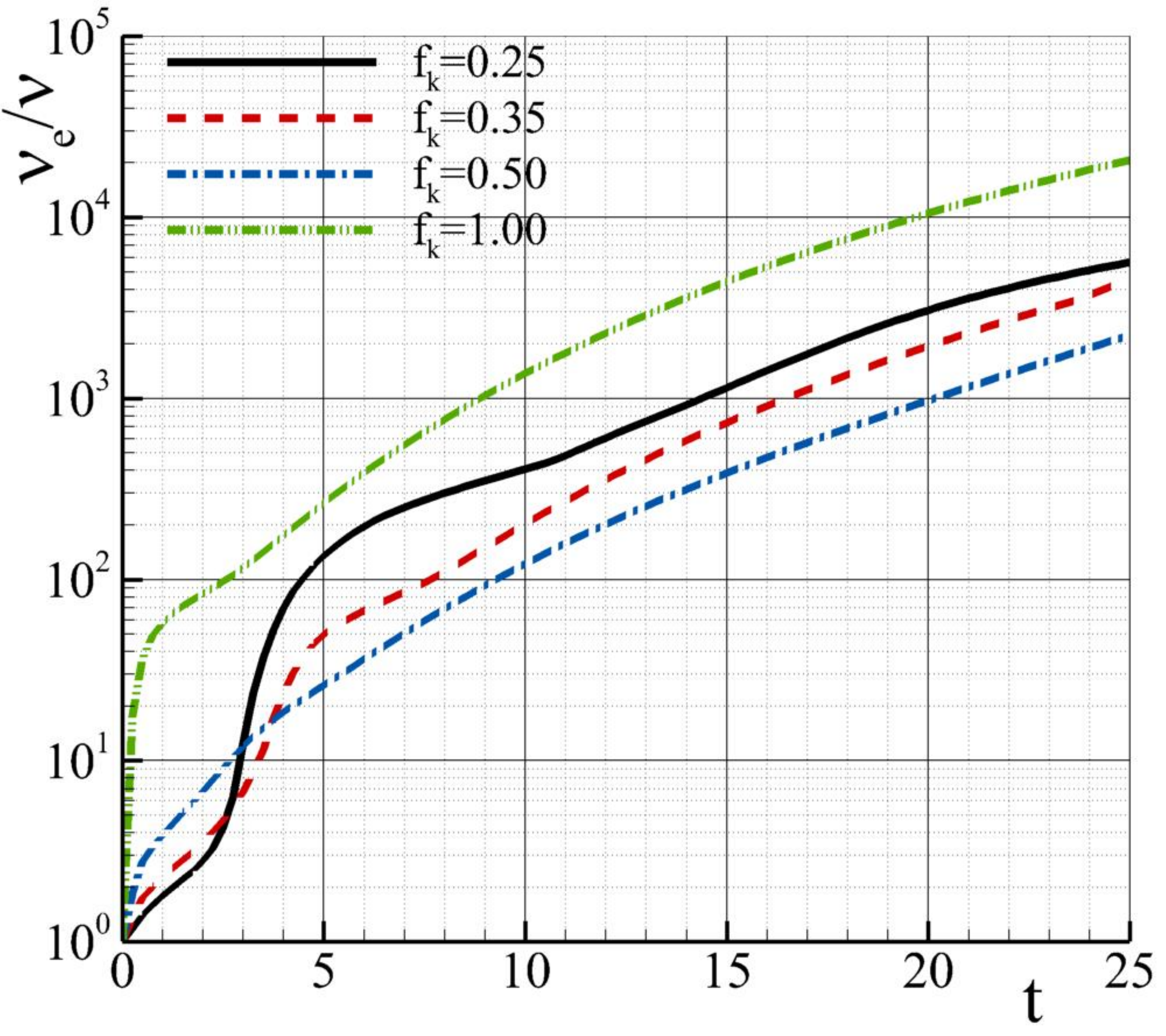}}
\caption{\color{blue}Temporal evolution of the maximum planar averaged effective-to-molecular kinematic viscosity ratio, $\nu_e/\nu$, predicted with different $f_k$ and $S_i$; (a) $S_1$, (b) $S_2$, (c) $S_3$.}
\label{fig:4.3.1_1}
\end{figure}

%
The efficient modeling of turbulent flows is closely dependent on the magnitude of the effective $\mathrm{Re}$,
\begin{equation}
\label{eq:4.3.1_1}
\mathrm{Re}_e \equiv \frac{V_oL_o}{\nu_e} \; ,
\end{equation}
where $L_o$ and $V_o$ are a reference length-scale and velocity, and $\nu_e$ is the effective kinematic viscosity. The latter quantity is defined as the sum of the molecular, $\nu$, turbulent, $\nu_u$, and numerical, $\nu_n$, kinematic viscosities,
\begin{equation}
\label{eq:4.3.1_2}
\nu_e \equiv \nu + \nu_u + \nu_n \; .
\end{equation}
The first is a fluid property, the second is due to the turbulence closure and represents the effect of the modeled scales on the resolved ones, and the third stems from the numerical discretization and solution procedure. $\nu_n$ and $\nu_u$ can significantly impact the accuracy of the simulations by excessively decreasing the $\mathrm{Re}_e$ of the computations. As demonstrated in \citeauthor{PEREIRA_JCP_2018} \cite{PEREIRA_JCP_2018,PEREIRA_IJHFF_2019,PEREIRA_OE_2019}, the magnitude of $\mathrm{Re}_e$ is crucial to the simulation of problems highly dependent on $\mathrm{Re}$ and flow regime. This is the case for transitional flows around circular cylinders \cite{WILLIAMSON_ARFM_1996,ZDRAVKOVICH_BOOK_1997}. To assess the importance of $\mathrm{Re}_e$, figure \ref{fig:4.3.1_1} depicts the ratio effective-to-molecular kinematic viscosity, $\mathrm{\nu_e}/\nu$, of the simulations disregarding $\nu_n$. This quantity measures how much the $\mathrm{Re_e}$ changes with $f_k$ and $S_i$. We stress that values of $\nu_u/\nu$ exceeding ${\cal{O}}(1)$ typically imply strong turbulent transport. 

As expected, $\nu_e/\nu$ decreases monotonically upon $f_k$ refinement. For example, at $t=25$, it decreases from $20,737$ with $f_k=1.00$ to $1376$ ($S_1$) and $408$ ($S_2$) with $f_k=0.25$. Once again, the exceptions are the computations based on strategy $S_3$. Although $\nu_e/\nu$ decreases with $f_k$ at early times, it increases significantly after $t\approx 3$. This inconsistent behavior, which is observed at $f_k<0.50$ and gets more pronounced upon $f_k$ refinement, is explained in Section \ref{sec:4.3.2}. The comparison of approaches $S_1$ and $S_2$ shows that the second strategy leads to smaller values of $\nu_e/\nu$ and, consequently, larger $\mathrm{Re}_e$. It is also worth noting that simulations performed with $f_k=0.25$ exhibit a smooth growth of $\nu_e/\nu$, whereas those using $f_k=1.00$ lead to a steep increase of this quantity until $t=1$, followed by a smoother growth rate. 

\begin{figure}[t!]
\centering
\subfloat[$f_k=0.25$ and $S_2$.]{\label{fig:4.3.1_2a}
\includegraphics[scale=0.21,trim=0 0 0 0,clip]{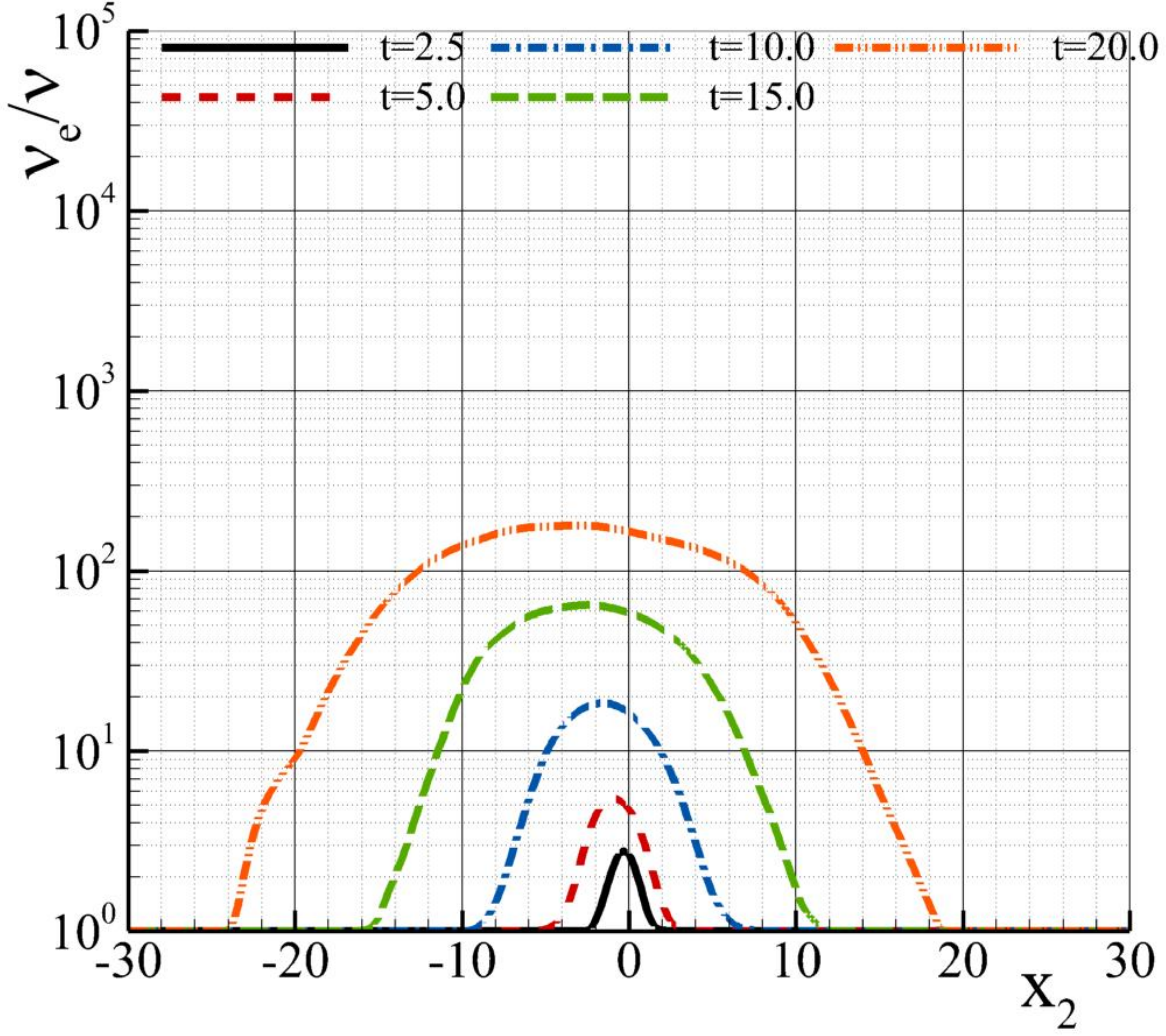}}
\\
\subfloat[$f_k=1.00$.]{\label{fig:4.3.1_2b}
\includegraphics[scale=0.21,trim=0 0 0 0,clip]{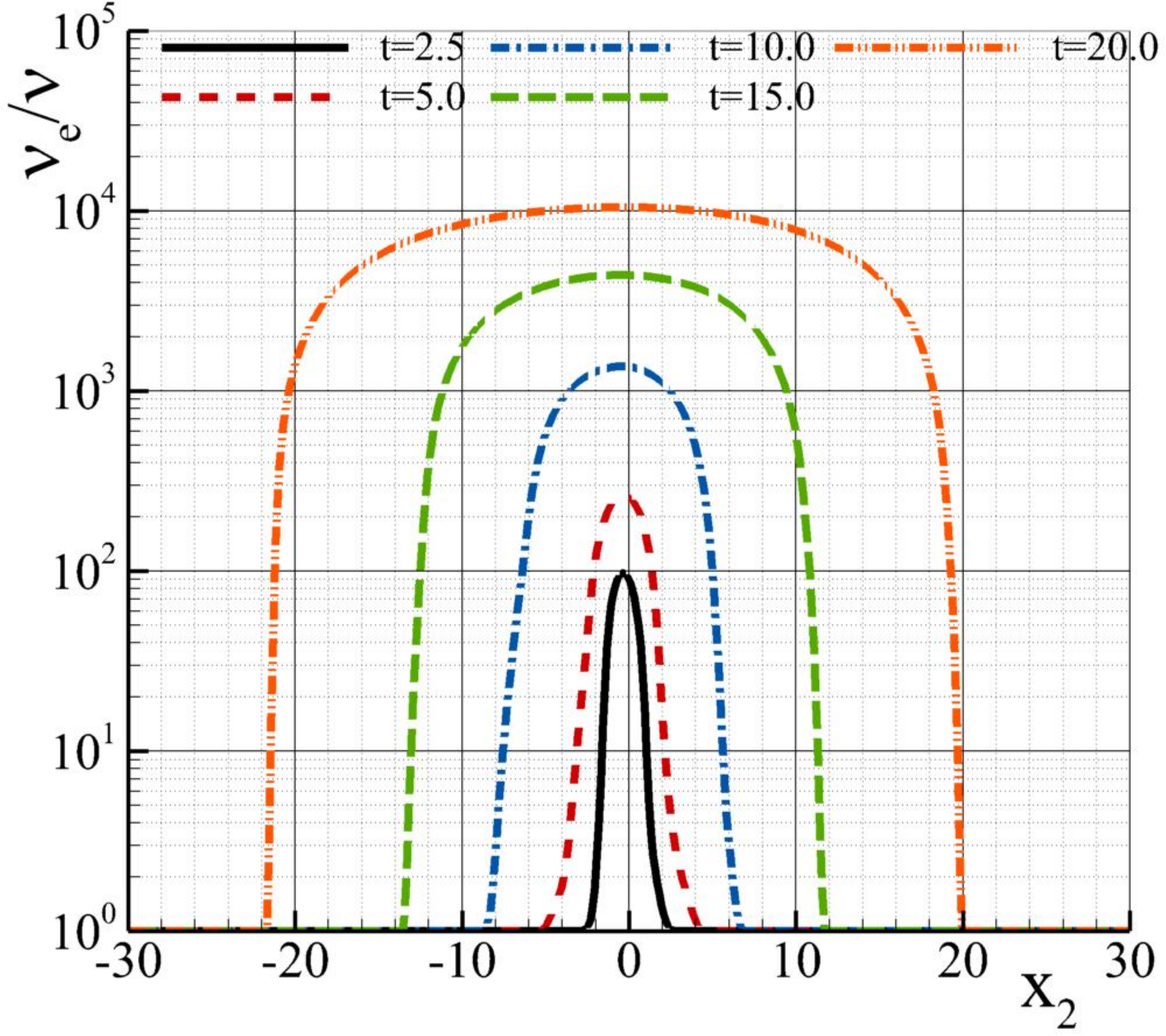}}
\caption{\color{blue}Temporal evolution of $x_2$-profiles of the planar averaged effective-to-molecular viscosity ratio, $\nu_e/\nu$, predicted with (a) $f_k=0.25$ and $S_2$, and (b) $f_k=1.00$.}
\label{fig:4.3.1_2}
\end{figure}

Figure \ref{fig:4.3.1_2} presents the $x_2$-profiles of the planar averaged $\nu_e/\nu$ predicted with $f_k=0.25$ utilizing $S_2$ and $f_k=1.00$ at different times. In addition to the large magnitudes also seen in figure \ref{fig:4.3.1_1}, the RANS profiles of $\nu_e/\nu$ exhibit a sharp increase at the interface of the mixing-layer. At $t=20.0$, this quantity grows from approximately $1$ to $10^4$. $\nu_e/\nu$ becomes more uniform inside the mixing-layer, explaining the shape of the profiles of $\overline{\chi}_1$ (figure \ref{fig:4.2_2c}). In contrast, the simulation using $f_k=0.25$ leads to smaller values of $\nu_e/\nu$, and significantly smoother profiles at the mixing-layer interfaces. In both cases, the peak of $\nu_e/\nu$ occurs in the light fluid side, i.e., at $x_2<0$. However, it moves further into the light fluid side with the decrease of $f_k$.

\begin{figure}[t!]
\centering
\subfloat[$S_1$.]{\label{fig:4.3.1_3a}
\includegraphics[scale=0.266,trim=0 0 0 0,clip]{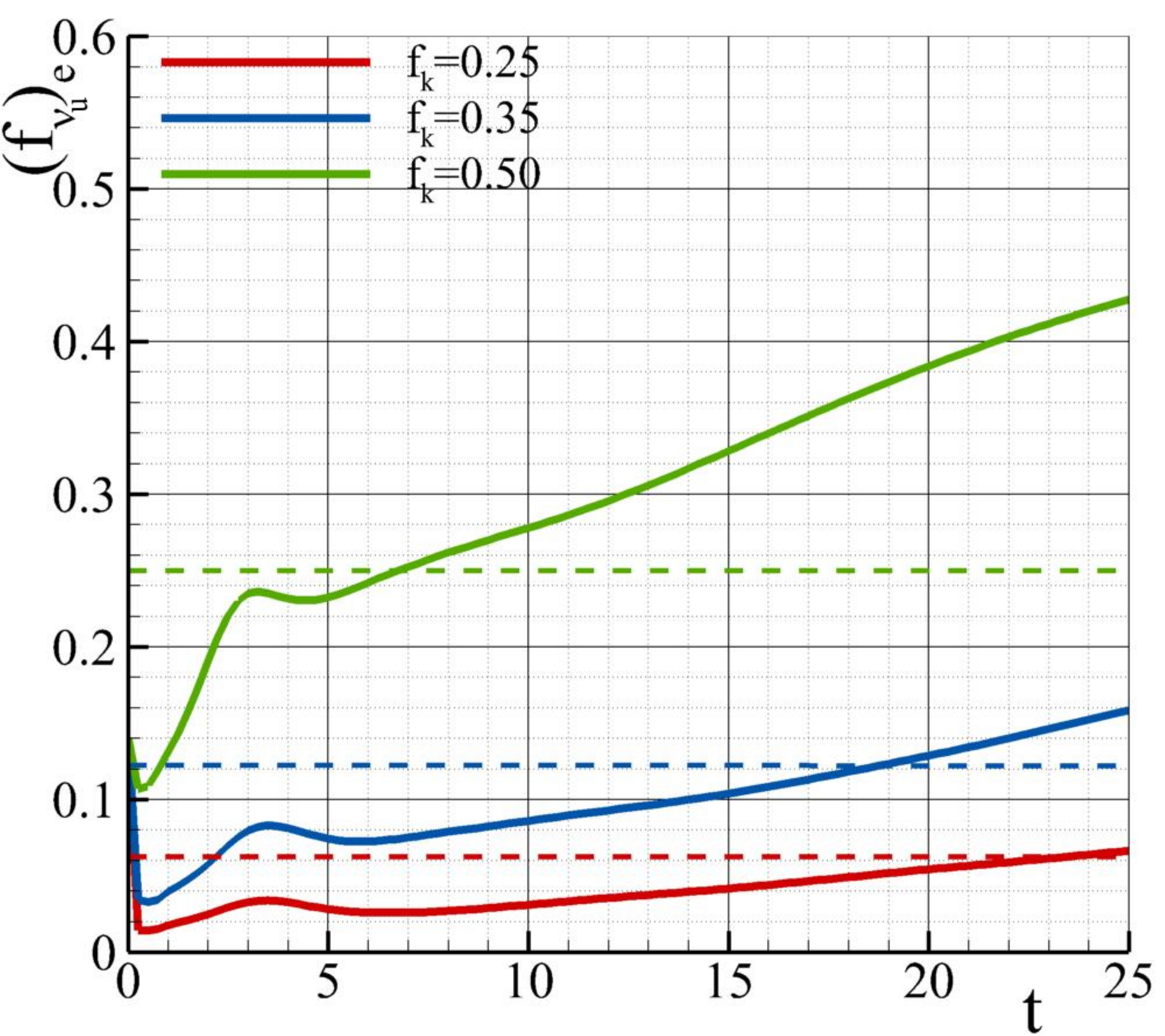}}
\\
\subfloat[$S_2$.]{\label{fig:4.3.1_3b}
\includegraphics[scale=0.266,trim=0 0 0 0,clip]{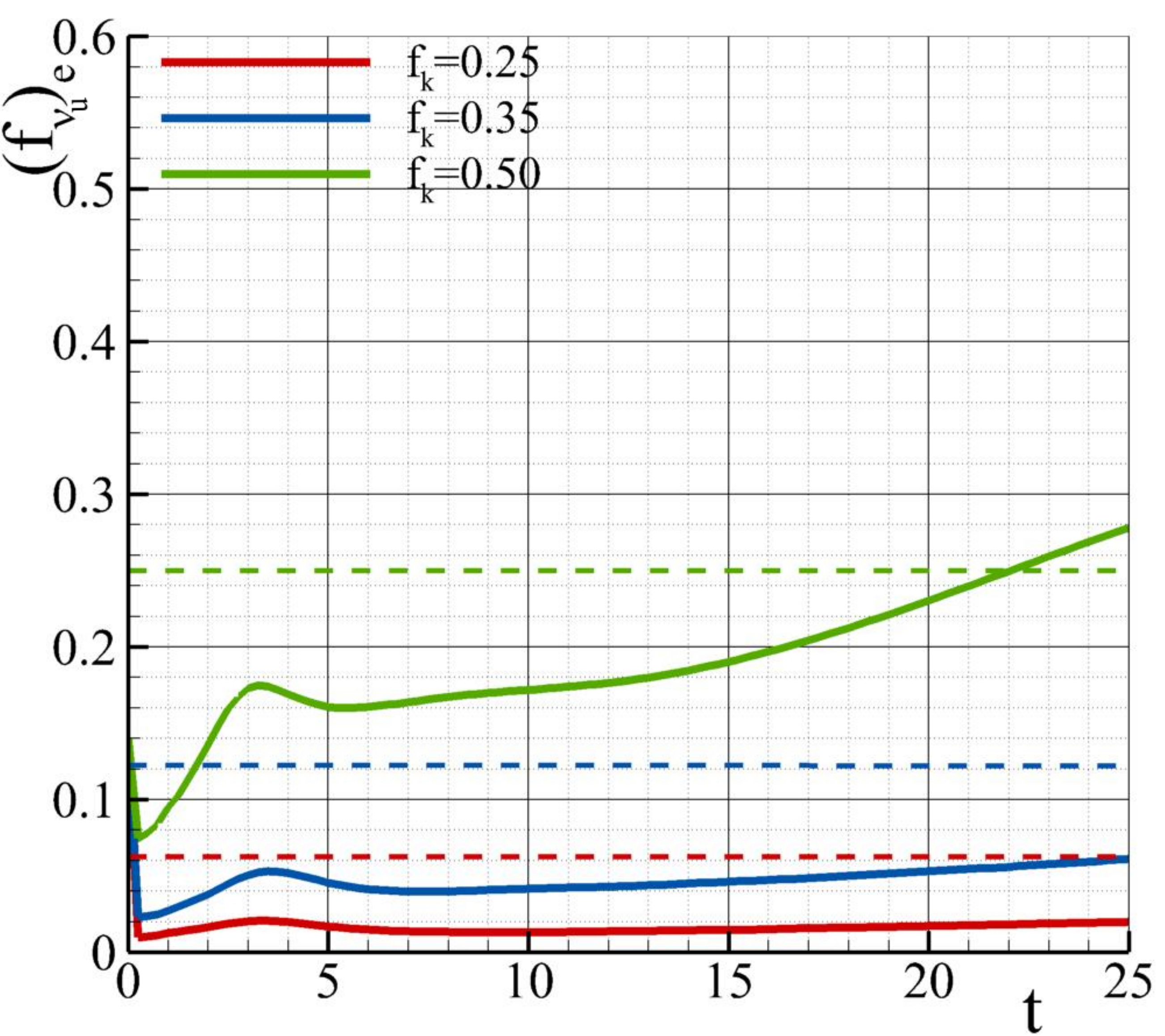}}
\\
\subfloat[$S_3$.]{\label{fig:4.3.1_3c}
\includegraphics[scale=0.266,trim=0 0 0 0,clip]{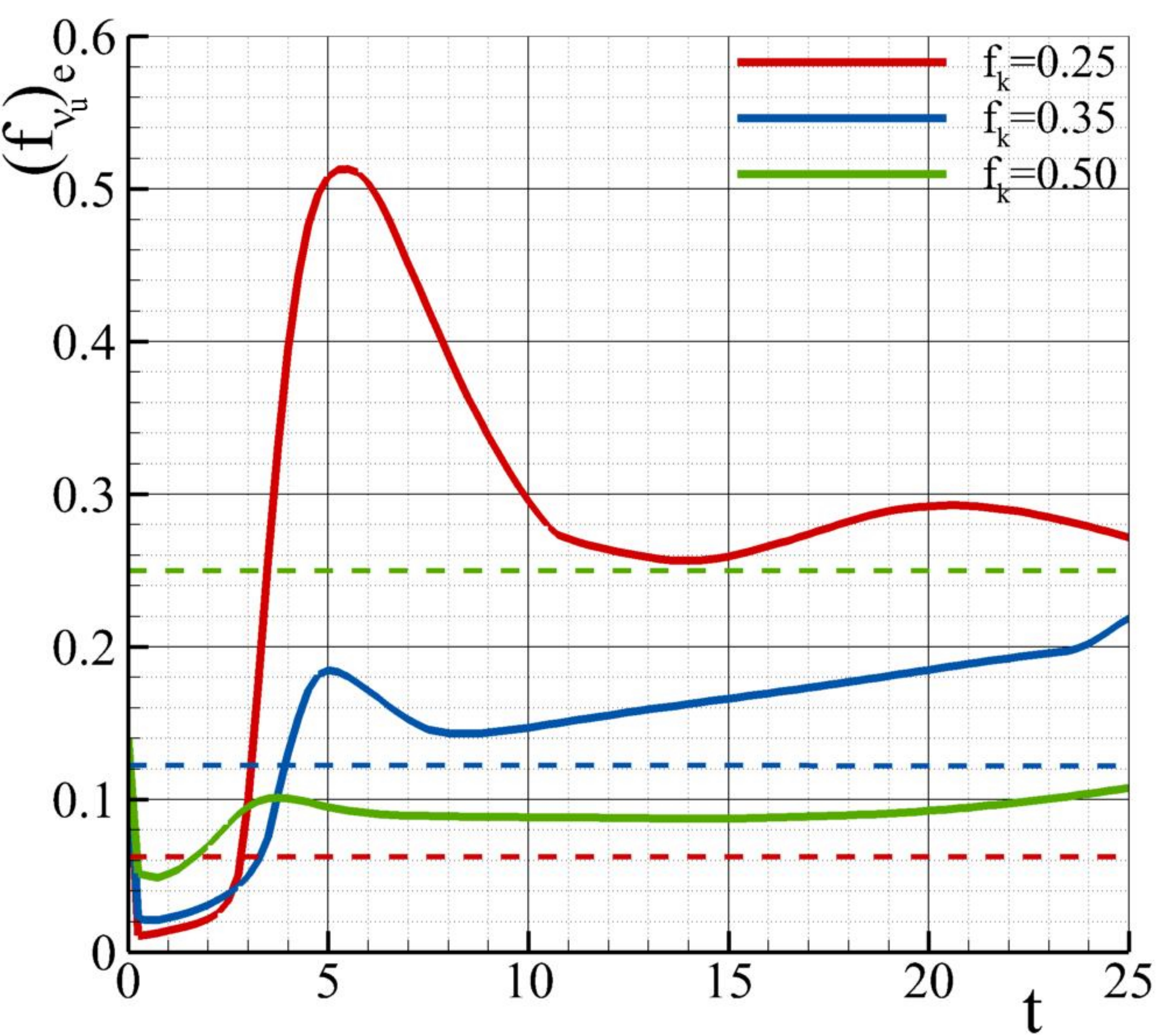}}
\caption{\color{blue}Temporal evolution of the maximum planar averaged effective unresolved-to-total ratio of turbulent viscosity ratio, $(f_{\nu_u})_e$, predicted with different $f_k$ and $S_i$. Dashed lines indicate prescribed values; (a) $S_1$, (b) $S_2$, (c) $S_3$.}
\label{fig:4.3.1_3}
\end{figure}

Next, figure \ref{fig:4.3.1_3} illustrates how the prescribed (dashed line) and effective (solid line) unresolved-to-total ratio of turbulent viscosity, $f_{\nu_u}$ and $(f_{\nu_u})_e$, evolve in time. From the definition of $\nu_u$ (equation \ref{eq:3_2}) and assuming $f_\varepsilon=1.00$, one can calculate these quantities as follows,
\begin{equation}
\label{eq:4.3.1_3}
f_{\nu_u} = f_k^2 \; ,
\end{equation}
\begin{equation}
\label{eq:4.3.1_4}
(f_{\nu_u})_e = \frac{\nu_u}{{(\nu_u)}_{f_k=1.00}} \; ,
\end{equation}
where ${(\nu_u)}_{f_k=1.00}$ is the value of $\nu_u$ computed with $f_k=1.00$. The plots show that $(f_{\nu_u})_e$ decreases monotonically upon $f_k$ refinement with strategies $S_1$ and $S_2$. This behavior becomes more pronounced utilizing the second approach. The data also indicate that the effective and prescribed $f_{\nu_u}$ do not match. The observed differences are caused by the fact that the RT flow is a transient problem and, as such, highly sensitive to memory effects, i.e., a misrepresentation of the flow physics at a time $t$ affects the predictions at a later time $t+\Delta t$ (cumulative modeling errors).  Hence, the well-known limitations of one-point closures to predict the spatio-temporal development of this class of flows leads to a misrepresentation of the unresolved turbulent stress tensor so that $(f_{\nu_u})_e \ne (f_{\nu_u})$. The comparison of strategies $S_1$ and $S_2$ shows that the latter predicts the smallest values of $(f_{\nu_u})_e$, and computations with $f_k\leq 0.35$ converge to nearly constant values of $(f_{\nu_u})_e$ during the simulated time. As expected (see figure \ref{fig:4.3.1_1}), approach $S_3$ leads to a significant increase of $(f_{\nu_u})_e$ upon physical resolution refinement. Also, figure \ref{fig:4.3.1_3c} shows a peak of $(f_{\nu_u})_e$ in the early stages of the nonlinear regime, followed by a substantial decrease. This nonphysical behavior is investigated below. 
%
%
%
%
\subsubsection{Evolution of $k_u$, $a_{2_u}$, and $b_u$}
\label{sec:4.3.2}
%

\begin{figure*}[t!]
\centering
\subfloat[$k_u$ at $f_k=0.25$ and $S_1$.]{\label{fig:4.3.2_1a}
\includegraphics[scale=0.198,trim=0 0 0 0,clip]{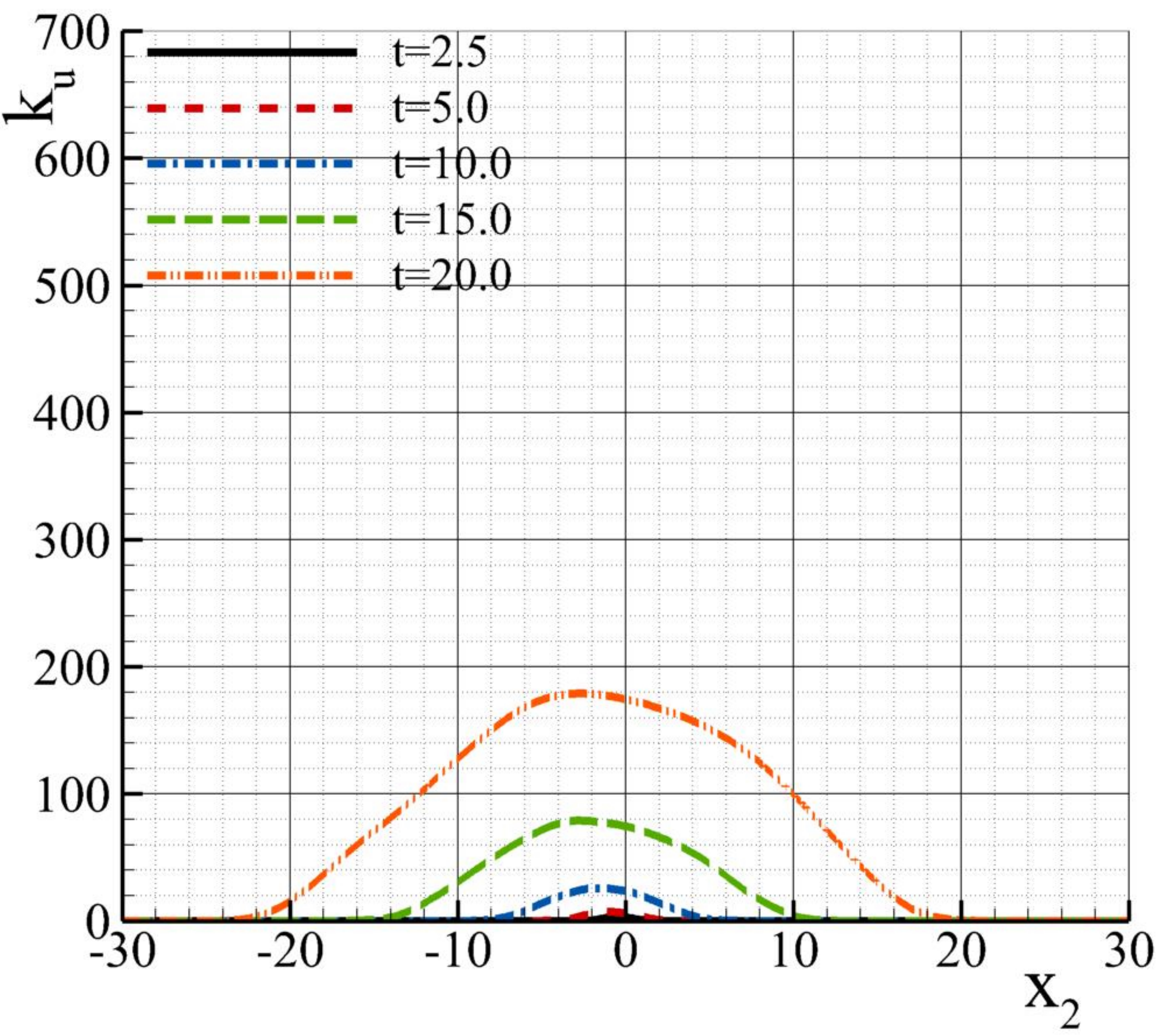}}
~
\subfloat[$a_{2_u}$ at $f_k=0.25$ and $S_1$.]{\label{fig:4.3.2_1b}
\includegraphics[scale=0.198,trim=0 0 0 0,clip]{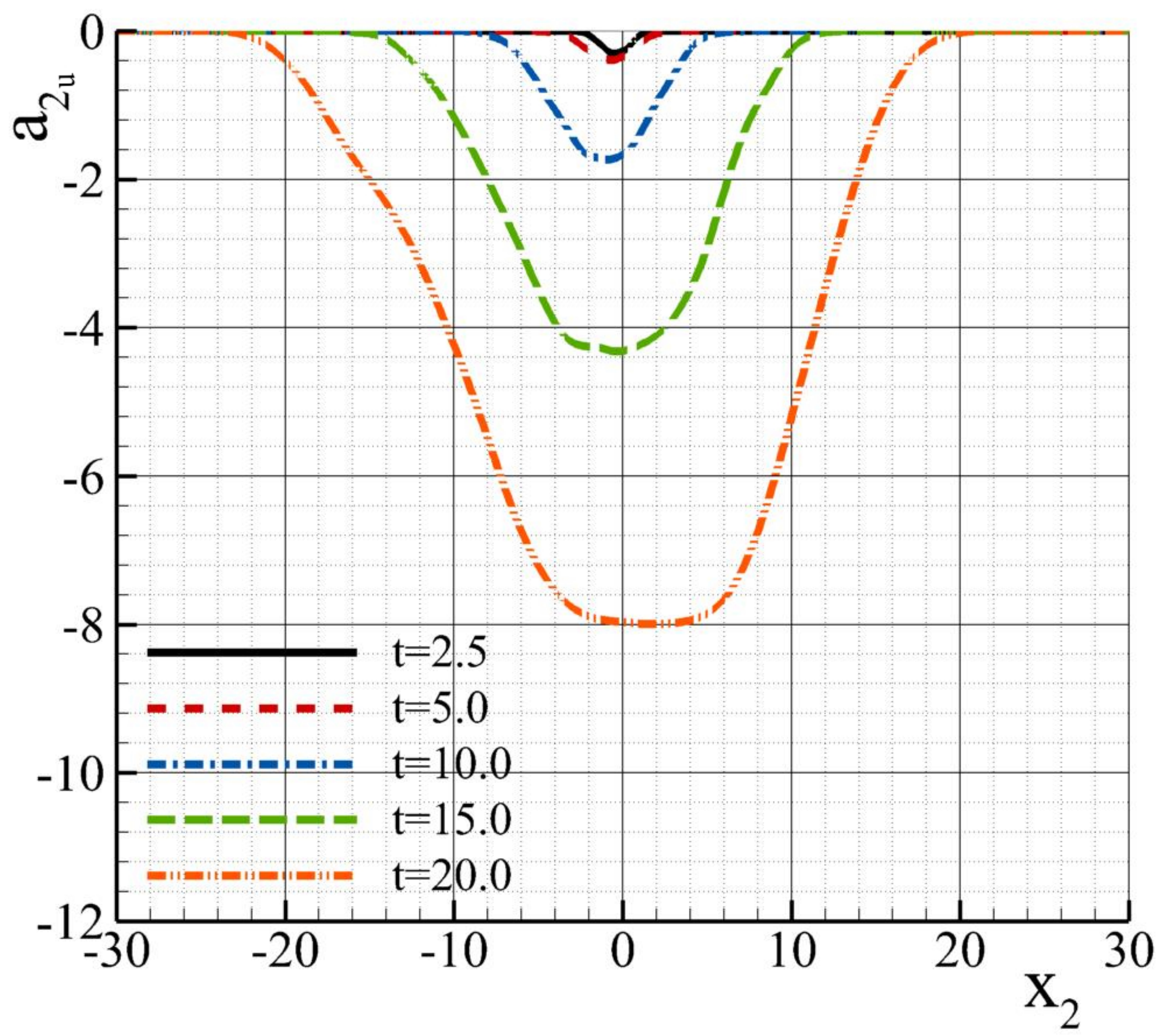}}
~
\subfloat[$b_{u}$ at $f_k=0.25$ and $S_1$.]{\label{fig:4.3.2_1c}
\includegraphics[scale=0.198,trim=0 0 0 0,clip]{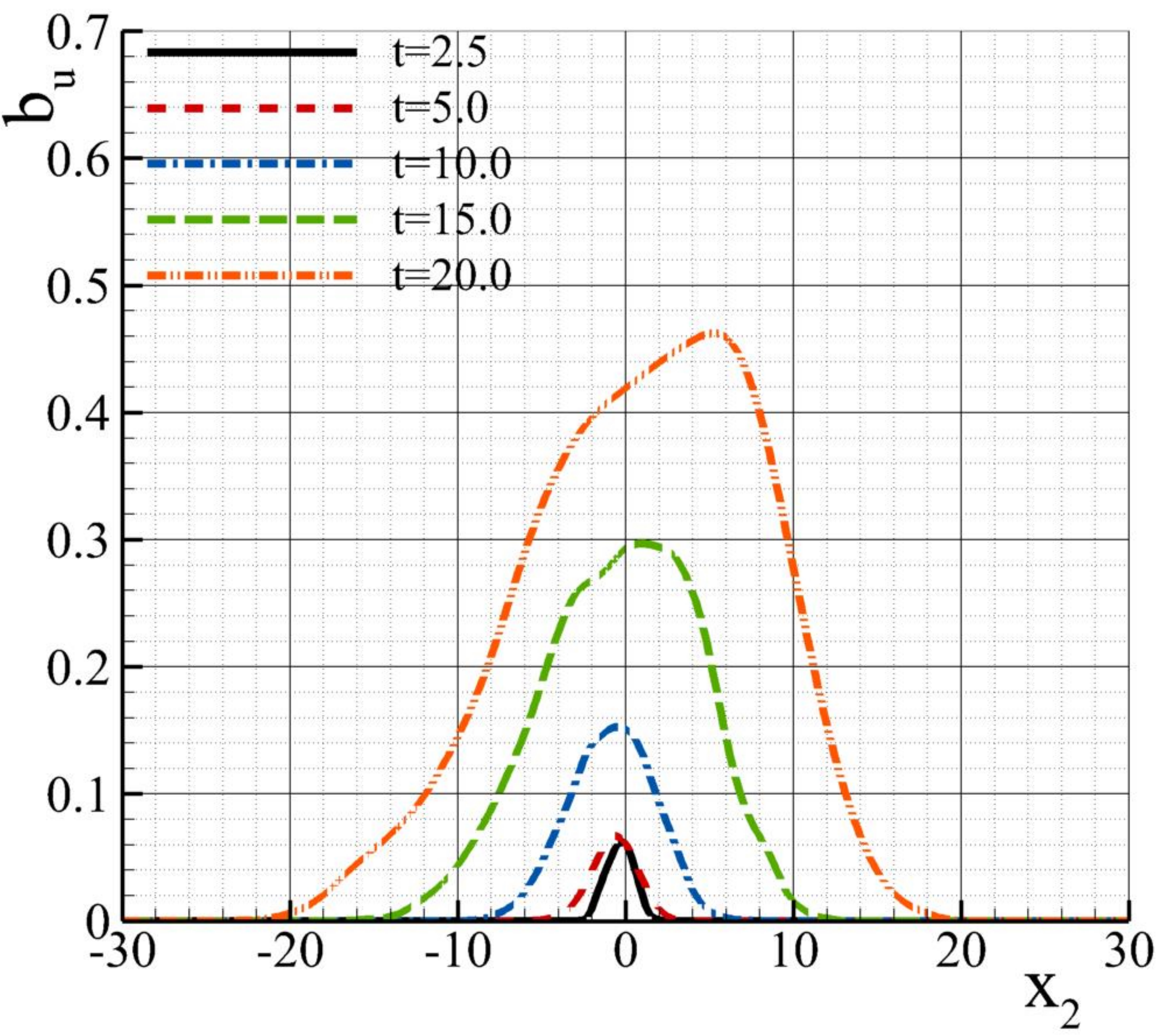}}
\\
\subfloat[$k_u$ at $f_k=0.25$ and $S_2$.]{\label{fig:4.3.2_1d}
\includegraphics[scale=0.198,trim=0 0 0 0,clip]{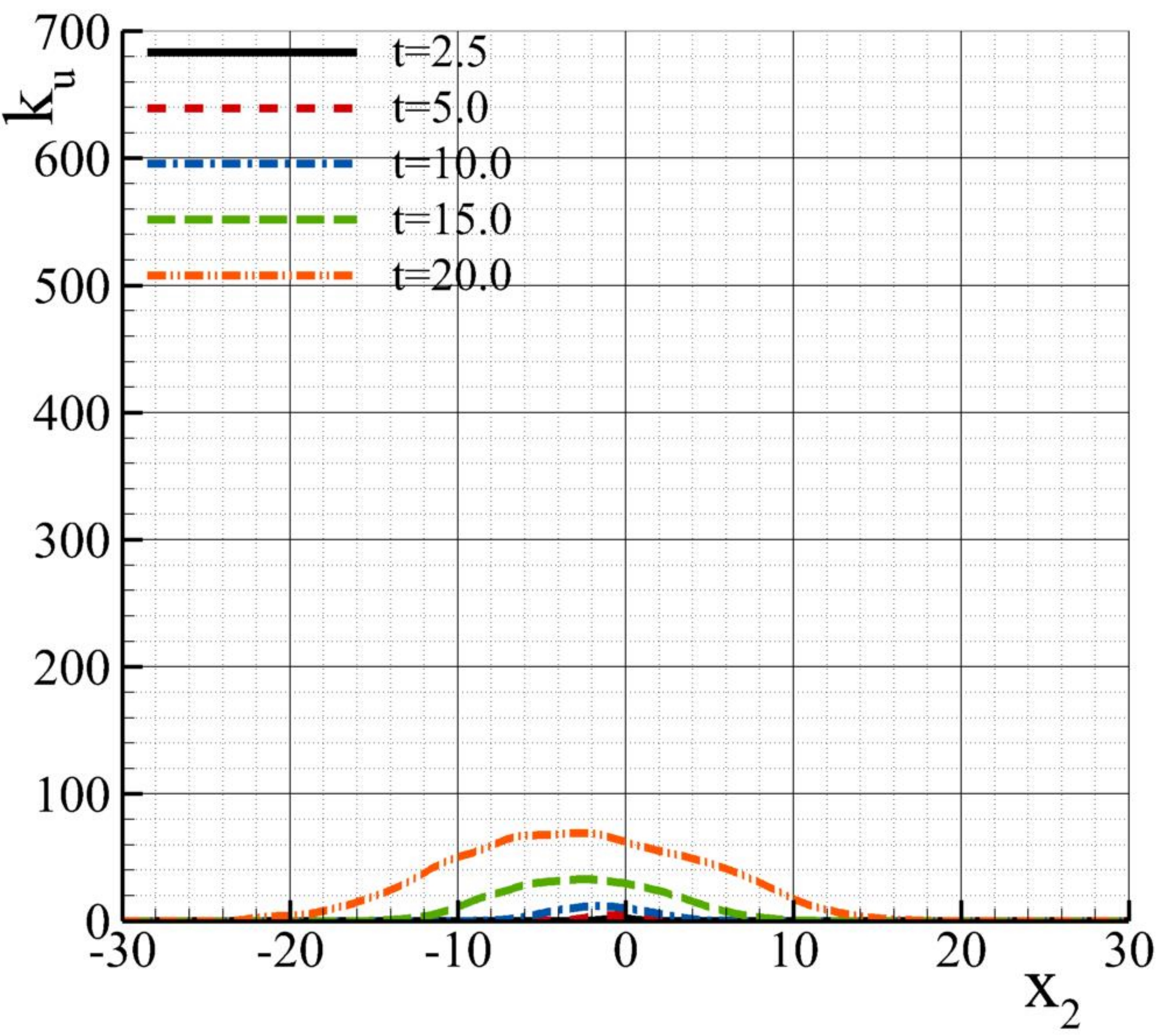}}
~
\subfloat[$a_{2_u}$ at $f_k=0.25$ and $S_2$.]{\label{fig:4.3.2_1e}
\includegraphics[scale=0.198,trim=0 0 0 0,clip]{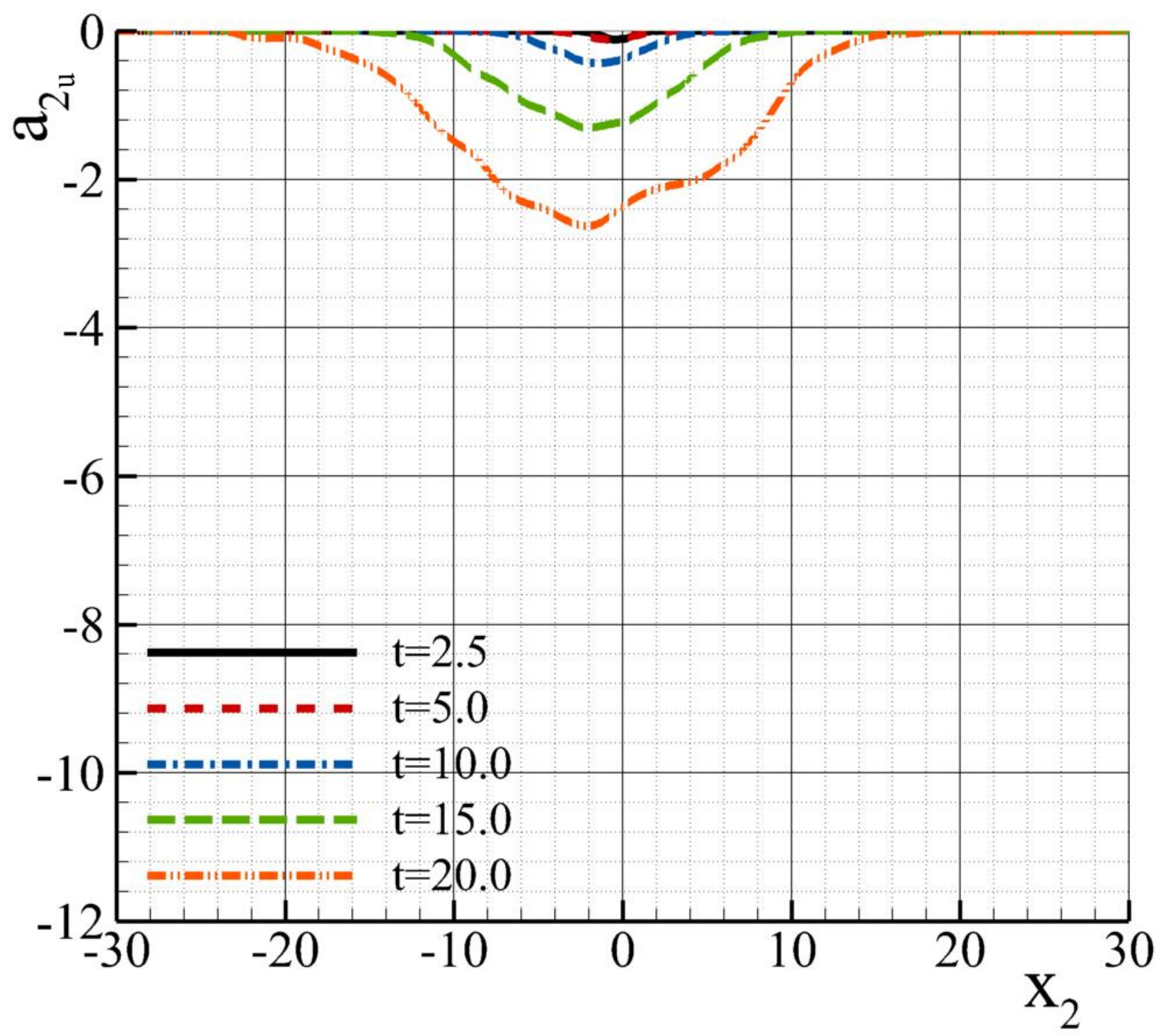}}
~
\subfloat[$b_{u}$ at $f_k=0.25$ and $S_2$.]{\label{fig:4.3.2_1f}
\includegraphics[scale=0.198,trim=0 0 0 0,clip]{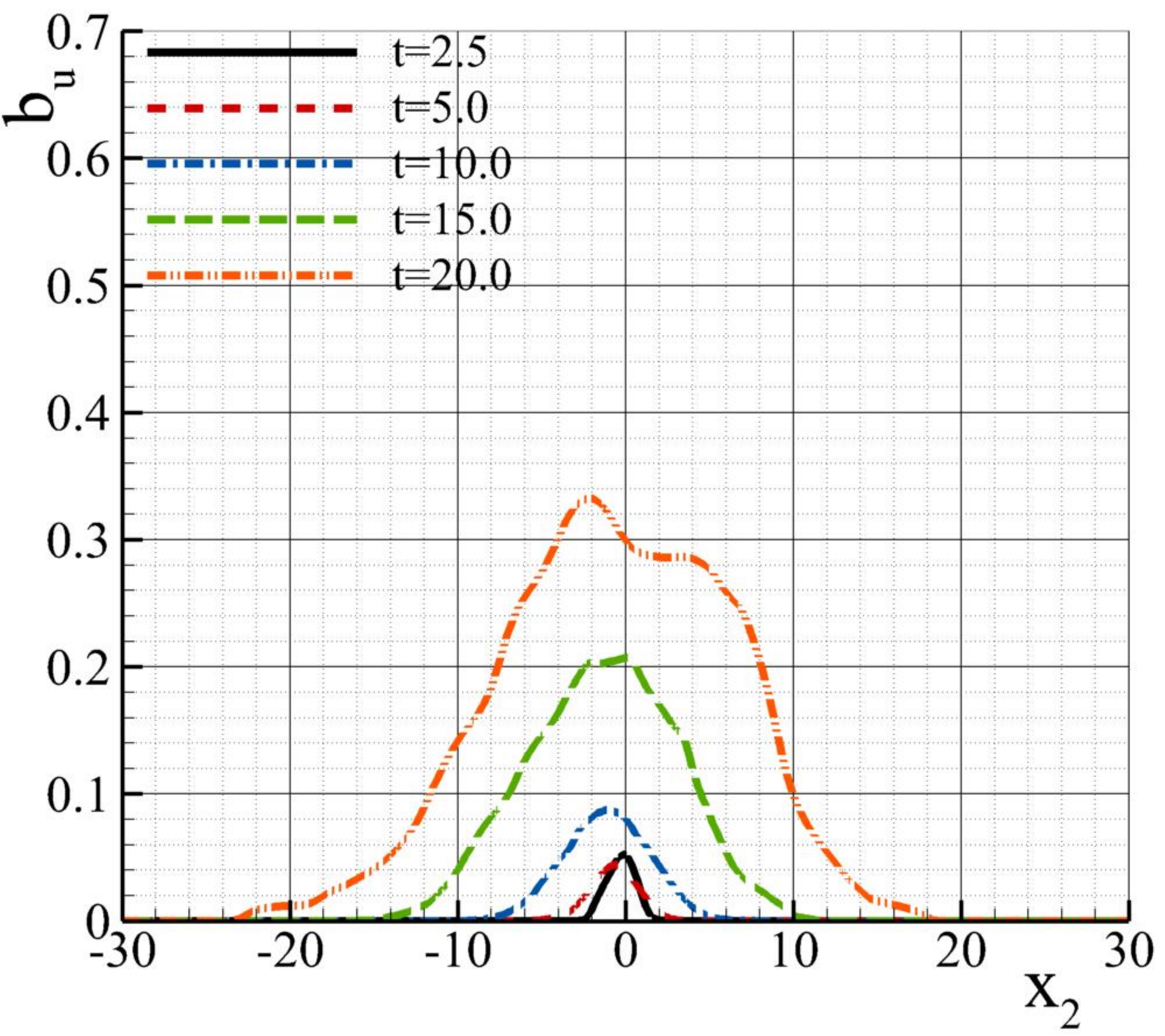}}
\\
\subfloat[$k_u$ at $f_k=0.25$ and $S_3$.]{\label{fig:4.3.2_1g}
\includegraphics[scale=0.198,trim=0 0 0 0,clip]{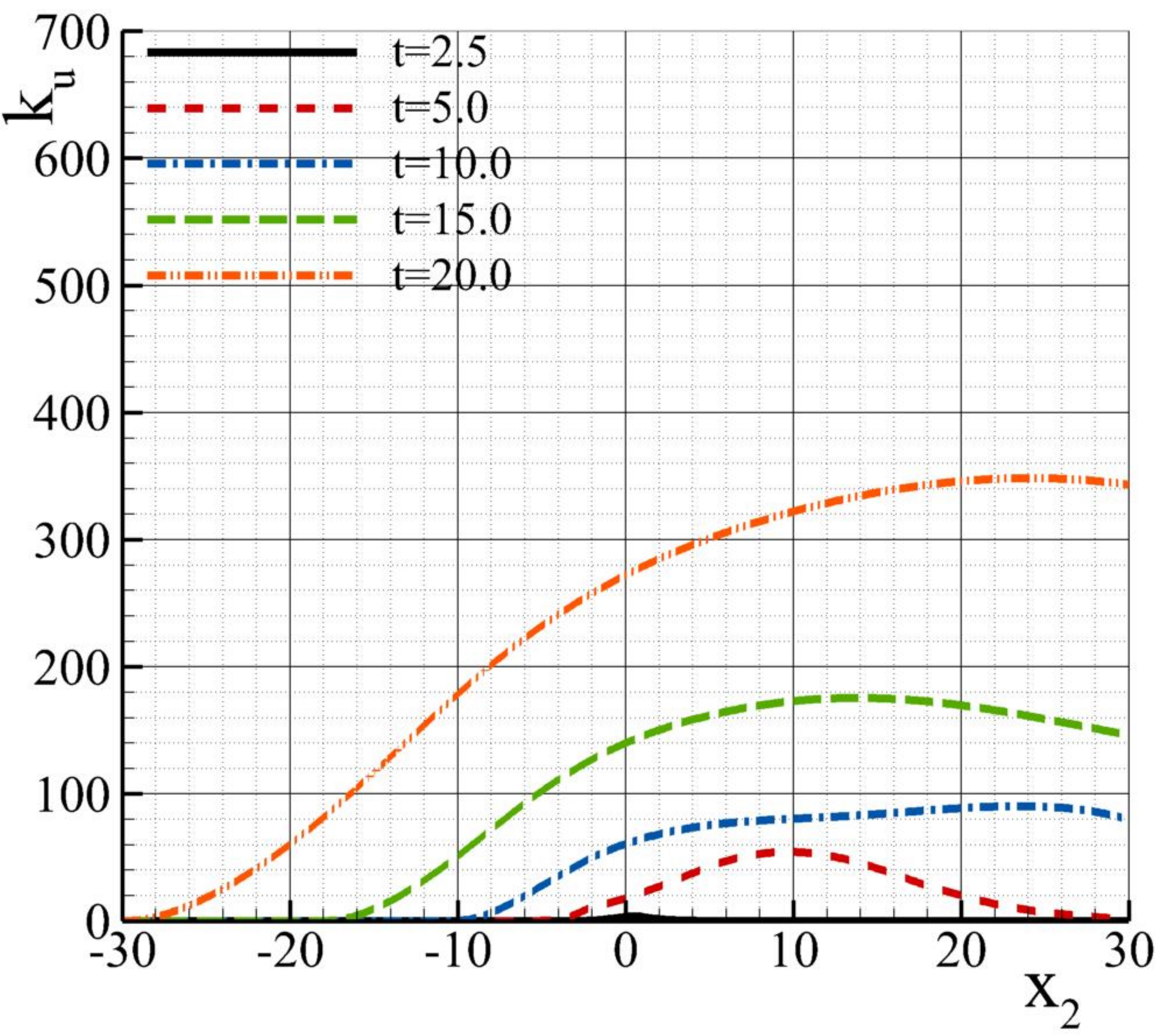}}
~
\subfloat[$a_{2_u}$ at $f_k=0.25$ and $S_3$.]{\label{fig:4.3.2_1h}
\includegraphics[scale=0.198,trim=0 0 0 0,clip]{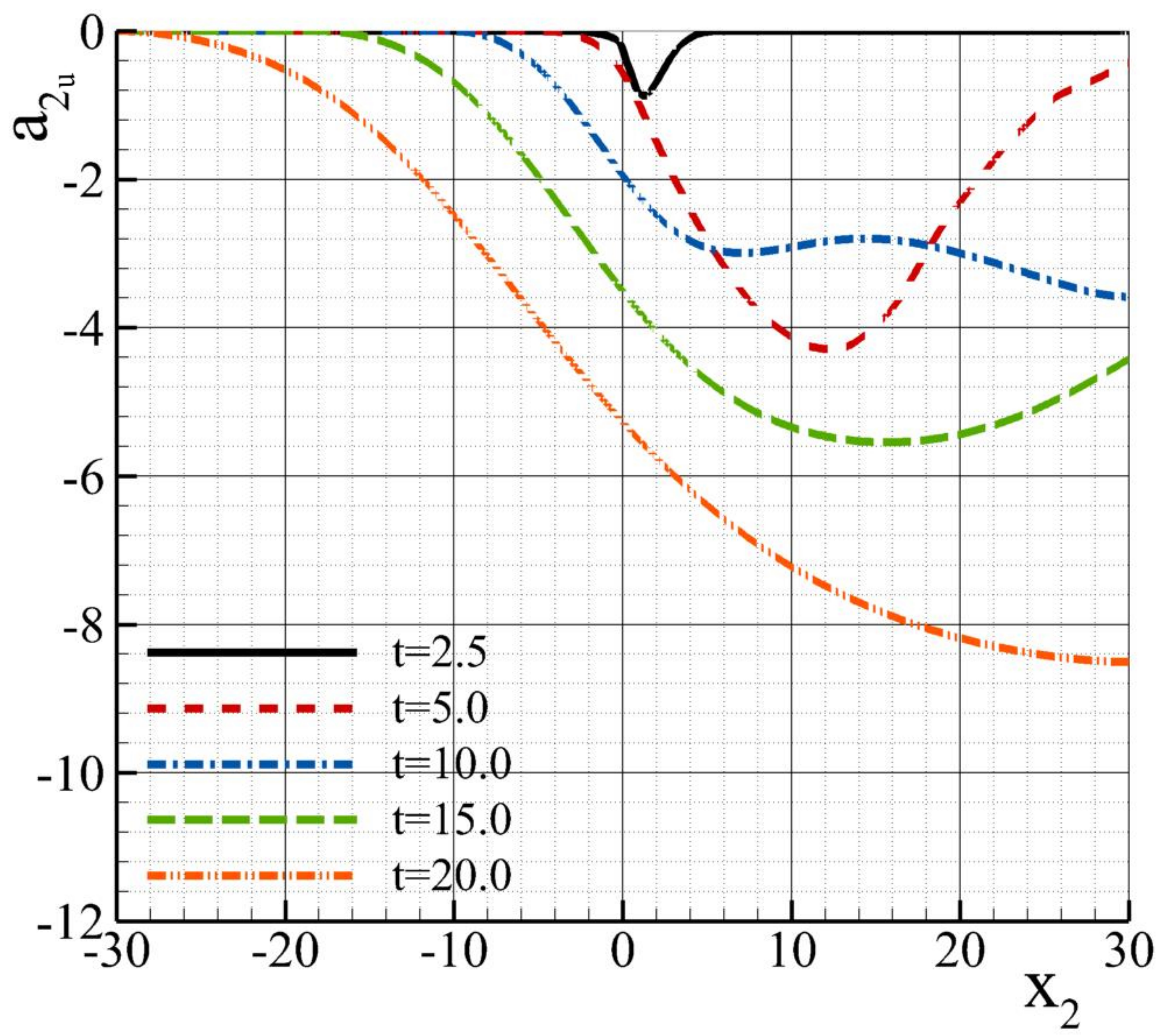}}
~
\subfloat[$b_{u}$ at $f_k=0.25$ and $S_3$.]{\label{fig:4.3.2_1i}
\includegraphics[scale=0.198,trim=0 0 0 0,clip]{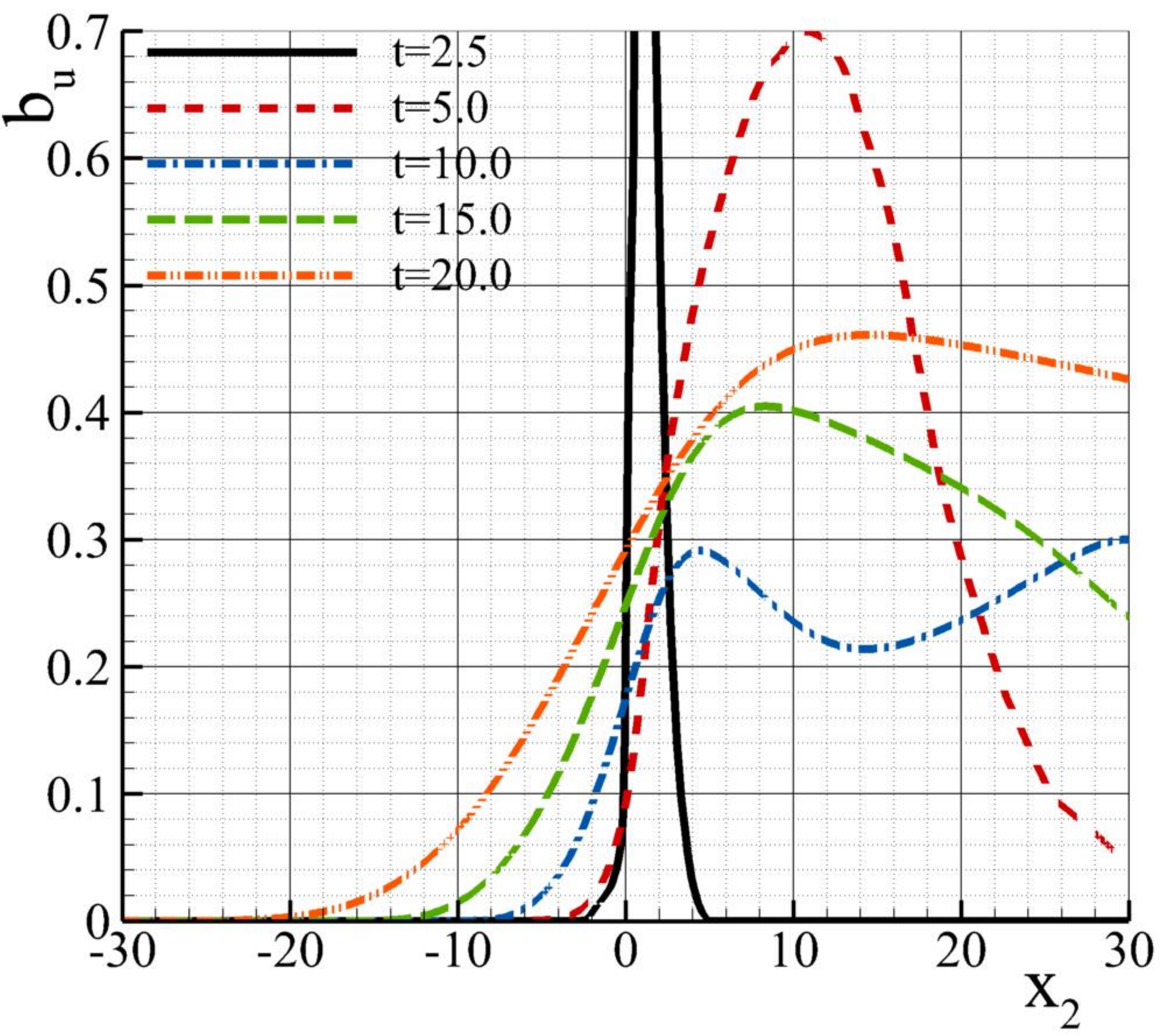}}
\\
\subfloat[$k_u$ at $f_k=1.00$.]{\label{fig:4.3.2_1j}
\includegraphics[scale=0.198,trim=0 0 0 0,clip]{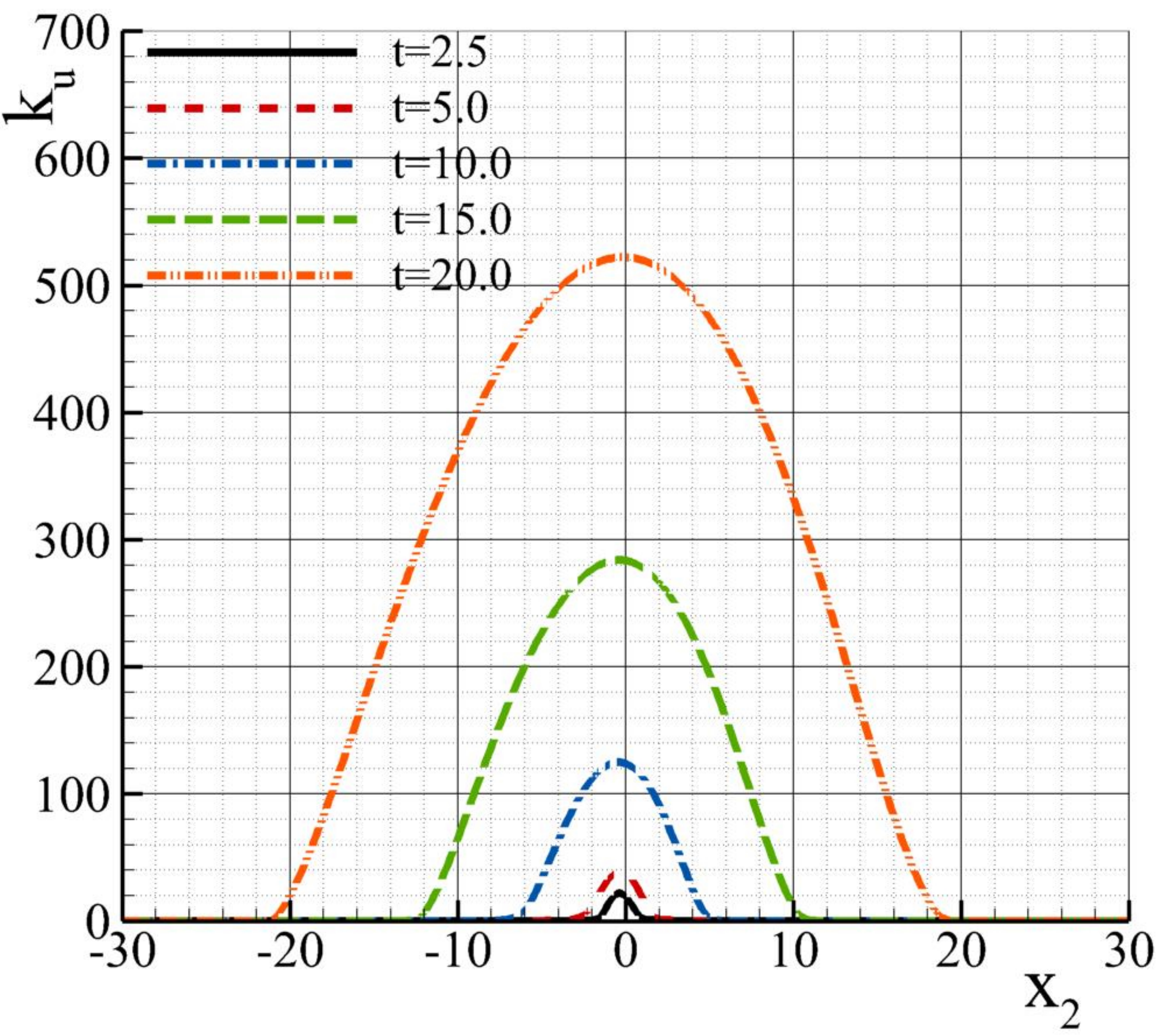}}
~
\subfloat[$a_{2_u}$ at $f_k=1.00$.]{\label{fig:4.3.2_1k}
\includegraphics[scale=0.198,trim=0 0 0 0,clip]{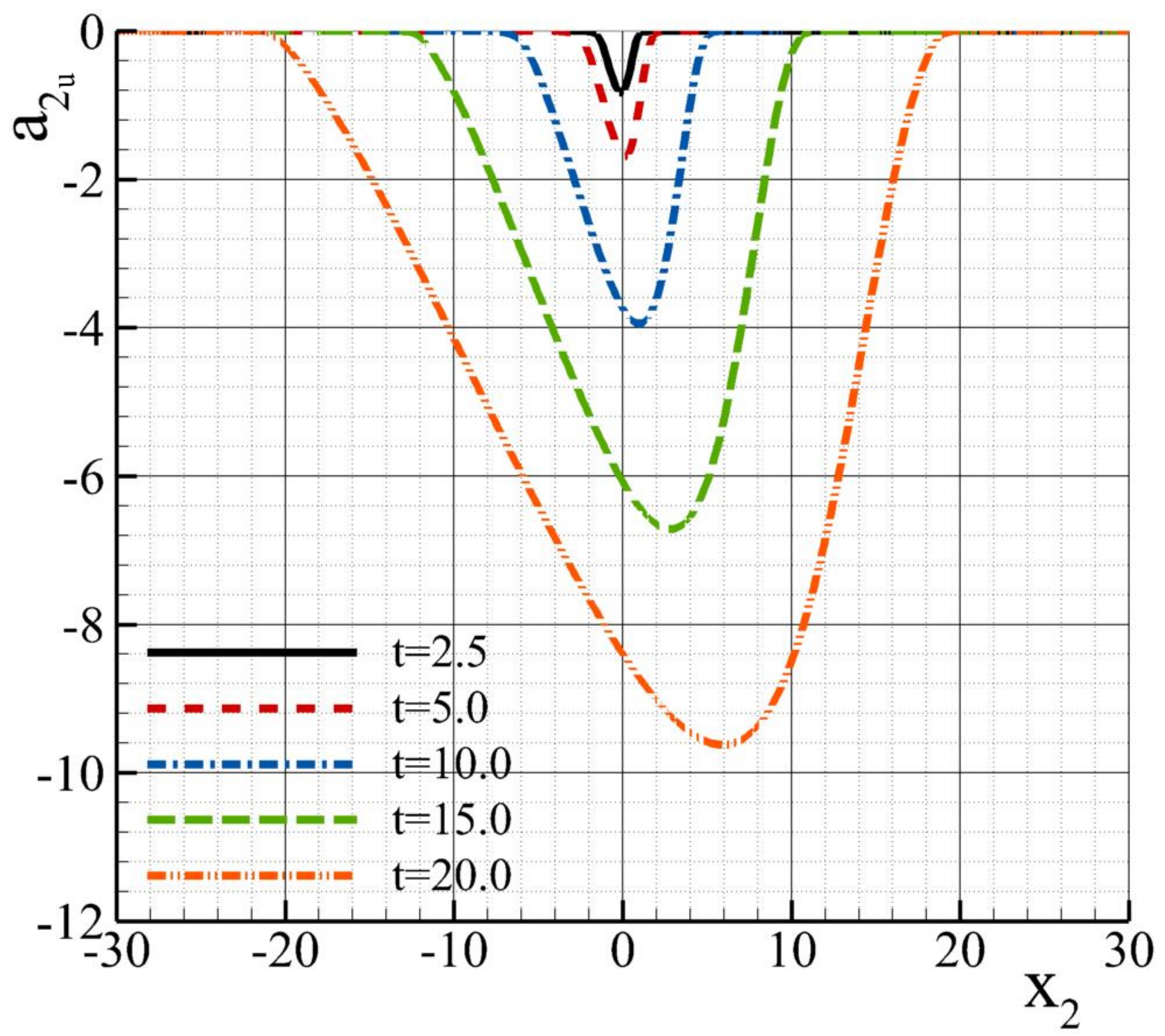}}
~
\subfloat[$b_{u}$ at $f_k=1.00$.]{\label{fig:4.3.2_1l}
\includegraphics[scale=0.198,trim=0 0 0 0,clip]{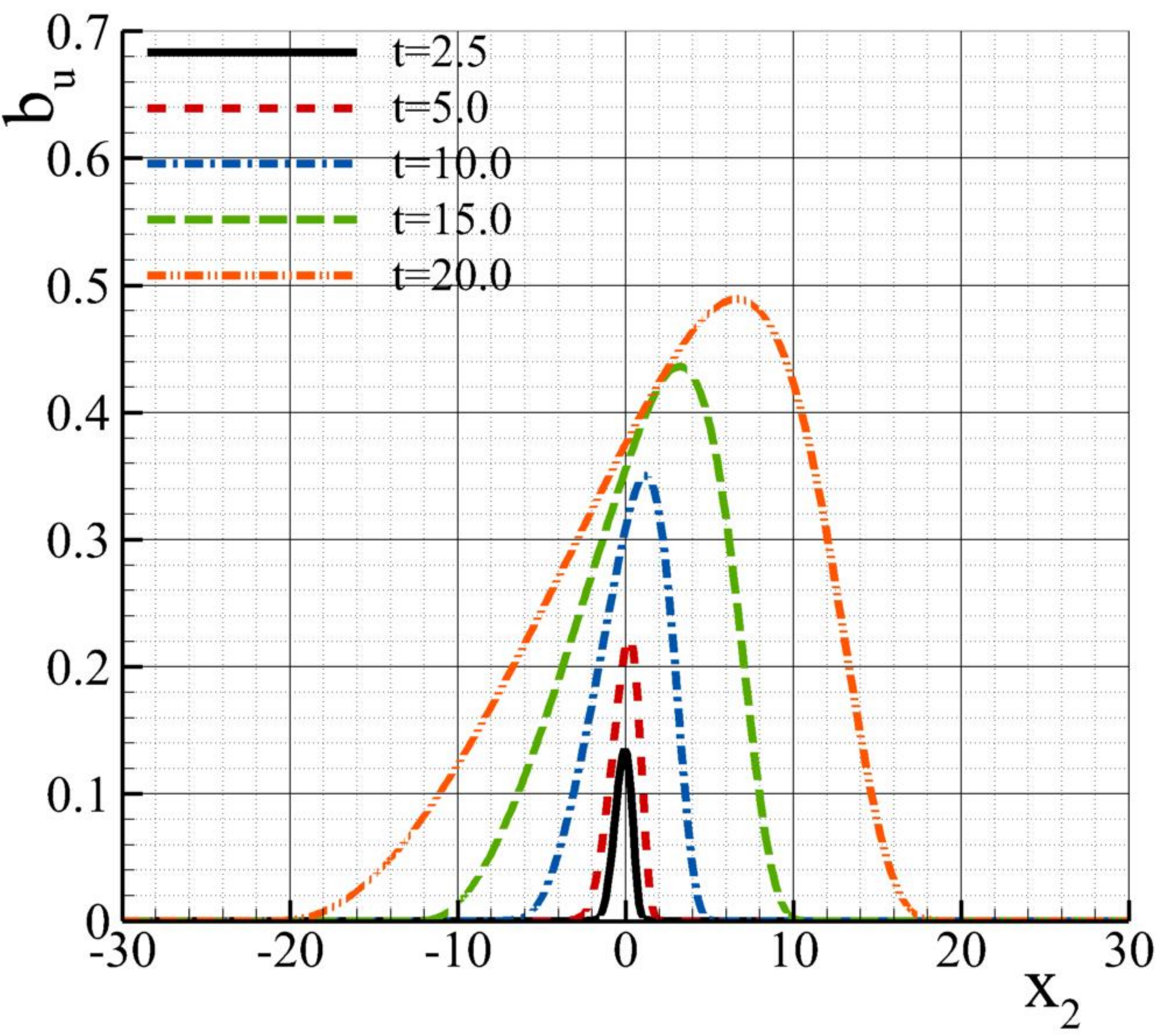}}
\caption{\color{blue}Temporal evolution of unresolved turbulent kinetic energy, $k_u$ (cm$^2$/s$^2$), vertical velocity mass flux, $a_{2_u}$ (cm/s), and density-specific volume correlation, $b_u$, predicted with different $f_k$ and $S_i$; (a)-(c) $k_u$, $a_{2_u}$, $b_{u}$ at $f_k=0.25$ and $S_1$, (d)-(f) $k_u$, $a_{2_u}$, $b_{u}$ at $f_k=0.25$ and $S_2$, (g)-(i) $k_u$, $a_{2_u}$, and $b_{u}$ at $f_k=0.25$ and $S_3$, (j)-(l) $k_u$, $a_{2_u}$, $b_{u}$ at $f_k=1.00$.}
\label{fig:4.3.2_1}
\end{figure*}

Now we turn our attention to the temporal evolution of the $x_2$-profiles of $k_u$, $a_{2_u}$, and $b_u$ predicted with different $f_k$ and $S_i$. These quantities are depicted in figure \ref{fig:4.3.2_1}. Focusing on $k_u$, the results illustrate that the profiles predicted with $f_k=1.00$ are approximately symmetric, and their center is slightly shifted to the light fluid side ($x_2<0$). The refinement of physical resolution to $f_k=0.25$ with $S_1$ and $S_2$ reduces the symmetry and magnitude of the profiles and moves the peak of $k_u$ further to the light fluid side. Compared to the RANS case ($f_k=1.00$), this behavior is caused by the broader range of resolved scales and consequent better prediction of $\tau^1(V_i,V_j)$ that prevents an over diffusive and homogeneous mixing-layer. The outcome of $S_3$ illustrates that this approach leads to significantly larger values of $k_u$ than the remaining $S_i$, these being nearly constant on the top of the mixing-layer. 

On the other hand, $a_{2_u}$ calculated with $f_k=1.00$ grows in time and exhibits its maximum value on the heavy side fluid ($x_2>0$). This behavior gets more pronounced with time, and it is in disagreement with the DNS literature  (e.g., \citeauthor{LIVESCU_JOT_2009} \cite{LIVESCU_JOT_2009,LIVESCU_PD_2021}) and our $a_{2_t}$ obtained with $f_k=0.00$ (not shown). These studies indicate that the peak should occur on the light fluid side, and its magnitude is smaller than for the present $f_k=1.00$. Nonetheless, our RANS results are consistent with those of \citeauthor{BANERJEE_PRE_2010} \cite{BANERJEE_PRE_2010} obtained with a simplified version of the BHR-LEVM closure. The refinement of $f_k$ with $S_2$ reduces the magnitude of $a_{2_u}$ and moves its maximum to the light fluid side. This tendency is not observed with strategy $S_1$ because $f_a=1.00$. The results using $S_3$ exhibit similar tendencies to those of $k_u$.

The profiles of $b_u$ obtained with $f_k=1.00$ indicate that this quantity grows during the simulated time, being close to $0.5$ at $t=20.0$. Also, its maximum value moves to the heavy fluid side in time. As for $a_{2_u}$, $b_u$ is overpredicted, and its peak should occur on the light fluid side \cite{LIVESCU_JOT_2009,LIVESCU_PD_2021}. Since $a_{u_i}$ and $b_u$ determine ${\cal{P}}_{b_u}$, these results explain the global performance of RANS computations and the overprediction of $\tau^1(V_i,V_j)$. The refinement of $f_k$ to $0.25$ reduces the magnitude of the profiles, especially those predicted with $S_2$. Recall that $f_b=1.00$ in approaches $S_1$ and $S_2$. The decrease of $f_k$ also makes the peak of $b_u$ move toward the light fluid side.  

One of the most significant results of figures \ref{fig:4.3.2_1c}, \ref{fig:4.3.2_1i}, and \ref{fig:4.3.2_1l} is the behavior of $b_u$ in simulations using $S_3$. They show a steep overproduction of $b_u$ at $t=2.5$, which later affects $a_{i_u}$ and $k_u$. Recall that $b_u$ governs the production of $a_{i_u}$, which, in turn, defines the production of $k_u$ by buoyancy effects (see Section \ref{sec:3}). We attribute the nonphysical behavior of $b_u$ predicted with $S_3$ to the selection of $f_{a_i}$ and $f_b$ for early flow stages assuming high-intensity turbulence. This makes $S_3$ physically inconsistent, causing modeling and numerical robustness issues. We believe that the further improvement of the selection of $f_{a}$ and $f_b$ may require dynamic strategies of setting $f_\phi$ so that these parameters can be consistently defined at different flow stages. 

In summary, the results of $k_u$, $a_{2_u}$, and $b_u$ illustrate the impact of the unresolved-to-total ratio of the dependent variables of the closure and the importance of a consistent selection of $f_\phi$. Considering the three approaches tested, the first two seem robust and accurate options, particularly $S_2$.
%
%
%
\subsubsection{Turbulence kinetic energy production}
\label{sec:4.3.3}
%
\begin{figure}[t!]
\centering
\subfloat[$S_1$.]{\label{fig:4.3.3_1a}
\includegraphics[scale=0.270,trim=0 0 0 0,clip]{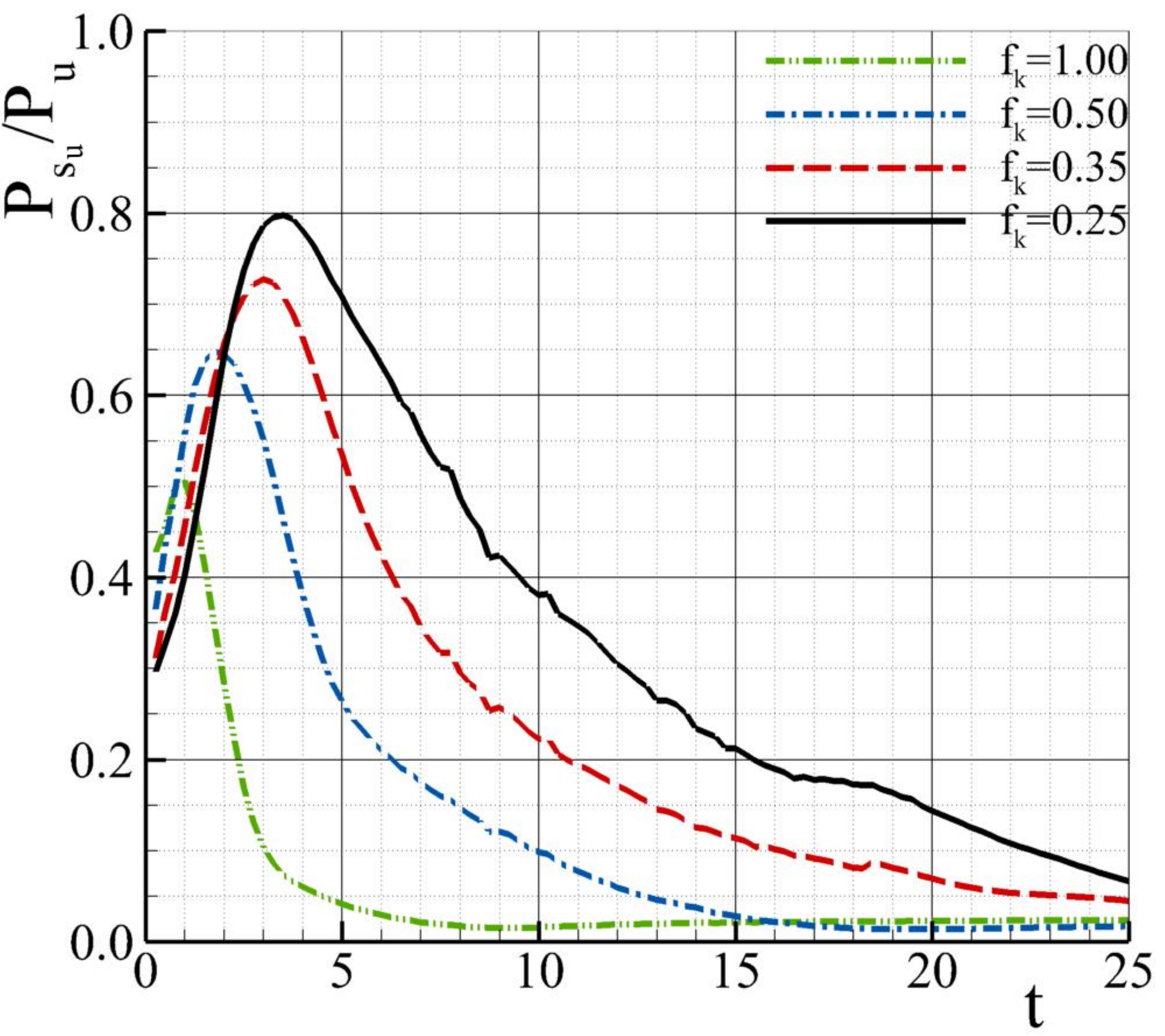}}
\\
\subfloat[$S_2$.]{\label{fig:4.3.3_1b}
\includegraphics[scale=0.270,trim=0 0 0 0,clip]{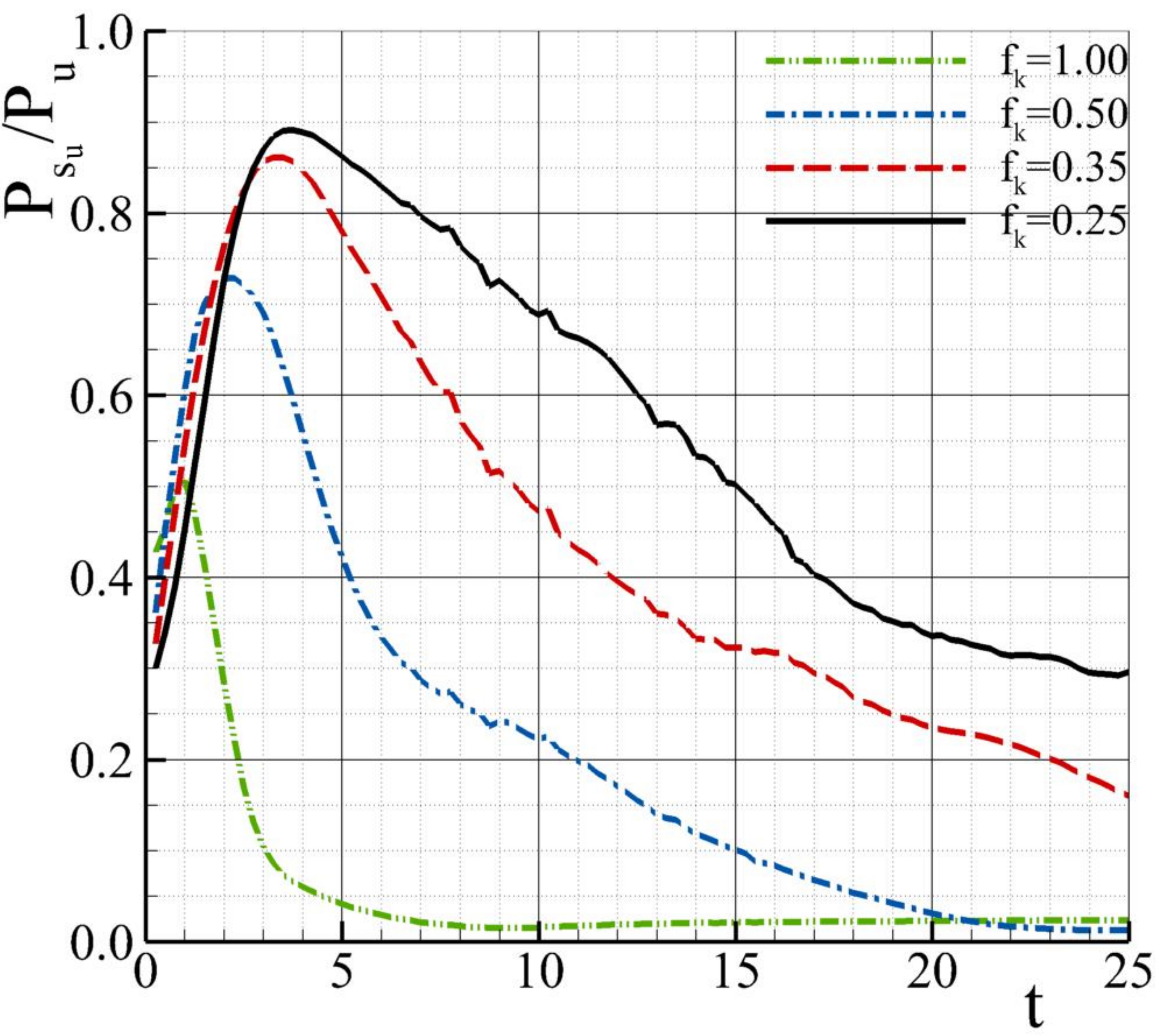}}
\\
\subfloat[$S_3$.]{\label{fig:4.3.3_1c}
\includegraphics[scale=0.270,trim=0 0 0 0,clip]{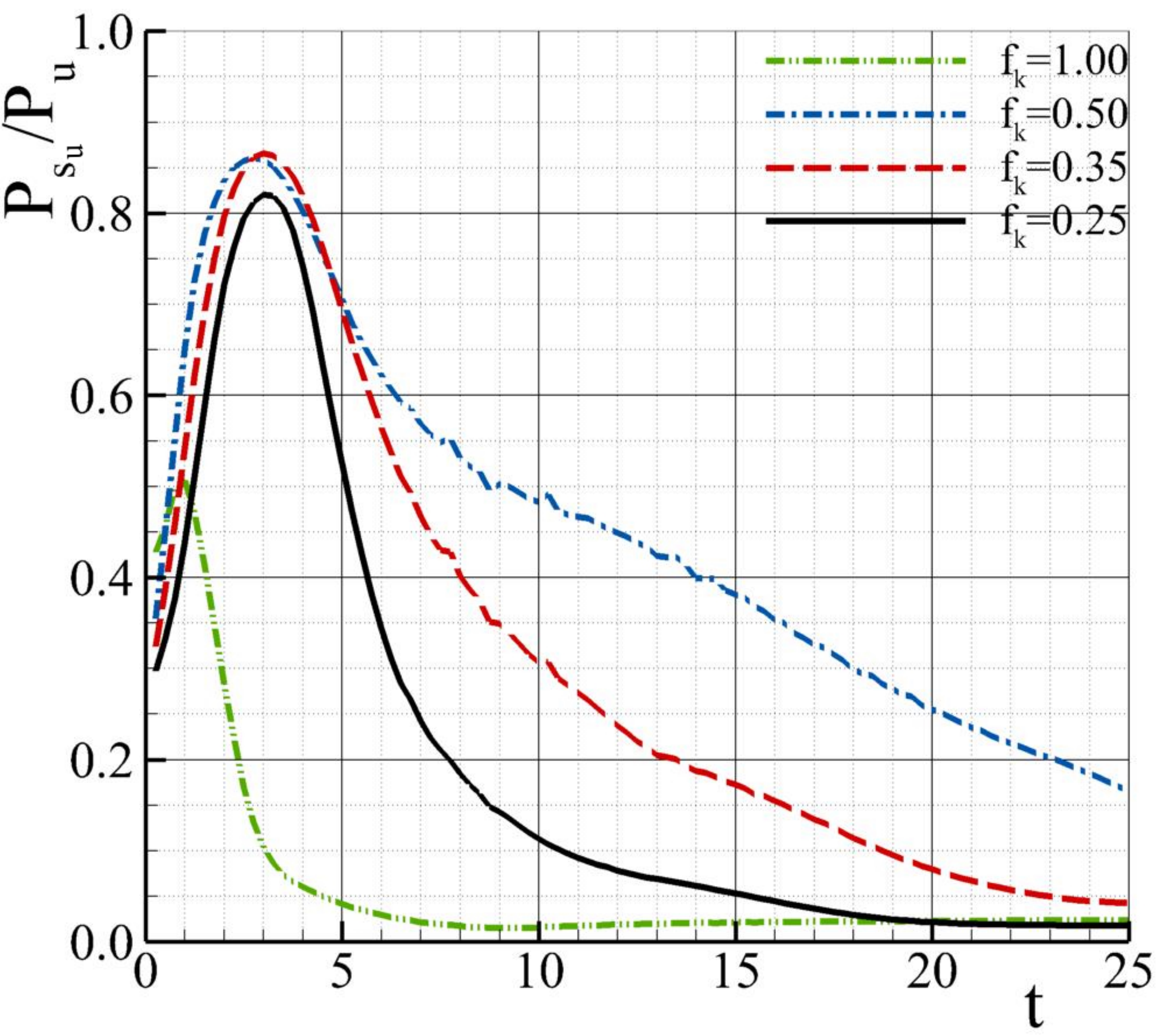}}
\caption{\color{blue}Temporal evolution of the ratio shear-to-total of production of $k_u$, ${\cal{P}}_{s_u}/{\cal{P}}_{u}$, predicted with different $f_k$ and $S_i$ at $x_2=0$; (a) $S_1$, (b) $S_2$, (c) $S_3$.}
\label{fig:4.3.3_1}
\end{figure}

One of the specificities and modeling challenges of variable-density problems is that the flow turbulence kinetic energy is produced by both shear and buoyancy mechanisms. To illustrate the complexity of modeling the second production mechanism, note that the BHR-LEVM closure requires four evolution equations ($a_{i_u}$ and $b_u$) to do it. We now investigate how the two mechanisms evolve in time and their dependence on physical resolution. 

Figure \ref{fig:4.3.3_1} presents the ratio shear-to-total (buoyancy and shear) of production of $k_u$, ${\cal{P}}_{s_u}/{\cal{P}}_{u}$, predicted with different $f_k$ and $S_i$ at $x_2=0$. The results of simulations employing strategies $S_1$ and $S_2$ show that the shear and buoyancy mechanisms have initially similar contributions to the overall production of $k_u$. However, ${\cal{P}}_{s_u}/{\cal{P}}_{u}$ increases rapidly in time and reaches a maximum during the predicted late linear/early nonlinear regime. Such a result is likely because the buoyancy effects are primarily occurring at the scales associated with larger structures. This first generates shear instabilities at small scales, causing ${\cal{P}}_{s_u}/{\cal{P}}_{u}$ to increase. Only later, as the fluids become mixed at the sub-filter scale, the modeled buoyancy production increases relative to the shear component.

As $f_k$ refines, the peak of ${\cal{P}}_{s_u}/{\cal{P}}_{u}$ grows and occurs later. It increases from $0.51$ with $f_k=1.00$ to $0.88$ with $f_k=0.25$ and $S_2$. After the peak, the relative contribution of ${\cal{P}}_{s_u}$ diminishes for all $f_k$, especially for coarser physical resolutions. At $t=25$, ${\cal{P}}_{s_u}/{\cal{P}}_{u}$ is $0.02$ with $f_k=1.00$ and $0.30$ with $f_k=0.25$ and $S_2$. Independently of the accuracy of the computations, these results indicate that shear effects become more important with finer $f_k$, and both production mechanisms are crucial to the total generation of turbulence kinetic energy. Although simulations using $S_3$ show similar tendencies at early times, the evolution of ${\cal{P}}_{s_u}/{\cal{P}}_{u}$ after its maximum is distinct. It is observed a steep decrease of ${\cal{P}}_{s_u}/{\cal{P}}_{u}$, which becomes more pronounced as $f_k$ decreases. Also, the values of this quantity increase from $f_k=0.25$ to $0.50$. We emphasize that the former results are observed in the entire mixing-layer and not only at $x_2=0$. 
%
%
%
\subsubsection{Coherent field}
\label{sec:4.3.4}
%

\begin{figure*}[t!]
\centering
\subfloat[$f_k=0.00$.]{\label{fig:4.3.4_1e}
\includegraphics[scale=0.08,trim=0 0 0 0,clip]{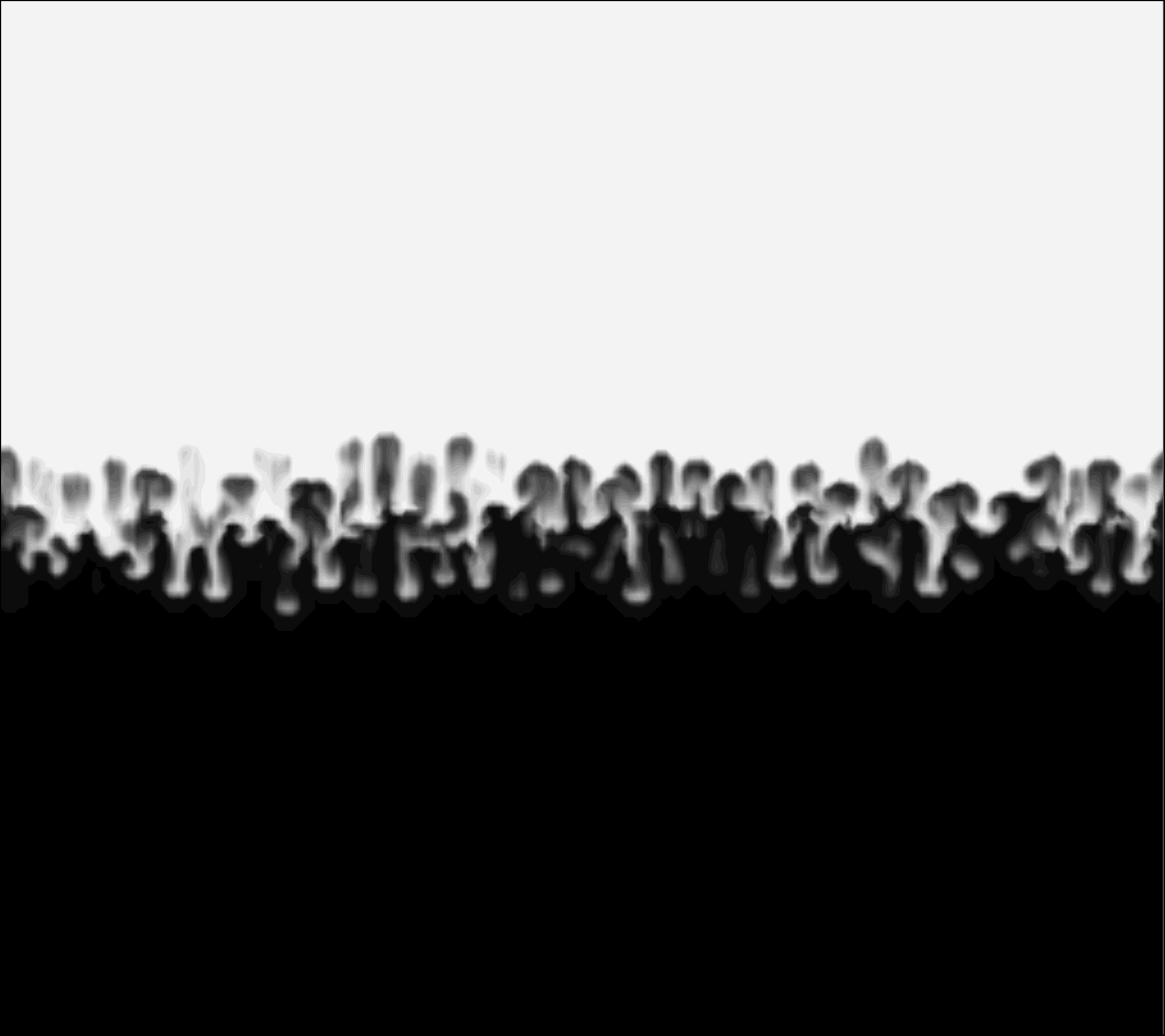}}
~
\subfloat[$f_k=0.25$ and $S_2$.]{\label{fig:4.3.4_1d}
\includegraphics[scale=0.08,trim=0 0 0 0,clip]{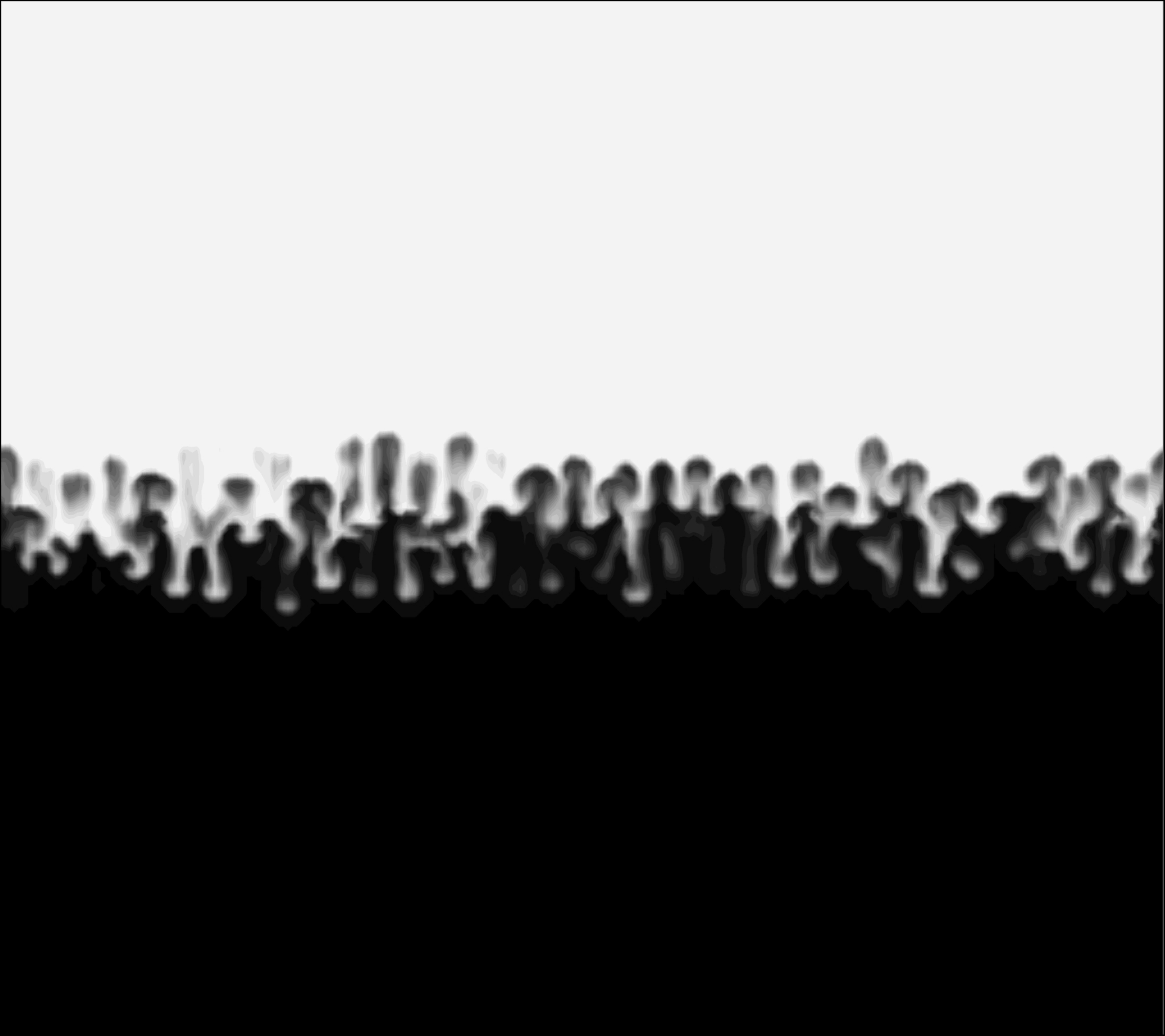}}
~
\subfloat[$f_k=0.35$ and $S_2$.]{\label{fig:4.3.4_1c}
\includegraphics[scale=0.08,trim=0 0 0 0,clip]{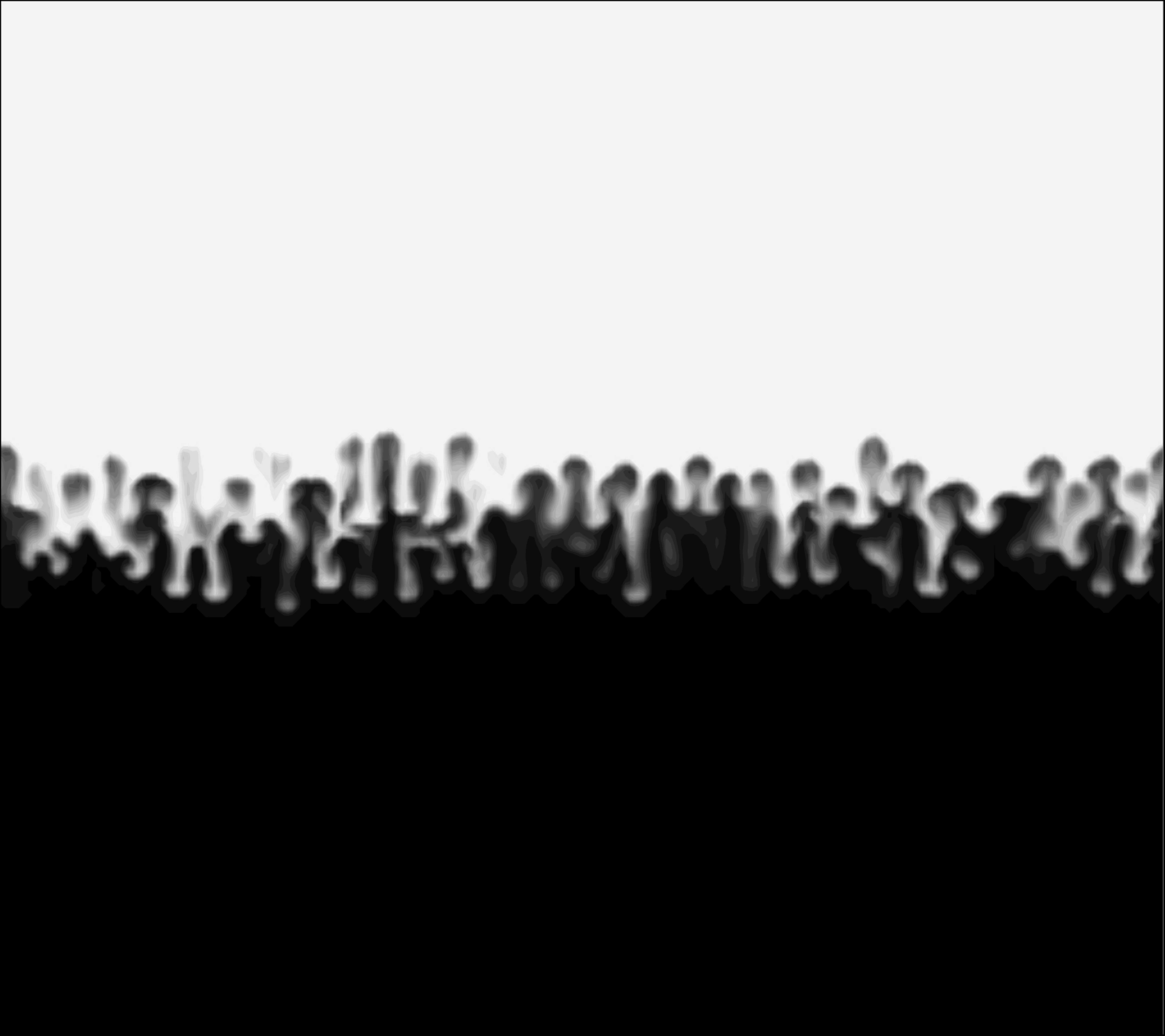}}
\\
\subfloat[$f_k=0.50$ and $S_2$.]{\label{fig:4.3.4_1b}
\includegraphics[scale=0.08,trim=0 0 0 0,clip]{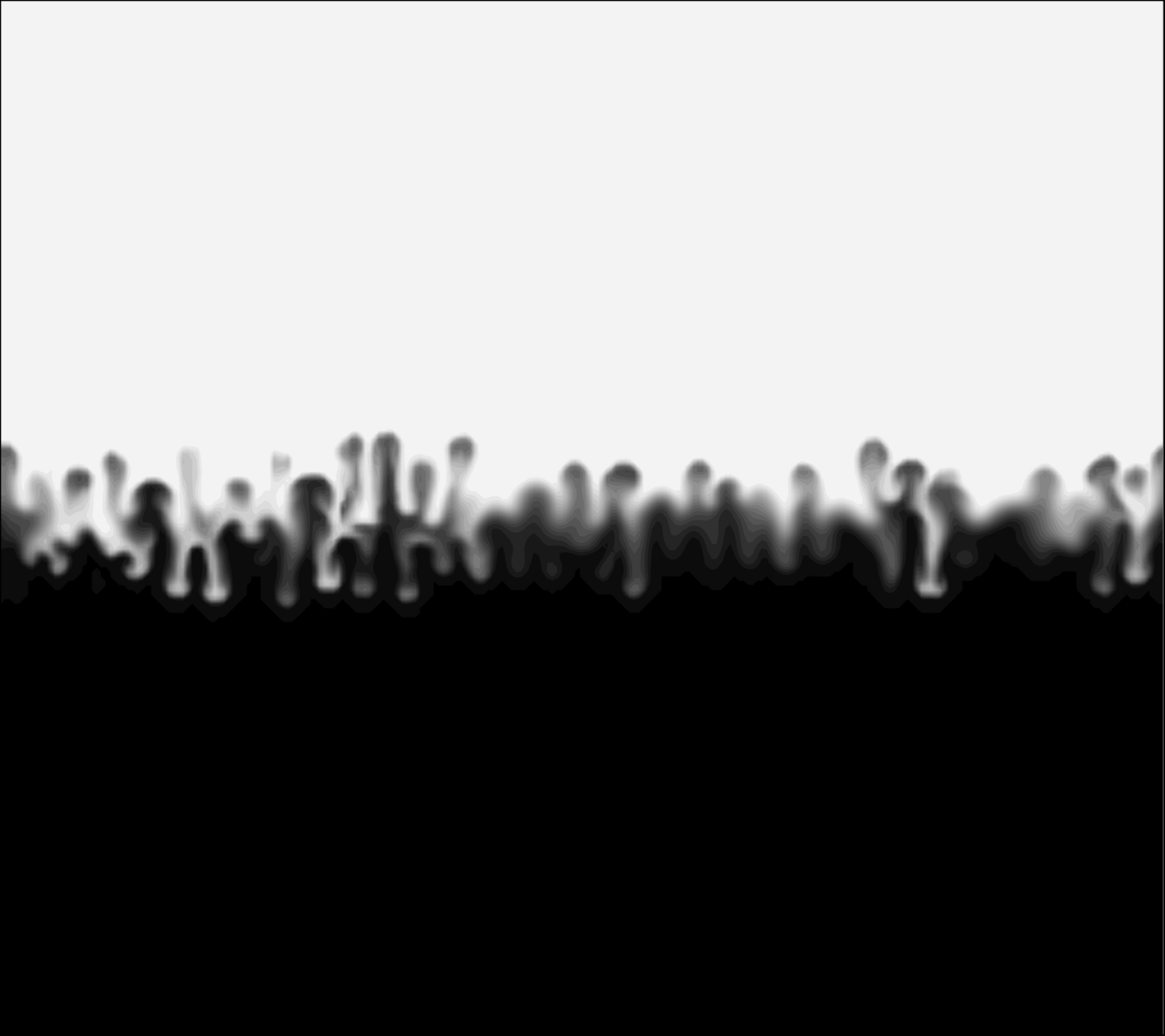}}
~
\subfloat[$f_k=1.00$.]{\label{fig:4.3.4_1a}
\includegraphics[scale=0.08,trim=0 0 0 0,clip]{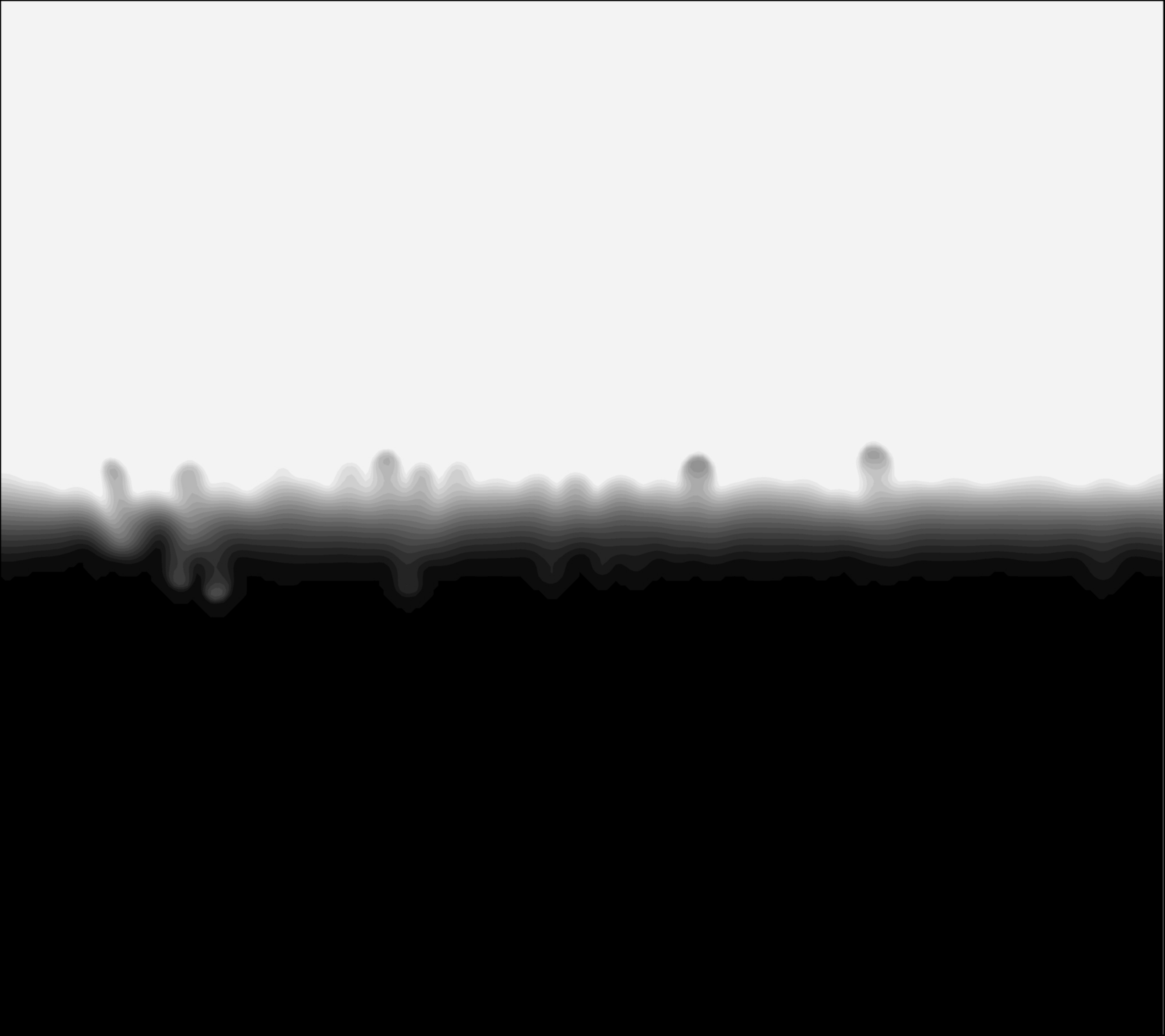}}
\caption{\color{blue}RT coherent structures at $t=2.5$ predicted with different $f_k$ and $S_2$ (2D slice). Vortical structures identified through the density field; (a) $f_k=0.00$, (b) $f_k=0.25$ and $S_2$, (c) $f_k=0.35$ and $S_2$, (d) $f_k=0.50$ and $S_2$, (e) $f_k=1.00$.}
\label{fig:4.3.4_1}
\end{figure*}

In Section \ref{sec:4.1}, figure \ref{fig:4.1_1} shows the coherent structures (bubbles and spikes) generated by the RT flow. After $t \approx 2.5$, these vortical structures start breaking down and lead to the onset of turbulence. Considering the results in the previous sections, simulations using $S_2$ and $f_k<0.50$ seem to accurately represent the statistics of the flow physics. Thus, we now analyze the impact of the physical resolution on the coherent structures driving the onset of turbulence and qualitatively assess the connection between the model's performance and the prediction of these coherent structures \cite{PEREIRA_JCP_2018,PEREIRA_IJHFF_2019,PEREIRA_OE_2019,PEREIRA_PRF_2021}.

Figure \ref{fig:4.3.4_1} depicts the coherent structures predicted with different $f_k$ and $S_2$ at $t=2.5$. The results illustrate that computations with $f_k \leq 0.35$ can accurately represent these vortical structures, and it is not possible to distinguish figure \ref{fig:4.3.4_1e} from \ref{fig:4.3.4_1d} or \ref{fig:4.3.4_1c}. In contrast, computations at lower physical resolutions ($f_k>0.35$) dissipate the coherent structures. Recall that at $t=2.5$, the present problem is at the tail of the linear regime and the flow is not turbulent. Therefore, the outcome of figure \ref{fig:4.3.4_1} indicates that low physical resolution simulations overpredict the unresolved turbulent stresses, dissipating the spikes and bubbles. Since the RT is a transient flow, memory effects are of paramount importance to the fidelity of the simulations. Misrepresenting the flow in early flow times compromises the accuracy of the calculations in the nonlinear regime. 

Most notably, the results of figure \ref{fig:4.3.4_1} and section \ref{sec:4.1} suggest that the spikes and bubbles leading to the onset of turbulence need to be resolved. Thus, they dictate the minimum physical resolution to obtain accurate RT simulations with the PANS BHR-LEVM closure. This requires $f_k\leq 0.35$  with $S_2$ and reiterates the findings of \citeauthor{PEREIRA_JCP_2021} \cite{PEREIRA_JCP_2021,PEREIRA_IJHFF_2019,PEREIRA_OE_2019,PEREIRA_PRF_2021}.

%
%
%
\subsubsection{Spectral features}
\label{sec:4.3.5}
%
\begin{figure*}[t!]
\centering
\subfloat[$t=10.0$.]{\label{fig:4.3.5_1e}
\includegraphics[scale=0.20,trim=0 0 0 0,clip]{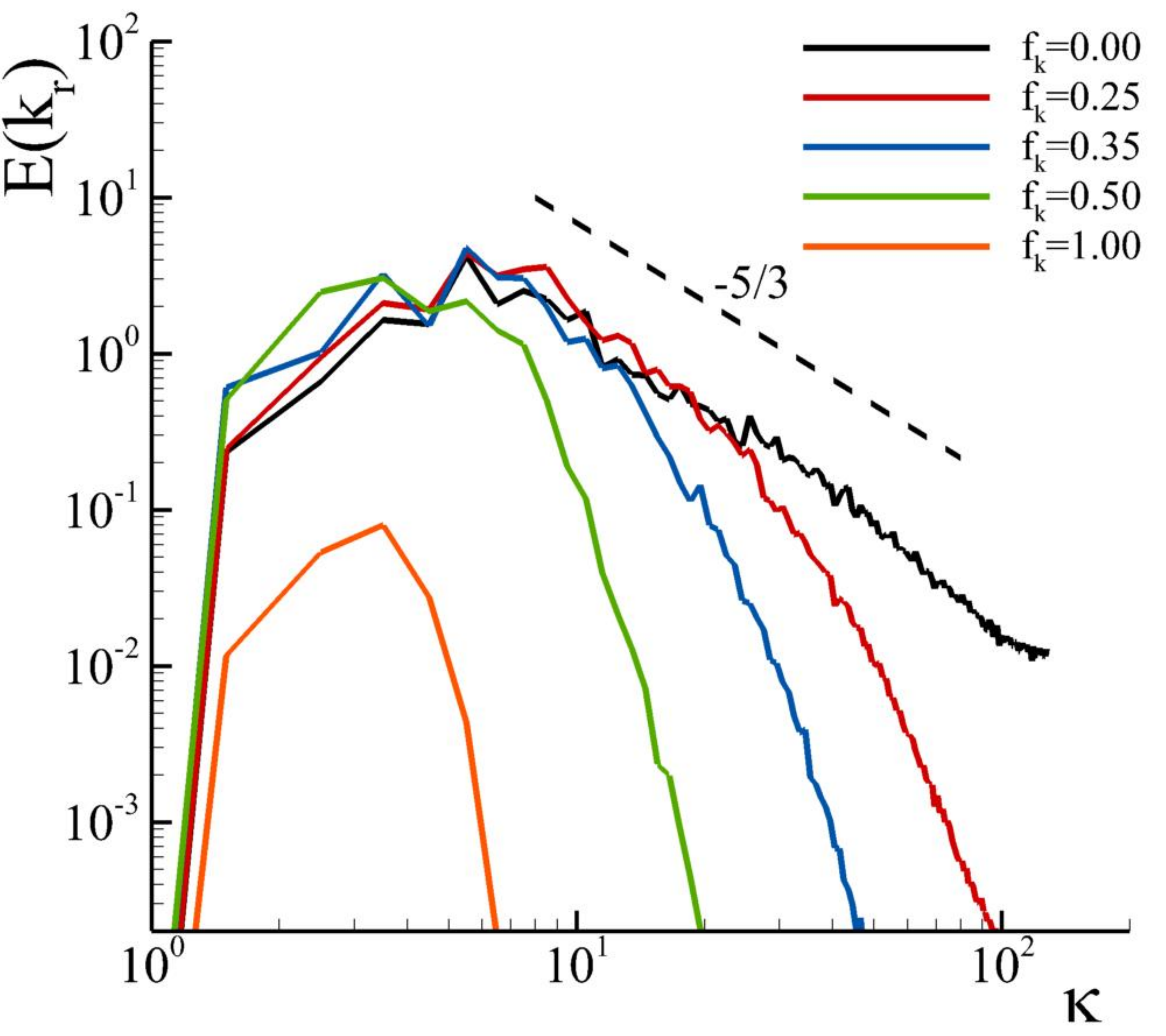}}
~
\subfloat[$t=15.0$.]{\label{fig:4.3.5_1d}
\includegraphics[scale=0.20,trim=0 0 0 0,clip]{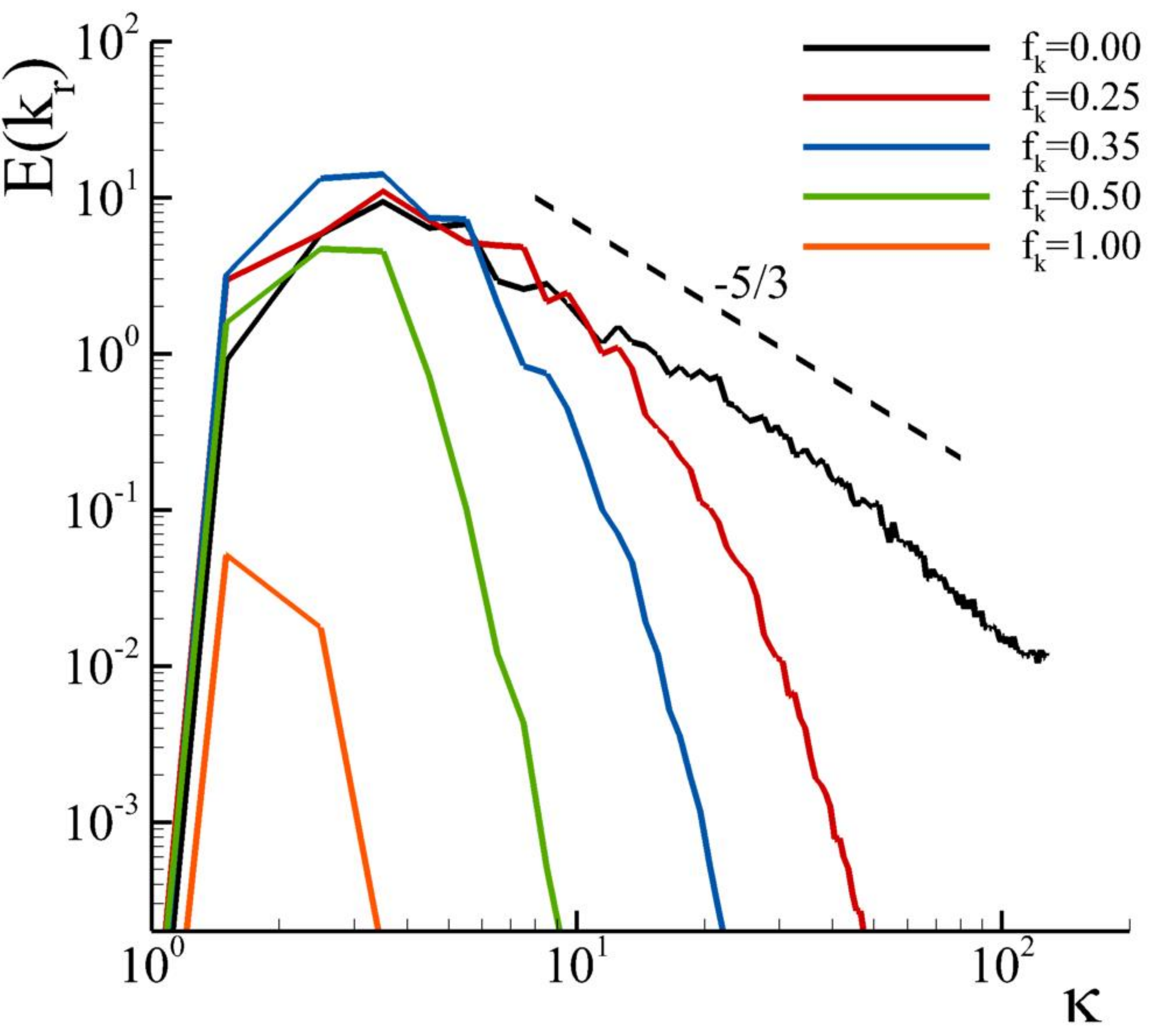}}
\\
\subfloat[$t=20.0$.]{\label{fig:4.3.5_1c}
\includegraphics[scale=0.20,trim=0 0 0 0,clip]{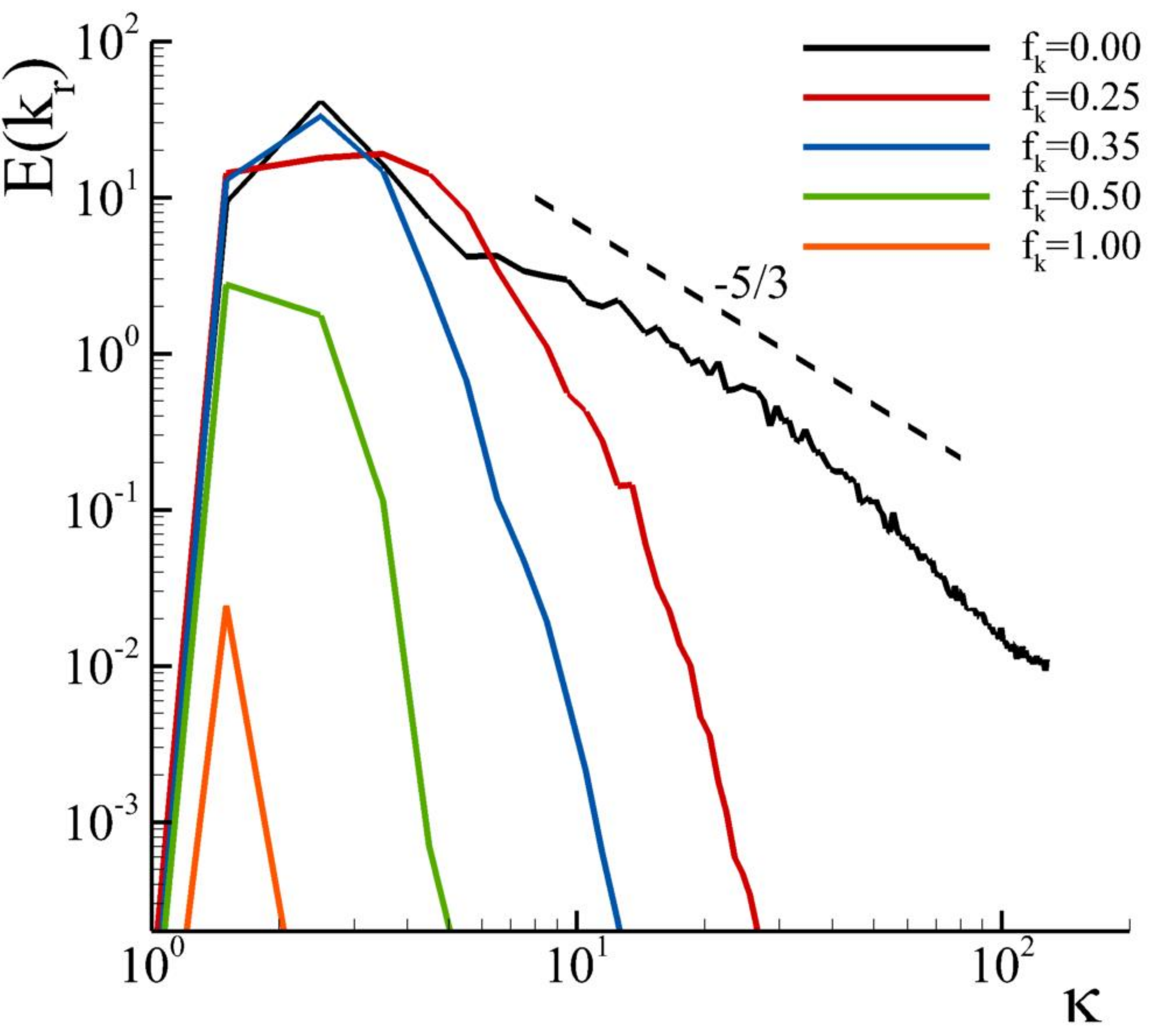}}
~
\subfloat[$t=25.0$.]{\label{fig:4.3.5_1b}
\includegraphics[scale=0.20,trim=0 0 0 0,clip]{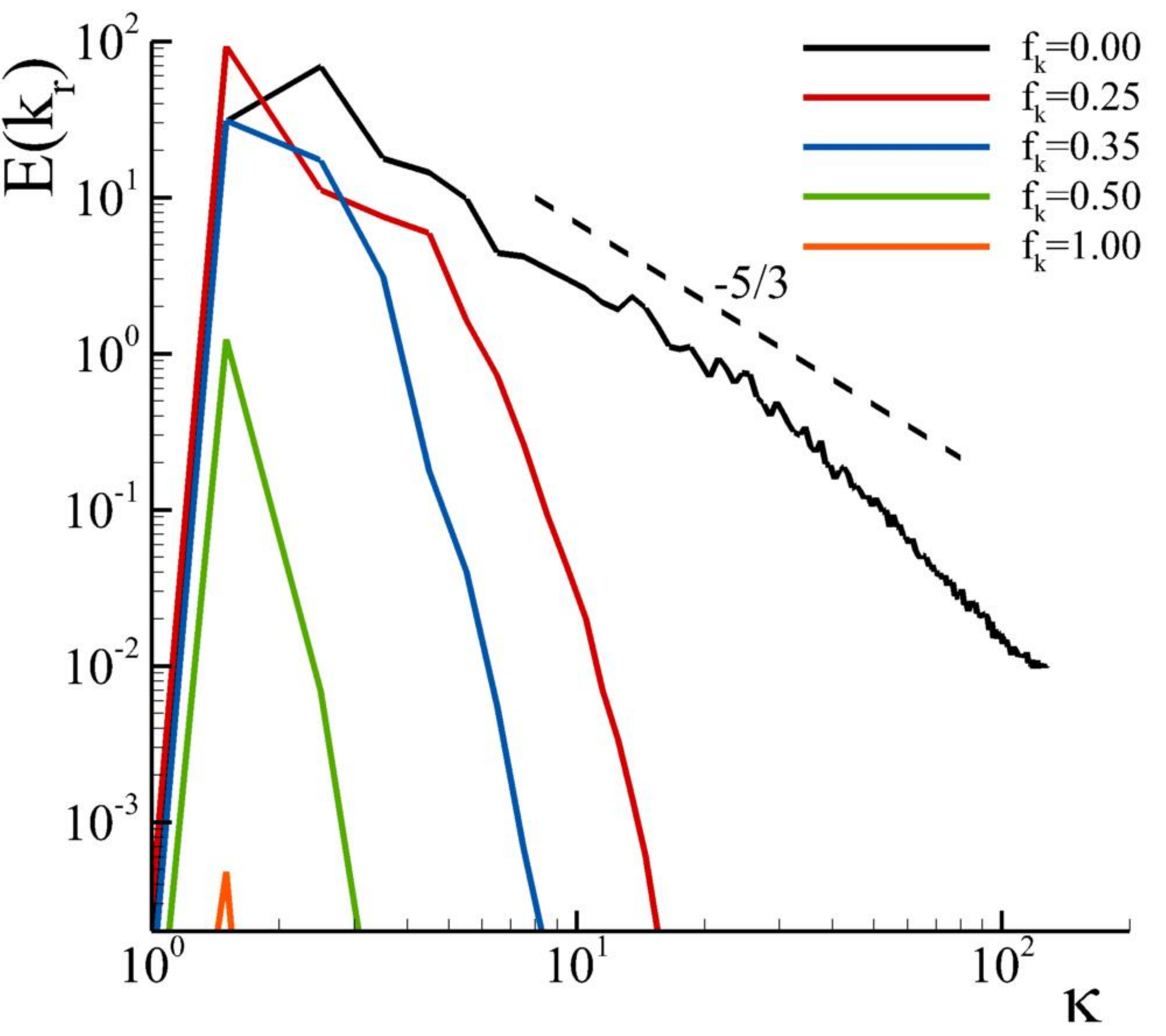}}
\caption{\color{blue}Resolved turbulence kinetic energy spectrum, $E_r(k)$ (cm$^2$/s$^2$), for simulations using different values of $f_k$ and distinct times; (a) $t=10.0$, (b) $t=15.0$, (c) $t=20.0$, (d) $t=25.0$.}
\label{fig:4.3.5_1}
\end{figure*}

Finally, we investigate the effect of the physical resolution on the spectral features of the RT flow. Figure \ref{fig:4.3.5_1} depicts the resolved turbulence kinetic energy spectra of the simulations at successively later times and different $f_k$. It considers the fluctuating velocity field given by equation \ref{eq:3_9} at the plane $x_2=0$.

Until $t=20$, the spectra obtained from $f_k\leq 0.35$ computations include an inertial range, in which the physical resolution determines the largest wavelength. As expected, it is observed that refining $f_k$ leads to longer inertial ranges but does not significantly affect the large, energy-containing scales; only the simulation at $f_k = 0.00$ resolves the viscous scales. Considering the results of this study, this demonstrates that it is possible to obtain accurate simulations without resolving all or most turbulent scales. In contrast, the spectra obtained from $f_k>0.35$ simulations are characterized by lower energy levels across the entire spectrum, including significant damping of the large-scale structures. Along with the decrease of physical resolution, this result stems from the premature onset and overprediction of turbulence. At later times, $t\ge 20$, $k_r$ starts decaying, this becoming more pronounced as $f_k \rightarrow 1.00$.

%
%
%
\section{Conclusions}
\label{sec:5}

This work evaluates the performance of PANS BHR-LEVM closure in predicting the spatio-temporal development of the RT flow. This benchmark problem of variable-density features transient and transitional flow, density variations due to multi-material effects, and turbulence kinetic energy produced by shear and buoyancy mechanisms. These phenomena make it challenging for modeling and simulation. The numerical simulations were conducted at different levels of physical resolution to investigate the importance of this parameter and ascertain three approaches ($S_i$) to define the unresolved-to-total ratio of the dependent variables of the closure, $f_\phi$.

The results demonstrate that the PANS BHR-LEVM can accurately predict the spatio-temporal development of the present flow by resolving only a fraction of the turbulent spectrum. With $S_2$, this only requires $f_k \leq 0.35$. Regarding the strategies to define the different $f_\phi$, $S_1$ and $S_2$ reveal the most consistent and accurate, particularly $S_2$. On the other hand, approach $S_3$ based on \textit{a priori} studies leads to modeling and numerical issues. This stems from the fact that this approach assumes fully developed turbulence. Thus, strategy $S_3$ seems better suited for a dynamic selection of $f_\phi$. Overall, the calculations illustrate the importance of a precise specification of $f_\phi$, and the potential of PANS modeling to improve the efficiency of SRS calculations.

This study also reiterates that the accuracy of the computations is closely dependent on the ability to resolve the flow phenomena not amenable to modeling by the closure. For the current flow and closure, one needs to resolve the vortical structures (spikes and bubbles) involved in the onset of turbulence. The scale-aware BHR-LEVM closure can reasonably represent the remaining flow scales.
%
%
%
%
\section*{Acknowledgments}

{\color{blue}We would like to thank G. Dimonte for sharing his RT $\alpha_b$ data. Also, we would like to thank the editor and reviewers for their suggestions that improved our paper}. Los Alamos National Laboratory (LANL) is operated by TRIAD National Security, LLC for the US DOE NNSA. This research was funded by LANL Mix and Burn project under the DOE ASC, Physics and Engineering Models program. We express our gratitude to the ASC program and the High-Performance Computing (HPC) division for their dedicated and continued support of this work.
%
%
%
%
\section*{Data Availability Statement}

The data that support the findings of this study are available from the corresponding author upon reasonable request.
%
%
%
%
\appendix
\section{Flow quantities \& scaling}
\label{sec:A}

Apart from $k_u$, $a_{2_u}$, $b_{u}$, and $E(k_r)$, all quantities analyzed in this paper are scaled (or normalized) by $g$, $\mathrm{At}$, $h_o$, $\rho_o$, and $\nu$. This aims to facilitate the comparison of the results with those of other studies, as well as their physical interpretation \cite{TENNEKES_BOOK_1972,POPE_BOOK_2000,ISRAEL_JFM_2021}. The quantities discussed in Section \ref{sec:4} not defined in text are calculated as follows:
\begin{itemize}
%
\item[--] time, $t$:
\begin{equation}
\label{eq:A_1}
t= t^* \sqrt{\frac{g\ \mathrm{At}}{h_o}}\; ,
\end{equation}
where the superscript $^*$ indicates the corresponding non-scaled quantity, and $h_o$ is the length-scale of the most unstable initial perturbation \cite{LIVESCU_PD_2021},
\begin{equation}
\label{eq:A_2}
h_o=\frac{2\pi}{32}\; .
\end{equation}
%
%
\item[--] spatial coordinates, $x_i$:
\begin{equation}
\label{eq:A_3}
x_i=\frac{x_i^*}{h_o}\; .
\end{equation}
%
%
\item[--] mixture fractions $\overline{\chi}_1$ and $\overline{\chi}_2$:
\begin{equation}
\label{eq:A_4}
\overline{\chi}_1=\left( \frac{\overline{\rho}-\rho_l}{\rho_o} \right) \; ,
\end{equation}
\begin{equation}
\label{eq:A_5}
\overline{\chi}_2=\left(\frac{\rho_h-\overline{\rho}}{\rho_o}\right)\; ,
\end{equation}
where,
\begin{equation}
\label{eq:A_6}
\rho_o=\rho_h-\rho_l\; ,
\end{equation}
and bars denote $x_1-x_3$ planar averaged quantities.
%
%
\item[--] mixing-layer height, $h$:
\begin{equation}
\label{eq:A_7}
h=x_2(\overline{\chi}_1={0.95})-x_2(\overline{\chi}_1={0.05})\; .
\end{equation}
%
%
\item[--] molecular mixing parameter, $\theta$:
\begin{equation}
\label{eq:A_8}
\theta=\frac{\int{\overline{\chi_1\ \chi_2}}\ d x_2}{\int{\overline{\chi}_1 \ \overline{\chi}_2\ d x_2}}\; .
\end{equation}
%
\item[--] averaged growth rate of the mixing-layer half-width, $\alpha_b$:
\begin{equation}
\label{eq:A_9}
\alpha_b= \frac{\dot{h}^2}{4 g h \mathrm{At} }\; .
\end{equation}
$\alpha_b$ is time averaged using the last ten time units of the simulation.
%
\item[--] Reynolds number, $\mathrm{Re}$:
\begin{equation}
\label{eq:A_10}
\mathrm{Re} = \frac{\dot{h}h}{\nu}\; ,
\end{equation}
where $\dot{h}$ is the time derivative of $h$.
\end{itemize}

\section{Grid refinement exercise at $f_k=0.00$}
\label{sec:B}

\begin{figure}[t!]
\centering
\subfloat[Grids $g_1$, $g_2$, and $g_3$.]{\label{fig:B_1a}
\includegraphics[scale=0.21,trim=0 0 0 0,clip]{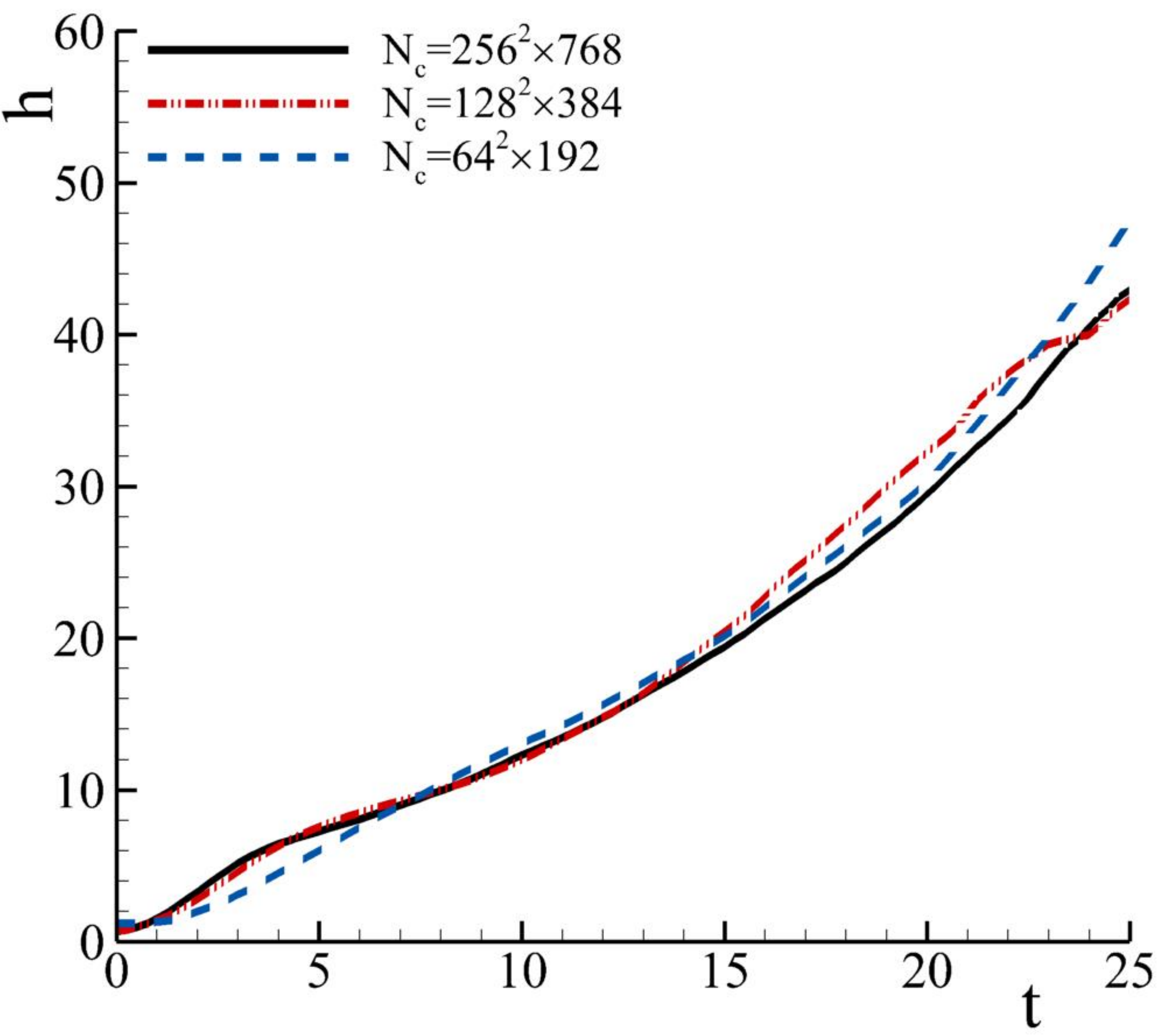}}
\\
\subfloat[Grids $g_0$ and $g_1$.]{\label{fig:B_1b}
\includegraphics[scale=0.21,trim=0 0 0 0,clip]{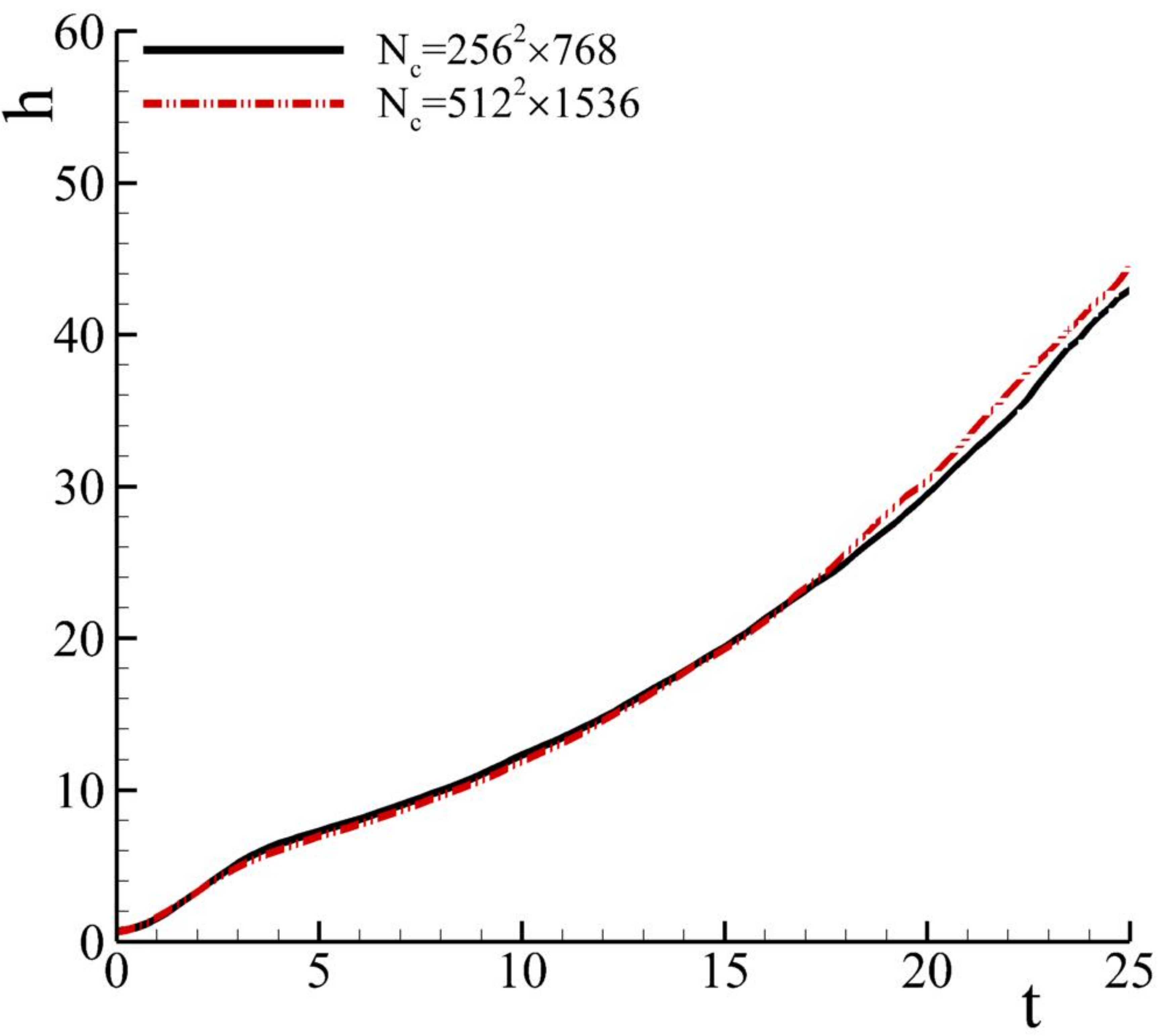}}
\caption{\color{blue}Temporal evolution of the mixing-layer height, $h$, predicted with $f_k=0.00$ and different spatio-temporal grid resolutions; (a) grids $g_1$, $g_2$, and $g_3$, (b) grids $g_0$ and $g_1$.}
\label{fig:B_1}
\end{figure}

To evaluate the effect of grid resolution on the simulations and guarantee that it does not affect the conclusions of the study, we performed grid refinement studies with $f_k=0.00$ and three mesh resolutions: {\color{blue}$64^2 \times 192$ ($g_3$), $128^2 \times 384$ ($g_2$), and $256^2 \times 768$ ($g_1$) cells. We also included results obtained on a finer grid with $512^2 \times 1536$ cells ($g_0$), CFL$=0.8$, $f_k=0.00$.} {\color{blue}Figure \ref{fig:B_1} presents the temporal evolution of the mixing-layer height, $h$, predicted with these four grids. It shows that the discrepancies between solutions using the two finest grid resolutions used in the PANS study ($g_2$ and $g_1$) are negligible until $t\approx 15$. After this time, the differences grow, but they are still relatively small. Note that there is an increase of two times} in resolution (time and space) between these two grids. In contrast, the coarsest grid resolution leads to more significant discrepancies and cannot predict the shape of $h(t)$. {\color{blue}Regarding the two finest grids, $g_1$ and $g_0$, these lead to nearly identical mixing-layer heights during almost the entire simulated time. Overall, this succinct analysis shows that the finest grid resolution used in the PANS computations ($g_1$)} is adequate for the objectives of this work.
%
%
%
\bibliography{references}
\end{document}